\documentclass[11pt]{article}
\usepackage{amsmath,amssymb,amscd,amsthm}
\makeatletter\@addtoreset {equation}{section}\makeatother

\newtheorem{theorem}{Theorem}
\newtheorem{lemma}{Lemma}
\newtheorem{remark}{Remark}

\newtheorem{corollary}{Corollary}

\newtheorem{definition}{Definition}
\newtheorem{proposition}{Proposition}
\def\R{\mathbb{R}}

\newenvironment{proof1}{
    \noindent {\it Proof }}{\hfill $\Box$}

\usepackage[dvips]{epsfig}
\usepackage{graphicx}

\begin{document}

\title{\bf Symmetry-breaking bifurcation in the nonlinear Schr\"{o}dinger equation with
symmetric potentials}

\author{E. Kirr$^1$, P.G. Kevrekidis$^2$,  and D.E. Pelinovsky$^3$  \\
{\small $^{1}$ Department of Mathematics, University of Illinois,
Urbana--Champaign, Urbana, IL 61801} \\
{\small $^{2}$ Department of Mathematics and Statistics,
University of Massachusetts, Amherst, MA 01003}\\
{\small $^{3}$ Department of Mathematics and Statistics, McMaster
University, Hamilton, Ontario, Canada, L8S 4K1} }

\maketitle

\begin{abstract}
We consider the focusing (attractive) nonlinear
Schr\"odinger (NLS) equation with an external, symmetric potential which vanishes at infinity and supports a linear bound state. We prove that the symmetric, nonlinear ground states must undergo a symmetry breaking bifurcation if the potential has a non-degenerate local maxima at zero. Under a generic assumption we
show that the bifurcation is either subcritical
or supercritical pitchfork. In the particular case of double-well potentials
with large separation, the power of nonlinearity determines the
subcritical or supercritical character of the bifurcation. The results
are obtained from a careful analysis of the spectral properties of the ground states at both
small and large values for the corresponding eigenvalue parameter.

We employ a novel technique combining concentration--compactness and spectral
properties of linearized Schr\"odinger type operators to show that the
symmetric ground states can either be uniquely continued for the entire interval
of the eigenvalue parameter or they undergo a symmetry--breaking pitchfork bifurcation due to the second
eigenvalue of the linearized operator crossing zero. In addition we prove
the appropriate scaling for the $L^q,\ 2\leq q\leq\infty$ and
$H^1$ norms of any stationary states in the limit of large values of the eigenvalue
parameter. The scaling and our novel technique imply that all ground states at
large eigenvalues must be localized near a critical point of the
potential and bifurcate from the soliton of the focusing NLS equation
without potential localized at the same point.

The theoretical results are illustrated numerically for a double-well potential
obtained after the splitting of a single-well potential. We compare the cases before and after the splitting, and numerically investigate bifurcation and stability properties of the ground states which are beyond the reach of our theoretical tools.
\end{abstract}

\section{Introduction}

Over the past few years, there has been a remarkable growth of
interest in the study of nonlinear Schr{\"o}dinger (NLS) equations
with external potentials. This has been fueled, to a considerable
extent, by the theoretical and experimental investigation of
Bose-Einstein condensates (BECs) \cite{book1,book2}. Localized
waveforms emerge within the atom-trapping potentials in such
ultracold systems \cite{ourbook}. Another major area of applications
for NLS equations is nonlinear optics, in particular photonic
crystals and optical waveguides \cite{kivshar,joan}.

One generic type of the external potential for the NLS equation that
has drawn considerable attention is the symmetric double-well
potential. This is due to its relative simplicity which often makes
it amenable to analytical considerations, but also due to the wealth
of phenomenology that even such a relatively simple system can
offer. Such potentials in the atomic physics setting of BECs have
already been experimental realized \cite{markus1} through the
combination of routinely available parabolic and periodic (optical
lattice) potentials. Among the interesting phenomena studied therein
were Josephson oscillations and tunneling for a small number of
atoms, or macroscopic quantum self-trapping and an asymmetric
partition of the atoms between the wells for sufficiently large
numbers of atoms. Double well potentials were also examined in the
context of nonlinear optics, e.g. in twin-core self-guided laser
beams in Kerr media \cite{HaeltermannPRL02}, optically induced
dual-core waveguiding structures in a photorefractive crystal
\cite{zhigang}, and trapped light beams in a structured annular core
of an optical fiber \cite{Longhi}.

In the present work, we address the NLS equation with a symmetric
potential as the prototypical mathematical model associated with the
above experimental settings. For simplicity we will focus on the case of one
space dimension. We write the equation in the normalized form
\begin{equation}\label{GP}
i u_t = -u_{xx} + V(x) u + \sigma |u|^{2p} u,
\end{equation}
where $u(x,t) : \R \times \R \to \mathbb{C}$ is the wave function,
$p > 0$ is the nonlinearity power, $\sigma \in \R$ determines the
defocusing (repulsive), respectively focusing (attractive),
character of the nonlinearity when $\sigma >0,$ respectively
$\sigma<0,$ and $V(x) : \mathbb{R}\mapsto\mathbb{R}$ is an external
real-valued, {\em symmetric} (even in $x$) potential satisfying:
\begin{itemize}
\item[(H1)] $V(x)\in L^{\infty}(\R),$
\item[(H2)] $\lim_{|x|\rightarrow\infty}V(x)=0,$
\item[(H3)] $V(-x)=V(x)$ for all $x\in\mathbb{R}.$
\end{itemize}
Hypothesis (H1) implies that $-\partial_x^2+V(x)$ is a self-adjoint operator on
$L^2(\R)$ with domain $H^2(\R).$ We will make the following spectral
assumption:
\begin{itemize}
\item[(H4)] $L_0=-\partial_x^2+V(x)$ has the lowest eigenvalue $-E_0<0.$
\end{itemize}
It is well known from the Sturm-Liouville theory that all eigenvalues of
$L_0$ are simple, the corresponding eigenfunctions can be chosen to be real valued and
the one corresponding
to the $k+1$-th eigenvalue has exactly $k$ zeroes, and, because of the symmetry (H3), is symmetric
(even in $x$) if $k$ is even and
anti-symmetric if $k$ is odd. We can choose a normalized eigenfunction, $\psi_0,$
corresponding to the eigenvalue $-E_0,$ which will satisfy:
\begin{equation}\label{linear-mode}
-\psi_0''(x) + V(x) \psi_0(x) + E_0 \psi_0(x) =
0, \quad \psi_0(x)>0,\ \psi_0(-x)=\psi_0(x),\ x \in \R,\ \|\psi_0\|_{L^2}=1.
\end{equation}

We are interested in understanding properties of stationary, symmetric and
asymmetric states of \eqref{GP}, i.e. solutions of the form $u(t,x)=e^{iEt}\phi(x),$ where $\phi$ satisfies the stationary NLS equation
\begin{equation}\label{stationary}
 -\phi''(x) + V(x) \phi(x) + \sigma |\phi(x)|^{2p} \phi(x)  + E \phi(x) = 0, \quad x \in \R,
\end{equation}
and $E \in \R$ is an arbitrary parameter. We recall the following
basic facts about solutions of the stationary NLS equation in one
dimension.

\begin{itemize}
\item[(i)] Via standard regularity theory, if $V(x)\in L^{\infty}(\R)$,
then any weak solution $\phi(x) \in H^1(\R)$ of the stationary
equation \eqref{stationary} belongs to $H^2(\mathbb R)$.

\item[(ii)] All solutions of the stationary equation \eqref{stationary}
in $H^2(\R) \hookrightarrow {\cal C}^1(\R)$ are real-valued up to
multiplication by $e^{i \theta}$, $\theta \in \R$.

\item[(iii)] If $E>0$, all solutions of \eqref{stationary}
in $H^2(\R)$ decay exponentially fast to zero as $|x| \to \infty$.
\end{itemize}

Numerically we will focus on a one parameter double well potential $V\equiv V_s$ constructed from splitting the single-well potential $V_0(x) = -{\rm sech}^2(x):$
\begin{equation}
\label{double-potential} V_s(x) = V_0(x + s) +
V_0(-x+s), \quad s \geq 0.
\end{equation}

The general theory of bifurcations from a simple eigenvalue of the
linearized operator \cite{Nir} implies that solutions with small $H^2$ norm of the
stationary equation (\ref{stationary}) exist for $E$ near $E_0.$ The symmetry hypothesis (H3) implies that these solutions are symmetric (even in $x$).
Variants of the local bifurcation analysis near $E = E_0,$ including
the fact that $E > E_0$ if $\sigma<0,$ and $E < E_0$ if $\sigma>0$,
have already appeared in \cite{sw:mc1,PW}, as well as in many recent
publications. We review this analysis in Section 2 to give readers a
complete picture.

Orbital stability \cite{Weinstein} of the stationary state $e^{iEt}\phi(x),\ \phi(x)\in\R ,$
is closely related to the linearization of the time-dependent NLS equation \eqref{GP} at the stationary state, which, in the direction $e^{iEt}[u_1(x,t)+ i u_2(x,t)],$ is given by:
$$
\partial_t\left[\begin{array}{c} u_1 \\ u_2 \end{array}\right] =
\left[\begin{array}{cc} 0 & L_-(\phi,E) \\ -L_+(\phi,E) & 0 \end{array}\right]
\left[\begin{array}{c} u_1 \\ u_2 \end{array}\right],
$$
where $L_\pm$ are self adjoint linear Schr\"{o}dinger operators with domains
$H^2(\R)\subset L^2(\R):$
\begin{eqnarray}
\label{operators} \left\{ \begin{array}{l}
L_+(\phi,E) = -\partial^2_x + E + V(x) + \sigma (2p+1) |\phi |^{2p}(x), \\
L_-(\phi,E) = -\partial^2_x + E + V(x) + \sigma |\phi |^{2p}(x).
\end{array} \right.
\end{eqnarray}
Sufficient conditions for orbital stability and orbital instability,
which we will use throughout this paper, were obtained in
\cite{Weinstein,GSS,Gr}.
\begin{definition} If $(\phi,E)$ solves \eqref{stationary} and zero is the lowest eigenvalue of $L_-(\phi,E)$ we call $\phi$ a ground state of \eqref{stationary}.
\end{definition}
\begin{remark} Note that, for any solution $(\phi,E)$ of \eqref{stationary}, zero is an eigenvalue of $L_-(\phi,E)$ with eigenfunction $\phi.$ Via standard theory of second order elliptic operators the above definition is equivalent to the one requiring a ground state to be strictly positive or strictly negative.
\end{remark}
In particular it is known that solutions $(\phi,E)$ of
the stationary NLS equation \eqref{stationary} with $|E-E_0|$
and $\|\phi\|_{H^1}$ small are orbitally stable ground states, see Section 2. We remark
as a side note that, for critical and supercritical nonlinearities, $p \geq 2,$ and more restrictive
hypotheses on the potential $V(x),$ one can show
asymptotic stability of these solutions in the space of one dimension
\cite{Cuccagna,Mizumachi}. For subcritical nonlinearities asymptotic stability is proven only in dimensions higher than one, see \cite{kz:as2d1,km:as3d, km:as45d}.

Kirr {\em et al.} \cite{Kirr} showed that the symmetric ground states undergo a symmetry--breaking
bifurcation at $E = E_* > E_0,$ in the focusing case $\sigma<0,$
with $p=1,$ or other cubic like nonlinearities, provided the first two
eigenvalues of $L_0 = -\partial_x^2 + V(x)$ are sufficiently close to
each other. In particular, the result is applicable to
double-well potentials such as \eqref{double-potential} for
sufficiently large separation parameter $s$ between the two wells.
Furthermore the authors show that the symmetric states become
unstable for $E > E_*,$ and, a new pair of orbitally stable, asymmetric ground states exist
for $E > E_*.$ The proofs rely on a
Lyapunov-Schmidt type projection onto the two eigenvectors
corresponding to the lowest eigenvalues of $L_0,$ which exists for small
$\|\phi\|_{H^1}$, combined with a normal form analysis of the
reduced system valid to all orders. Marzuola \& Weinstein \cite{MW10}
used a time dependent
normal form valid for finite time to extract interesting
properties of the dynamics of solutions of the NLS equation \eqref{GP} near the
bifurcation point $E=E_*.$ We also mention that \cite{Sacchetti2}
uses a similar finite time, normal form technique to study the
solutions and predict bifurcations of the first excited
(anti-symmetric) state for defocusing NLS equation ($\sigma > 0$) with a
symmetric double well potential which is essentially brought in the
large separation regime by passing to the
semi-classical limit and assuming $\sigma\searrow 0$ at a specific
rate.

To our knowledge there are very few results for
bifurcations of NLS stationary states in non-perturbative
regimes. Rose \& Weinstein \cite{RW:88} use variational methods to show
that the stationary NLS equation \eqref{stationary} with $\sigma <0,$ and
potential satisfying (H1), (H2) and (H4), has at least one solution for any $E>E_0.$
Jeanjean \& Stuart \cite{js:ubs} prove that for $\sigma<0,$ the symmetric states
bifurcating from the lowest eigenvalue of $L_0$ can be
uniquely continued for all $E>E_0,$ hence there are no bifurcations along this
branch, provided $V(x)$ is monotonically increasing for $x>0,$
and $C^1$ in addition to satisfying (H1)-(H4). In particular, the
result applies to the potential \eqref{double-potential} if $s \leq s_*$, where
\begin{equation}
\label{critical-s}
s_*=\frac{1}{2}{\rm arccosh}(2)={\rm arccosh}(\sqrt{3}/\sqrt{2})\approx 0.6585,
\end{equation}
because for $x>0$ and $0\leq s \leq s_*:$
$$
V_s'(x) = V_0'(x-s)+V_0'(x+s)=\frac{\sinh(2x)}{\cosh^3(x-s)\cosh^3(x+s)}[2+\cosh(2x)\cosh(2s)-\cosh^2(2s)] > 0.
$$
Results on continuation of branches of stationary states in the defocusing case
$\sigma>0,$ but without reference to existence or non-existence of bifurcation points can be
found in \cite{jls:bsd,jls:gsd}. In \cite{AFGST:02} the authors
rely on variational techniques to deduce symmetry--breaking of the ground
states in Hartree equations. Their method can be adapted to our problem
and implies the emergence of asymmetric, ground state branches in the focusing case $\sigma <0$ provided the nonlinearity is subcritical, $p<2,$ and $V(x)$ is continuous, bounded, and has at least two
separated minima. In particular, assuming $p<2,$ asymmetric ground state branches will appear for the
potential \eqref{double-potential} as soon as it becomes a double well,
i.e., for $s>s_*,$ but the method cannot tell whether the asymmetric branches are connected to the symmetric branch of ground states bifurcating from the lowest eigenvalue of $L_0.$ Jackson \& Weinstein \cite{JW:bifdp} use a topological shooting method for the case $p=1$ and Dirac type double-well potential, i.e. $V_0(x)=\delta(x)$ in \eqref{double-potential}, to show that the asymmetric branches emerge from the symmetric ones via a pitchfork bifurcation and they all coexists past a certain value of $E.$

Our main result extends the ones in \cite{Kirr} to non-perturbative
regimes and the ones in \cite{AFGST:02} to critical and supercritical
nonlinearities, $p\geq 2,$ while proving that the asymmetric ground states emerge from the symmetric ones via a pitchfork bifurcation. The main theorem is formulated as follows.

\begin{theorem}\label{th:main}
Consider the stationary NLS equation \eqref{stationary} in the focusing case $\sigma <0,$ with
$V(x)$ satisfying {\rm (H1)-(H4).} Then the $C^1$ curve $E\mapsto\phi=\psi_E\in H^2,\ E>E_0,\ \psi_E(x)>0$ of
symmetric, real valued solutions bifurcating from the zero solution at
$E=E_0,$ undergoes another bifurcation at a finite $E=E_*>E_0,\ \psi_{E_*}\in H^2,$
provided $V(x)$ has a non-degenerate maxima at $x=0,$ and
 \begin{itemize}
 \item[{\rm (H5)}] $x V'(x) \in L^\infty(\R).$
 \end{itemize}
The bifurcation is due to the second eigenvalue $\lambda(E)$ of
$L_+(\psi_E,E)$ crossing zero at $E=E_*.$ Moreover, if $p\geq 1/2,$ and the following
non-degeneracy condition holds:
$$\frac{d\lambda}{dE}(E_*)\not=0,$$
then the bifurcation is of pitchfork type: the set of real valued solutions
$(\phi,E) \in H^2\times\R$ in a neighborhood of $(\psi_{E_*},E_*)$ consists of exactly two orthogonal $C^1$ curves:
the symmetric branch $E\mapsto\phi=\psi_E$ which continues past $E=E_*,$ but becomes orbitally unstable, and an
asymmetric branch $(\phi(a),E(a)):$
$$E(a)=E_*+\frac{Q}{2}a^2+o(a^2),\ \phi(a)=\psi_{E_*}+a\phi_*+O(a^2),\ a\in\R,\ |a|\ {\rm small} $$
where $\phi_*$ is the eigenfunction corresponding to the second eigenvalue of $L_+(\psi_{E_*},E_*),$ and
$Q$ can be calculated from $\psi_{E_*}$ and $\phi_*,$ see \eqref{defQ}.
The asymmetric solutions are orbitally stable if
$Q>0$ and $\|\phi(a)\|_{L^2}$ is increasing as
$E(a)$ increases, but they are orbitally unstable if $\|\phi(a)\|_{L^2}$ is decreasing
with $E(a)$ or if $Q<0.$
\end{theorem}

In particular, the result applies to the potential
\eqref{double-potential} for $s>s_*={\rm arccosh}(\sqrt{3}/\sqrt{2})$ because
$$V_s'(0)=V_0'(s)-V_0'(s)=0,\ V_s''(0) = 2V_0''(s) = 12 {\rm sech}^4(s) - 8 {\rm sech}^2(s) < 0, \quad s > s_*, $$
and implies that a pitchfork bifurcation occurs along the branch of
symmetric states. Recall that for $s\leq s_* $ this branch can be uniquely continued for
all $E>E_0$ due to the result in \cite{js:ubs}, see \eqref{critical-s}. Moreover, in the
large separation limit $s\rightarrow\infty,$ the branch of asymmetric states
is orbitally stable near the pitchfork bifurcation if
$p < p_*$, where
\begin{equation}
\label{critical-p} p_*= \frac{1}{2} \left( 3+\sqrt{13}\right)
\approx 3.3028,
\end{equation}
and orbitally unstable for $p > p_*,$ see Corollary \ref{cor:dwbif}. The threshold power $p_*$ of the nonlinearity was
predicted in \cite{Sacchetti2} but we justify this result with rigorous analysis.

We emphasize that hypotheses (H1) and (H5) can be relaxed to $V(x), xV'(x) \in L^q(\R)$
for some $q\geq 1$, at the expense of slightly complicating the proofs in this paper. Moreover, our results
extend to more than one dimension $x\in\R^n,\ n\geq 2,$ and other symmetries in $\R^n,$ provided that the
symmetries still prevent the solutions to concentrate at $|x|=\infty,$ see Remark \ref{rm:main1}. Note that for classifying the
bifurcation, we will have to assume that the second eigenvalue of $L_+(\psi_E,E)$ is simple. To completely remove any symmetry
assumptions, or the spectral assumption (H4), or the simplicity of the second eigenvalue of $L_+,$ is
a much more difficult task, see \cite{kn:bif} for partial results.

The proof of the main result relies on Theorems \ref{cor:max},
\ref{th:comp}, \ref{th:brelarge}, and \ref{th:bfelarge} which, viewed individually, are important themselves. Properly
generalized they could completely describe the set of all solutions of
the stationary NLS equation \eqref{stationary} for any dimension,
arbitrary potentials and more general nonlinearities in terms of the
critical points of the potential. In Section \ref{se:small} we prove
the following dichotomy: the branch of stationary solutions
$(\psi_E,E)$ bifurcating from the lowest eigenvalue $-E_0$ of
$-\partial_x^2+V(x)$ can either (a) be uniquely continued for all
$E>E_0$ or (b) there exists a finite $E_*>E_0$ such that zero is an
accumulation point for the discrete spectrum of the linearized
operator $L_+(\psi_E,E),$ as $E\nearrow E_*$. The result essentially
eliminates the possibility that $\|\psi_E\|_{H^2}$ diverges to
infinity as $E$ approaches $E_*$ with
$\|L_+^{-1}(\psi_E,E)\|_{L^2\mapsto L^2}$ remaining uniformly bounded,
and relies on the differential estimates for the mass (charge) and energy of the NLS equation \eqref{GP}.

In Section \ref{se:bif} we use a novel technique combining concentration--compactness, see for example \cite{caz:bk}, and the spectral properties of the linearized operator $L_+,$ to show that, in case (b), the states $\psi_E$ must converge in $H^2$ to a nonzero state $\psi_{E_*}$ for $E=E_*.$ By continuity we deduce that the linearized operator $L_+(\psi_{E_*},E_*)$ has zero as a simple eigenvalue,
then we use a Lyapunov-Schmidt decomposition and the Morse Lemma, see for example \cite{Nir}, to show that a pitchfork bifurcation occurs at
$(\psi_{E_*},E_*).$ The symmetry hypothesis (H3) implies that $\psi_E(x)$ are even in $x$ which is essential in showing that the limit $\lim_{E\nearrow E_*}\psi_E$ exists. It turns out that without assuming (H3) $\psi_E$ may drift to infinity as $E$ approaches $E_*,$ i.e. there exists $y_E\in\R,\ \lim_{E\nearrow E_*}y_E=\pm\infty,$ such that $\lim_{E\nearrow E_*}\psi_E(\cdot-y_E)=\psi_{E_*}$ where $\psi_{E_*}$ is now a solution of
the stationary NLS equation \eqref{stationary} with $V(x)\equiv 0$ and $E=E_*,$ see \cite{kn:bif} for a more detailed discussion of this phenomenon.

In Section \ref{se:large}, we obtain new, rigorous results on the behavior of all stationary states in the focusing case $\sigma <0,$ for large $E.$ In Theorem \ref{th:brelarge} we combine Pohozaev type identities with differential estimates for mass, energy and the $L^{2p+2}$ norm of the stationary solutions to prove how the relevant norms of the solutions scale with $E$ as $E \to \infty$.
In particular, we obtain the behavior of the $L^2$ norm of the symmetric states, which was numerically and heuristically predicted in \cite{RW:88}. Moreover, by combining these estimates with our novel concentration--compactness/spectral technique we show that, modulo a re-scaling, the symmetric branches of solutions, along which $L_+$ has only one negative eigenvalue, converge to a non-trivial solution of the constant-coefficient
stationary NLS equation,
\begin{equation}\label{nlss}
-\phi''(x) +\sigma |\phi(x)|^{2p} \phi(x)  + \phi(x) = 0, \quad \phi\in H^2(\R).
\end{equation}
Since the set of solutions of the latter in dimension one is well known, we adapt and extend the bifurcation analysis of Floer \& Weinstein \cite{FW} to our problem and  obtain detailed information on {\em all} stationary solutions, which, modulo a re-scaling, bifurcate from a non-trivial solution of equation \eqref{nlss}. We show that such solutions can only be localized near a critical point of the potential $V(x),$ and they are always orbitally unstable for supercritical nonlinearities, $p > 2.$ If the critical point is a non-degenerate minimum, respectively maximum, then there is exactly one branch of solutions localized at that point, and these solutions are orbitally stable if and only if we are in the critical and subcritical regimes, $p\leq 2$. We note that compared to the semi-classical analysis in \cite{FW}, completed with orbital stability analysis in \cite[Example C]{GSS}, we are forced to make a precise analysis up to order four, instead of two, in the relevant small parameter. In addition, we prove non-existence of solutions localized near regular points of the potential $V(x)$, uniqueness of solutions localized near non-degenerate minima and maxima, and we recover the stability for the critical nonlinearity $p=2.$ All these results can now be extended without modifications to the problem studied in \cite{FW}.

In Section \ref{se:num}, we illustrate the main theoretical results numerically for the potential \eqref{double-potential}, subcritical nonlinearity $p = 1$ and
supercritical nonlinearities $p = 3,\ p = 5$ in the focusing case $\sigma < 0$. We note that $p = 3 < p_*$
and $p = 5 > p_*$, where $p_*$ is defined by (\ref{critical-p}).
We will show that both subcritical and supercritical pitchfork
bifurcations occur depending on the value of $p$. For this potential we will also show numerically that, except for the bifurcation predicted by our main result, there are no other bifurcations along any of the ground state branches, a result beyond the grasp of our current theoretical techniques.

In what follows, we shall use notations ${\cal
O}(\varepsilon)$ and $o(\varepsilon)$ as $\varepsilon \to 0$ in the
sense
$$
\Lambda = {\cal O}(\varepsilon) \quad \Leftrightarrow   \quad
\lim_{\varepsilon \to 0} \varepsilon^{-1} \Lambda(\varepsilon) =
\Lambda_{\infty} \in \R \quad
\mbox{\rm and} \quad
\Lambda = o(\varepsilon) \quad \Leftrightarrow \quad
\lim_{\varepsilon \to 0} \varepsilon^{-1} \Lambda(\varepsilon) = 0.
$$
We will denote $\varepsilon$-independent constants by $C$, which may
change from one line to another line. We will also use  the standard
Hilbert space $L^2(\mathbb R)$ of the real valued square integrable
functions on a real line and the Sobolev space $H^2(\mathbb
R)\subset L^2(\mathbb{R})$ of the real valued functions on $\mathbb
R$ which are square integrable together with their first and second
order weak derivatives.

{\bf Acknowledgments.}
PGK is partially supported by NSF-DMS-0349023
(CAREER), NSF-DMS-0806762 and the Alexander-von-Humboldt Foundation.
EWK is partially supported by NSF-DMS-0707800.
DEP is partially supported by the NSERC. The authors are
grateful to V. Natarajan and M.I. Weinstein for fruitful discussions, as well as
to C. Wang for assistance with some of the numerical computations.

\section{Local bifurcations of symmetric ground states}\label{se:small}

In this section we trace the manifold of symmetric ground states of the stationary problem
(\ref{stationary}) from its local bifurcation from the linear
eigenmode of $L_0 = -\partial_x^2 + V(x)$ near $E = E_0$ up to its
next bifurcation. We will show that the symmetric state exists in
an interval to the right of $ E_0$ if $\sigma < 0$ and in an
interval to the left of $E_0$ if $\sigma > 0$. In the case of
$\sigma < 0$, we will further find necessary and sufficient
conditions for the symmetric state to be extended for all values
of $E>E_0$ or suffer a symmetry--breaking bifurcation.

Let us rewrite the stationary equation \eqref{stationary} for
real-valued solutions $\phi(x)$ as the root-finding equation for the
functional $F(\phi,E) : H^2(\R) \times \mathbb R\mapsto L^2(\R)$
given by
\begin{equation}\label{def:F}
F(\phi,E) = (- \partial_x^2 + V(x) + E)\phi + \sigma |\phi
|^{2p}\phi.
\end{equation}
We recall the following result describing the existence of
symmetric ground states near $E_0$.

\begin{proposition}\label{th:ex} Let $-E_0<0$ be the smallest eigenvalue of
$L_0 = -\partial_x^2 + V(x)$. There exist $\epsilon>0$ and $\delta > 0$
such that for each $E$ on the interval ${\cal I}_{\epsilon},$
${\cal I}_{\epsilon} = (E_0-\epsilon,E_0)$ for $\sigma > 0$,
${\cal I}_{\epsilon} = (E_0,E_0+\epsilon)$ for $\sigma < 0$, the
stationary equation \eqref{stationary} has exactly two nonzero, real
valued solutions $\pm\psi_E(x)\in H^2(\R),$ satisfying $\|\psi_E\|_{H^2}<\delta.$ Moreover
$$
\|\psi_E \|_{H^2} \leq C |E - E_0|^{\frac{1}{2p}}.
$$
for some $C>0,$ the map $E \mapsto \psi_E$ is $C^1$ from ${\cal
I}_{\epsilon}$ to $H^2$ and $\psi_E(x)=\psi_E(-x)$ for each
$x\in\mathbb{R}$ and $E\in{\cal I}_{\epsilon}.$
\end{proposition}

\begin{proof}
We sketch the main steps. For any $p
> 0$, the functional $F : H^2(\R) \times \R \mapsto L^2(\R)$ is $C^1$ ,
i.e. it is continuous with continuous Frechet derivative
$$
D_{\phi} F(\phi,E) = -\partial_x^2 + V(x) + E + (2p+1) \sigma
|\phi|^{2p}.
$$
We note that $D_{\phi} F(0,E) = L_0 + E$. Let $\psi_0$ be the
$L^2$-normalized eigenfunction of $D_{\phi} F(0,E) = L_0 + E_0$
corresponding to its zero eigenvalue. Then $D_\phi F(0,E_0)$ is a
Fredholm operator of index zero with ${\rm Ker}(L_0 + E_0) = {\rm
span}\{ \psi_0 \}$ and ${\rm Ran}(L_0 + E_0) = \left[ {\rm Ker}(L_0
+ E_0) \right]^{\perp}$. Let
$$
P_\parallel \phi = \langle \psi_0,\phi\rangle_{L^2} \psi_0,\quad
P_\perp \phi = \phi - \langle \psi_0,\phi\rangle_{L^2} \psi_0
$$
be the two orthogonal projections associated to this
Lyapunov-Schmidt decomposition. Then $F(\phi,E)=0$ with $\phi = a
\psi_0 + \varphi$, where $a = \langle \psi_0, \phi \rangle$ and
$\varphi = P_{\perp} \phi$, is equivalent to two equations
\begin{eqnarray}
P_\perp (L_0 + E) P_{\perp} \varphi + \sigma P_{\perp} |a \psi_0 + \varphi|^{2p}(a \psi_0 + \varphi ) &=&0 \label{eq:perp}\\
(E - E_0) a + \sigma \langle \psi_0,|a \psi_0 + \varphi|^{2p}(a
\psi_0 + \varphi )\rangle &=&0\label{eq:parallel}
\end{eqnarray}
Since $P_\perp (L_0 + E) P_{\perp}$ has a bounded inverse for any
$E$ near $E_0$, the Implicit Function Theorem states that there
exists a unique $C^1$ map $\R^2 \ni (a,E) \mapsto \varphi =
\varphi_*(a,E) \in H^2$ for sufficiently small $|a|$ and $|E-E_0|$
such that $\varphi$ solves equation (\ref{eq:perp}) and
$$
\exists C > 0 : \quad \| \varphi_*(a,E) \|_{H^2} \leq C |a|^{2p+1}.
$$
Hence \eqref{eq:parallel} becomes a scalar equation in variables
$(a,E) \in \R^2$ given by
\begin{equation}
\label{parallel-again} a (E-E_0) + \sigma \langle \psi_0,|a \psi_0 +
\varphi|^{2p}(a \psi_0 + \varphi ) \rangle_{L^2} = 0.
\end{equation}
Dividing (\ref{parallel-again}) by $a$ and invoking again the
Implicit Function Theorem for functions, we obtain the existence of
a unique continuous map $\R \ni a \mapsto E = E_*(a) \in \R$ for
sufficiently small $|a|$ such that $E$ solves (\ref{parallel-again})
and
$$
\exists C > 0 : \quad | E_*(a) - E_0 + \sigma \| \psi_0
\|_{L^{2p+2}}^{2p+2} |a|^{2p} | \leq C |a|^{4p}.
$$
Therefore, $E > E_0$ if $\sigma < 0$ and $E < E_0$ if $\sigma >
0$. Moreover, the map $a\mapsto E_*(a)$ is $C^1$ and invertible
for $a>0,$ rendering the map $E \mapsto \psi_E :=
a\psi_0+\varphi_*(a,E_*(a))$ to be $C^1.$ The negative branch
$-\psi_E$ is obtained from the negative values of $a.$ Moreover,
since for any solution $\psi_E(x)$ of \eqref{stationary} we have
that $\psi_E(-x)$ is also a solution, uniqueness and continuity in
$a$ imply that $\psi_E(-x)=\psi_E(x).$
\end{proof}

\begin{remark}\label{smallsol} A
result similar to Proposition \ref{th:ex} can be obtained for the
anti-symmetric state of the stationary equation (\ref{stationary}),
which bifurcates from the second eigenvalue $-E_1$ of $L_0 =
-\partial_x^2 + V(x)$, if the second eigenvalue of $L_0$ exists.
Moreover, $(0,E) \in H^2(\R) \times \R$ is the only solution of the
stationary NLS equation \eqref{stationary} in a small neighborhood
of $(0,E_*) \in H^2(\R) \times \R$ for any $E_*>0$, if $-E_*$ is not
an eigenvalue of $L_0.$
\end{remark}

Let us introduce operators $L_+$ and $L_-$ along the branch of
symmetric states $(\psi_E, E)$ according to definition
(\ref{operators}) with $\phi = \psi_E$. Since they depend $C^1$ on
$E\in{\cal I}_{\epsilon}$ and continuously at $E_0,$ their
isolated eigenvalues depend $C^1$ on $E\in{\cal I}_{\epsilon}$ and
continuously at $E_0$.

Since $L_- \psi_E = 0$ and $L_- = L_0 + E_0$ for $E = E_0$, $0$ is
the lowest eigenvalue of $L_-$ for all $E\in{\cal I}_{\epsilon}$.
On the other hand, we have
$$
L_+-L_- = \sigma 2p\psi_E^{2p}.
$$
Hence, $L_+<L_-$ if $\sigma <0$, while $L_+>L_-$ if $\sigma >0,$
and, via eigenvalue comparison principle, the lowest eigenvalue of
$L_+$ is strictly negative if $\sigma <0$ and strictly positive if
$\sigma >0.$ Consequently, $0$ is not in the discrete spectrum of
$L_+$ nor in the essential spectrum for $E \in {\cal
I}_{\epsilon}$. The latter follows from $V(x)\in L^\infty(\R)$,
$\lim_{|x|\rightarrow\infty}V(x)=0,$ and $|\psi_E|^{2p}\in
L^2(\R),$ since $\psi_E\in H^2(\R)\hookrightarrow L^q(\R)$ for any
$2\leq q\leq\infty.$ Together they imply that $L_+-E$ is a
relatively compact perturbation of $-\partial^2_x,$ hence, via
Weyl's theorem, the essential spectrum of $L_+$ is the
$[E,\infty)$ interval.

The following result shows that we can continue the branch of
symmetric states $(\psi_E, E)$ as long as $0$ is not in the
spectrum of $L_+.$

\begin{lemma}\label{th:cont}
Let $\psi_{E_1}(x) \in H^2(\R)$ be a real valued solution of the
stationary equation \eqref{stationary} for $E = E_1$ and assume $0$
is not in the spectrum of $L_+(\psi_{E_1},E_1).$ Then there exist $\epsilon>0$ and
$\delta > 0$ such that for each $E\in (E_1-\epsilon,E_1+\epsilon)$ the
stationary equation \eqref{stationary}  has a unique, real valued,
nonzero solution $\psi_E(x) \in H^2(\R)$ satisfying
$$
\| \psi_E -\psi_{E_1}\|_{H^2} \leq \delta.
$$
Moreover the map $E \mapsto \psi_E$ is $C^1$ from
$(E_1-\epsilon,E_1+\epsilon)$ to $H^2$.
\end{lemma}

\begin{proof}
The result follows from the Implicit Function Theorem for $F(\phi,E)
= 0$ at $(\psi_{E_1},E_1)$.
\end{proof}

Combining the Proposition and Lemma, we get the following maximal
result for the branch of symmetric modes bifurcating from the point
$E_0$. For definiteness, we state and prove this result only for the
focusing case $\sigma < 0$.

\begin{theorem}\label{cor:max} If $\sigma <0,$
then the branch of solutions $(\psi_E,E)$ of {\em\eqref{stationary}}
which bifurcates from the lowest eigenvalue $-E_0$ of $L_0$ can be
uniquely continued to a maximal interval $(E_0,E_*)$ such that
either:
 \begin{itemize}
 \item[(a)] $E_*=\infty;$
 \item[]or
 \item[(b)] $E_* < \infty$ and there exists a sequence $\{E_n\}_{n\in\mathbb N}\subset (E_0,
E_*)$ such that $\lim_{n\rightarrow\infty} E_n = E_*$ and $L_+(\psi_{E_n},E_n)$ has an eigenvalue $\lambda_n$ satisfying
$\lim_{n\rightarrow\infty} \lambda_n = 0$.
 \end{itemize}
\end{theorem}

\begin{proof} Define:
 \begin{eqnarray}
 E_*&=&\sup\{\tilde E : \quad \tilde E>E_0,\ E\mapsto\psi_E\ \mbox{is a}\ C^1\
 \mbox{extension on}\ (E_0,\tilde E)\ \mbox{ of the map in Proposition \ref{th:ex}},\nonumber\\
   &&\mbox{for which}\ 0\ \mbox{is not in the spectrum of } L_+ \}\nonumber
 \end{eqnarray}
Proposition \ref{th:ex} and the discussion following it guarantees
that the set above is not empty. Assume neither (a) nor (b) hold for
$E_*>E_0$ defined above. Then we can fix any $E_1,\ E_0<E_1<E_*$ and
find that, for $E\in [E_1,E_*),$ the spectrum of $L_+(\psi_E,E)$
(which is real valued since $L_+$ is self-adjoint) has no points in
the interval $[-d,d]$ for some $0 < d < E_1$. Indeed, as discussed
after Proposition \ref{th:ex} the essential spectrum of $L_+$ at $E$
is $[E,\infty),$ and if no $d > 0$ exists, there must be a sequence
of eigenvalues $\lambda_n$ for $L_+$ at $E_n\in [E_1,E_*)$ such that
$\lim_{n\rightarrow\infty}\lambda_n=0.$ But $[E_1,E_*]$ is compact
since (a) does not hold, hence there exists a subsequence $E_{n_k}$
of $E_n$ converging to $E_2\in [E_1,E_*]$. However, $E_2\not=E_*$
because we assumed that (b) does not hold, so $E_2\in [E_1,E_*)$ and
by continuous dependence of the eigenvalues of $L_+$ on $E\in
[E_1,E_*)$ we get that $0$ is an eigenvalue of $L_+$ at $E_2<E_*$
which contradicts the choice of $E_*.$

Consequently $L_+^{-1}:L^2(\R)\mapsto L^2(\R)$ is bounded with
uniform bound $K=1/d$ on $[E_1,E_*).$ Moreover, by differentiating
\eqref{stationary} with respect to $E$ we have:
\begin{equation}\label{eq:dpsie}
L_+ \partial_E\psi_E = -\psi_E \quad \Rightarrow \quad
\partial_E\psi_E= - L_+^{-1}\psi_E, \quad E \in (E_0,E_*),
\end{equation}
hence
\begin{equation}\label{estdpsie}
\| \partial_E \psi_E \|_{L^2} \le K \|\psi_E\|_{L^2}, \quad E \in
[E_1,E_*).
\end{equation}
and, by Cauchy-Schwarz inequality:
$$
\frac{d}{dE}\|\psi_E\|^2_{L^2}=2\langle \partial_E \psi_E
,\psi_E\rangle\leq 2K\|\psi_E\|^2_{L^2}, \quad E\in [E_1,E_*).
$$
The latter implies
\begin{equation}
\label{another-bound} \|\psi_E\|^2_{L^2}\leq
\|\psi_{E_1}\|^2_{L^2}e^{2K(E-E_1)}, \quad E\in [E_1,E_*),
\end{equation}
which combined with $E_*<\infty$ and bound \eqref{estdpsie} gives
that both $\partial_E \psi_E$ and $\psi_E$ have uniformly bounded
$L^2$ norms on $[E_1,E_*)$. By the Mean Value Theorem there exists
$\psi_{E_*}(x) \in L^2(\R)$ such that
\begin{equation}\label{eq:l2conv}
\lim_{E\nearrow E_*}\|\psi_E-\psi_{E_*}\|_{L^2}=0.
\end{equation}

We claim that $\psi_{E_*}(\R) \in H^1(\R)$ is a weak solution of the
stationary equation \eqref{stationary} with $E=E_*$, hence it is in
$H^2(\mathbb R)$ since $V \in L^{\infty}(\R)$. Indeed, consider the
energy functional
\begin{equation}\label{def:energy}
{\cal E}(E)=\int_\mathbb{R}| \nabla\psi_E(x)|^2dx+
\int_{\mathbb{R}}V(x)|\psi_E
(x)|^2dx+\frac{\sigma}{p+1}\int_\mathbb{R}|\psi_E (x)|^{2p+2}dx.
\end{equation}
Note that because, $\psi_E$ is a weak solution of the stationary
equation \eqref{stationary} for any $E\in (E_0,E_*)$ we have
\begin{equation}\label{eq:denergy}
\frac{d{\cal E}}{dE}=-2E \langle\psi_E,\partial_E \psi_E
\rangle_{L^2}
\end{equation}
and, via Cauchy-Schwarz inequality:
$$
\left| \frac{d{\cal E}}{dE} \right| \le E \|\psi_E\|_{L^2}\|
\partial_E \psi_E \|_{L^2}.
$$
Hence, from the uniform bounds (\ref{estdpsie}) and
(\ref{another-bound}), the derivative of ${\cal E}(E)$, and hence
${\cal E}(E)$, is uniformly bounded on $[E_1,E_*)$. On the other
hand, from the weak formulation of solutions of
\eqref{stationary}, we get
\begin{equation}\label{eq:spstat}
\| \nabla\psi_E \|_{L^2}^2+ \int_{\mathbb
R}V(x)|\psi_E(x)|^2dx+\sigma\|\psi_E\|_{L^{2p+2}}^{2p+2}+
E\|\psi_E\|_{L^2}^2=0
\end{equation}
Subtracting the latter from $(p+1){\cal E}$ we get that there exists
an $M>0$ such that
\begin{equation}\label{eq:h1bounds}
\left|\ p\| \nabla\psi_E \|_{L^2}^2+p \int_{\mathbb
R}V(x)|\psi_E(x)|^2dx\ \right|\le M,\qquad {\rm for\ all}\ E\in
[E_1,E_*).
\end{equation}
From H\" older inequality we obtain:
$$
\left|\int_{\mathbb R}V(x)|\psi_E(x)|^2dx\right|\le\|V\|_{L^\infty}
\|\psi_E\|_{L^{2}}.
$$
Using the inequality in \eqref{eq:h1bounds} we deduce that $\|\nabla
\psi_E \|_{L^2}$ has to be uniformly bounded. Consequently, there
exists $M>0$ such that
\begin{equation}\label{eq:h1b}
\|\psi_E\|_{H^1}\le M,\qquad {\rm for\ all}\ E\in [E_1,E_*).
\end{equation}
Because of the embedding of $H^1(\R)$ into $L^{\infty}(\R)$ and the
interpolation
$$
\| f \|_{L^q} \leq \| f \|^{2/q}_{L^2} \| f \|^{1 -
2/q}_{L^{\infty}}, \quad q \geq 2,
$$
bound \eqref{eq:h1b} together with convergence \eqref{eq:l2conv}
imply that as $E\nearrow E_*$ we have:
$$
\left. \begin{array}{l} \psi_E \rightarrow \psi_{E_*},\quad {\rm in}\ L^2(\R)\\
\psi_E\rightharpoonup\psi_{E_*},\quad {\rm in}\ H^1(\R) \end{array}
\right\} \quad \Rightarrow \quad \psi_E\rightarrow\psi_{E_*},\quad
{\rm in}\ L^q(\R),\ q\ge 2.
$$

Now, by passing to the limit in the weak formulation of the
stationary equation \eqref{stationary}, we conclude that
$\psi_{E_*}(x) \in H^1(\R)$ is a weak solution. Moreover, the
linearized operator $L_+$ depends continuously on $E$ on the
interval $[E_1,E_*]$. By the standard perturbation theory, the
discrete spectrum of $L_+$ depends continuously on $E$. Since $0$ is
not in the spectrum of $L_+$ for $E \in [E_1,E_*)$ and we assumed
that (b) does not hold we deduce that $0$ is not an eigenvalue of
$L_+$ at $E_*.$ Moreover, since the essential spectrum of $L_+$ is
$[E,\infty)$, $0$ is not in the spectrum of $L_+$ at $E = E_*$.
Applying now Lemma \ref{th:cont} we can continue the $C^1$ branch
$(\psi_E,E)$ past $E=E_*$ which contradicts the choice of $E_*.$

The theorem is now completely proven. \end{proof}

\begin{remark}\label{rm:lplus} Let $\sigma < 0$ and $L_+$ be computed at
the branch points $(\psi_E,E)$ for $E\in (E_0,E_*)$, where $E_*$ is
given by Theorem \ref{cor:max}. Then, $L_+$ has exactly one
(strictly) negative eigenvalue. This follows from the fact that the
eigenvalues of $L_+$ depend $C^1$ on $E\in (E_0,E_*),$ $0$ is not in
the spectrum of $L_+$ and for $E$ near $E_0,$ $L_+$ has exactly one
strictly negative eigenvalue, see the discussion after Proposition
\ref{th:ex}.
\end{remark}

\section{Symmetry-breaking transitions to asymmetric states}\label{se:bif}

In this section we show that, in the focusing case $\sigma < 0$, the
second alternative in Theorem \ref{cor:max} occurs if and only if
the second eigenvalue of $L_+,\ \lambda(E)$ crosses the zero value
at $E = E_*.$ Moreover, at $E=E_*,$ under the generic assumption of
$\lambda'(E_*)\not=0,$ the branch of symmetric states suffers a
symmetry--breaking bifurcation of pitchfork type with a new branch
of asymmetric states emerging. The new branch consists of solutions
of \eqref{stationary} that are neither even nor odd in $x.$
Depending on the sign of a quantities $Q$ and $R,$ see Theorem
\ref{th:bif}, which can be numerically computed, the asymmetric
solutions are either orbitally stable or orbitally unstable with
respect to the full dynamics of the NLS equation \eqref{GP}. For
double well potentials with large separation, e.g.
\eqref{double-potential} with $s\rightarrow\infty,$ the orbital
stability of the asymmetric branch is determined by the power of the
nonlinearity, see Corollary \ref{cor:dwbif}. Regarding the branch of
symmetric states, it continues past the bifurcation point but it is
always orbitally unstable.

Note that sufficient conditions for the symmetry-breaking
bifurcation at $E_*$ were presented by Kirr {\em et al.} \cite{Kirr}
under the assumption that $E_*$ was sufficiently close to $E_0$ and
$\|\psi_E\|_{H^1}$ was small for all $E \in (E_0,E_*)$. Moreover, only the
case of cubic nonlinearity ($p = 1$) was considered. We now present
a generalization of this result which allows large values of $|E_* -
E_0|$, large norms $\|\psi_E\|_{H^1}$, and any power nonlinearities
$p \geq 1/2$.

Remark \ref{rm:lplus} showed that, in the focusing case
$\sigma<0,$ the operator $L_+$ on the branch of symmetric states
has exactly one negative eigenvalue. We first show that this
eigenvalue cannot approach zero as $E$ approaches $E_*$.

\begin{lemma}\label{le:fev} Let $\sigma < 0$ and consider
the branch of solutions $(\psi_E,E)$ of {\em\eqref{stationary}}
which bifurcates from the lowest eigenvalue $-E_0$ of $L_0.$ Let
$(E_0,E_*)$ be the maximal interval on which this branch can be
uniquely continued. If $\lambda_0(E)$ is the lowest eigenvalue of
$L_+$ along this branch then there exist $\delta,\ d>0$ such that:
$$\lambda_0(E)\leq -d<0,\quad \mbox{for all}\ E\in [E_*-\delta,E_*)\ \mbox{if}\ E_*<\infty,\
\mbox{or for all}\ E > E_0,\ \mbox{if}\ E_*=\infty.$$
\end{lemma}

\begin{proof} Assume the contrary, that there exists a sequence
$E_n\nearrow E_*$ such that $\lambda_0(E_n)\nearrow 0.$ From the
min-max principle we have for all integers $n:$
 \begin{eqnarray}
 \lambda_0(E_n)&=&\inf_{\xi\in H^2,\ \|\xi\|_{L^2}=1}\langle \xi, L_+\xi\rangle\leq \frac{1}{\|\psi_{E_n}\|^2_{L^2}}\langle \psi_{E_n}, L_+\psi_{E_n}\rangle \nonumber\\
 &&= \frac{1}{\|\psi_{E_n}\|^2_{L^2}}\langle \psi_{E_n}, L_-\psi_{E_n}+2p\ \sigma |\psi_{E_n}|^{2p}\psi_{E_n}\rangle
 = 2p\ \sigma \frac{\|\psi_{E_n}\|^{2p+2}_{L^{2p+2}}}{\|\psi_{E_n}\|^2_{L^2}} < 0,\nonumber
 \end{eqnarray}
where we used the definitions \eqref{operators} and $L_-\psi_E=0$
for all $E\in (E_0,E_*)$. From the above inequalities and
$\lambda_0(E_n)\nearrow 0$ we conclude that:
$$\lim_{n\rightarrow\infty}\frac{\|\psi_{E_n}\|^{2p+2}_{L^{2p+2}}}{\|\psi_{E_n}\|^2_{L^2}}=0.$$
But $(\psi_{E_n},E_n)$ solves the stationary NLS equation
\eqref{stationary}. Plugging in and taking the $L^2$ scalar
product with $\frac{\psi_{E_n}}{\|\psi_{E_n}\|^2_{L^2}}$ we get:
$$-E_n-\sigma \frac{\|\psi_{E_n}\|^{2p+2}_{L^{2p+2}}}{\|\psi_{E_n}\|^2_{L^2}}=\left\langle \frac{\psi_{E_n}}{\|\psi_{E_n}\|_{L^2}},\ (-\partial_x^2+V(x))\frac{\psi_{E_n}}{\|\psi_{E_n}\|_{L^2}}\right\rangle\geq -E_0,$$
where the last inequality follows from $-E_0$ being the lowest
eigenvalue of $L_0 = -\partial_x^2 + V(x).$ Passing to the limit
when $n\rightarrow\infty$ we get the contradiction $E_0\geq
E_*.$\end{proof}

Lemma \ref{le:fev} combined with concentration compactness method
and the spectral theory of Schr\" odinger type operators enables
us to deduce the following important result regarding the behavior
of $\psi_E$ for $E$ near $E_*$.

\begin{theorem}\label{th:comp} Let $\sigma < 0$ and consider
the branch of solutions $(\psi_E,E)$ of {\em\eqref{stationary}}
which bifurcates from the lowest eigenvalue $-E_0$ of $L_0.$ Let
$(E_0,E_*)$ be the maximal interval on which this branch can be
uniquely continued. Denote
$$N(E)=\|\psi_E\|^2_{L^2},\quad\ E\in (E_0,E_*).$$ If $E_*<\infty ,$ then:
 \begin{itemize}
 \item[(i)] $N(E)$ is bounded on $[E_0,E_*)$,
 $N_* := \lim_{E\rightarrow E_*}N(E)$ exists, and $0< N_* <\infty;$
 \item[(ii)] there exists $\psi_{E_*}\in H^2(\R)$ such that $(\psi_{E_*},E_*)$
 solves the stationary NLS equation \eqref{stationary} and:
 $$\lim_{E\rightarrow E_*}\|\psi_E-\psi_{E_*}\|_{H^2}=0.$$
 \end{itemize}
\end{theorem}

\begin{proof} For part (i) denote $\xi^0_E\in H^2(\R),\ \|\xi^0_E\|_{L^2}=1$
the normalized eigenfunction of $L_+$ corresponding to its lowest eigenvalue,
$\lambda_0(E).$ We use the orthogonal decomposition:
$$
\psi_E=\langle \xi^0_E,\psi_E\rangle\ \xi^0_E+\psi_E^\perp,\quad \mbox{where}\
\langle \xi^0_E,\psi_E^\perp\rangle=0,\ E\in (E_0,E_*).
$$
Then
\begin{eqnarray}
\nonumber
 \frac{1}{2}\frac{dN}{dE} = \langle \psi_E,\partial_E\psi_E\rangle &=&
 -\langle \psi_E,L_+^{-1}\psi_E\rangle=-\frac{|\langle \xi^0_E,\psi_E\rangle|^2}{\lambda_0(E)}-\langle \psi_E^\perp,L_+^{-1}\psi_E^\perp\rangle \\
 \label{eq:dne}
 &&\leq \frac{|\langle \xi^0_E,\psi_E\rangle|^2}{-\lambda_0(E)}\leq
 \frac{N(E)}{-\lambda_0(E)},
\end{eqnarray}
where we used \eqref{eq:dpsie}, $\lambda_0(E)<0,$ and the fact
that $L_+ >0$ on the orthogonal complement of $\xi^0_E,$ see
Remark \ref{rm:lplus}. The above inequality implies that
$$
N(E)\leq N(E_1)e^{\int_{E_1}^E\frac{dE}{-\lambda_0(E)}},\quad \mbox{for any}\ E_0<E_1<E<E_*.
$$
Using now the bound $\frac{1}{-\lambda_0}\leq\frac{1}{d},\ E\in
[E_*-\delta, E_*)$ given by Lemma \ref{le:fev}, together with the
fact that $N(E)$ is continuous on $[E_0,E_*-\delta],$ see
Proposition \ref{th:ex} and Theorem \ref{cor:max}, we deduce that
there exists $N>0$ such that:
$$
0\leq N(E)\leq N,\quad \mbox{for all}\ E\in [E_0,E_*).
$$

To show that $N(E)$ converges actually to a finite value as
$E\nearrow E_*$ we go back to the bound \eqref{eq:dne} and
integrate it from $E_1=E_*-\delta$ to any $E,\ E_1<E<E_*.$ We get
 \begin{eqnarray}\int_{E_1}^E\langle \psi_E^\perp,L_+^{-1}\psi_E^\perp\rangle dE&=&\int_{E_1}^E\frac{|\langle \xi^0_E,\psi_E\rangle|^2}{-\lambda_0(E)}dE+\frac{1}{2}N(E_1)-\frac{1}{2}N(E)\label{eq:nestar}\\
 &&\leq \frac{N}{d}\delta+\frac{1}{2}N(E_1)\nonumber
 \end{eqnarray}
But the integrand on the left hand side is non-negative, hence the
uniform bound implies that the integral on the left hand side
converges as $E\nearrow E_*.$ Since the same holds for the integral on
the right hand side, we deduce from \eqref{eq:nestar} that $N(E)$
must converge to a finite limit as $E\nearrow E_*.$

Moreover, since both integrals in \eqref{eq:nestar} are now
convergent on $[E_1,E_*)$ we deduce that the derivative of $N(E)$
is absolutely convergent on the same interval, see \eqref{eq:dne},
consequently the derivative of the energy functional, see
\eqref{def:energy} and \eqref{eq:denergy}, is absolutely
convergent and the energy functional remains uniformly bounded on
$[E_1,E_*).$ By repeating the argument in the proof of Theorem
\ref{cor:max} we get uniform bounds in $H^1$ norm, see
\eqref{eq:h1b}, i.e. there exists $M>0$ such that:
\begin{equation}\label{h1b}
\|\psi_E\|_{H^1}\le M,\qquad {\rm for\ all}\ E\in [E_1,E_*).
\end{equation}

Now, if $\lim_{E\rightarrow E_*}N(E)=0,$ then, equivalently,
$\lim_{E\rightarrow E_*}\|\psi_E\|_{L^2}=0.$ Because of the bound
\eqref{h1b}, Sobolev imbedding and interpolation in $L^q$ spaces,
we get $\lim_{E\rightarrow E_*}\|\psi_E\|_{L^q}=0,\ q\geq 2.$
Hence $L_+$ depends continuously on $E\in [E_1,E_*],$ and $L_+$ at
$E_*$ becomes $L_0+E_*$. Then the eigenvalues of $L_+$ should
converge to the eigenvalues of $L_0+E_*$ as $E\nearrow E_*,$ so
$0$ is an eigenvalue of $L_0+E_*,$ because we are in case (b) of
Theorem \ref{cor:max}. Equivalently $-E_*<-E_0$ is an eigenvalue
of $L_0,$ which leads to a contradiction.

Part (i) is now completely proven. For part (ii) we will use the
following lemma.

\begin{lemma}\label{lm:lqtoh2} Under the assumptions of Theorem
\ref{th:comp}, if there exist a sequence on the branch
$(\psi_{E_n},E_n)$ and a function $\psi_{E_*}\in H^1(\R)$ such that
 \begin{eqnarray}
 E_n&\nearrow &E_*\label{econv}\\
 \psi_{E_n}&\stackrel{H^1}{\rightharpoonup}&\psi_{E_*}\label{weakconv}\\
 \psi_{E_n}&\stackrel{L^q}{\rightarrow}&\psi_{E_*},\quad \mbox{for all}\ 2<q\leq\infty, \label{lqconv}
 \end{eqnarray}
then $\psi_{E_n}\stackrel{H^1}{\rightarrow}\psi_{E_*}.$ Moreover,
$\psi_{E_*}$ is a solution of \eqref{stationary} and:
$$\lim_{E\rightarrow E_*}\|\psi_E-\psi_{E_*}\|_{H^2}=0.$$
\end{lemma}

\begin{proof} The equation \eqref{stationary} satisfied by the elements on the branch can be rewritten:
 \begin{equation}\label{h2stationary}
 \psi_{E_n} = (-\partial_x^2+E_n)^{-1} [-V(x) \psi_{E_n}(x) - \sigma |\psi_{E_n}(x)|^{2p}\psi_{E_n}],
 \end{equation}
and, from \eqref{lqconv}, we have
$$|\psi_{E_n}(x)|^{2p}\psi_{E_n}\stackrel{L^2}{\rightarrow} |\psi_{E_*}(x)|^{2p}\psi_{E_*}.$$
To show that $V(x) \psi_{E_n}(x)$ also converges in $L^2(\R)$ we use
the following compactness argument. Fix $\epsilon>0$ and choose
$R>0$ sufficiently large such that $$\|V(x)\|_{L^\infty (\{|x|\geq
R\})}<\frac{\epsilon}{4M},$$ where $M>0$ is a bound for the sequence
$\|\psi_{E_n}\|_{H^1},$ and such a bound exists because the sequence
$\psi_{E_n}$ is weakly convergent in $H^1(\R).$ Then, by Rellich's
compactness theorem,
$$\lim_{n\rightarrow\infty}\|\psi_{E_n}-\psi_{E_*}\|_{L^2(\{|x|< R\})}=0,$$
hence there exists $n_0$ such that
$$\|\psi_{E_n}-\psi_{E_*}\|_{L^2(\{|x|< R\})}<\frac{\epsilon}{2\|V\|_{L^\infty(\R)}},\qquad \mbox{for } n\geq n_0,$$
and, consequently,
 \begin{eqnarray}
 \|V\psi_{E_n}-V\psi_{E_*}\|_{L^2(\R)}&\leq &\|V\|_{L^\infty(\R)}\|\psi_{E_n}-\psi_{E_*}\|_{L^2(\{|x|< R\})}+\|V\|_{L^\infty(\{|x|\geq R\})}\|\psi_{E_n}-\psi_{E_*}\|_{L^2(\R)}\nonumber\\
 &< &\frac{\epsilon}{2}+\frac{\epsilon}{4M}2M\leq \epsilon,\qquad \mbox{for } n\geq n_0\nonumber
 \end{eqnarray}
which proves $V\psi_{E_n}\stackrel{L^2}{\rightarrow}V\psi_{E_*}.$
Now both terms in the bracket of the right hand side of
\eqref{h2stationary} converge in $L^2(\R),$ and
$(-\partial_x^2+E_n)^{-1}\rightarrow (-\partial_x^2+E_*)^{-1}$ in
the space of bounded linear operators from $L^2(\R)$ to $H^2(\R),$
we obtain $\psi_{E_n}$ convergent in $H^2(\R).$ But $H^2(\R)$ is
continuously imbedded in $L^q(\R),$ for $2\leq q\leq\infty,$ hence
the limit of $\psi_{E_n}$ in $H^2$ should be the same as its limit
in $L^q,\ 2< q\leq\infty.$ Therefore, \eqref{lqconv} implies that
$\psi_{E_n}$ converges to $\psi_{E_*}$ in $H^2(\R).$ Moreover, by
passing to the limit in \eqref{h2stationary}, we get
$$
\psi_{E_*}=(-\partial_x^2+E_*)^{-1}[-V(x) \psi_{E_*}(x) - \sigma
|\psi_{E_*}(x)|^{2p}\psi_{E_*}],
$$
or equivalently, $(\psi_{E_*},E_*)$ is a solution of the
stationary NLS equation \eqref{stationary}.

In addition,  the eigenvalues of $L_+$ calculated at
$(\psi_{E_n},E_n)$ will converge to the eigenvalues of $L_+$
calculated at $(\psi_{E_*},E_*).$ In particular, at
$(\psi_{E_*},E_*),$ $L_+$ will have the lowest eigenvalue strictly
negative, followed by $0$ as the second eigenvalue, see Lemma
\ref{le:fev} and case (b) in Theorem \ref{cor:max}. By
Sturm-Liouville theory $0$ is a simple eigenvalue with corresponding
eigenfunction being odd. Hence $L_+$ at $(\psi_{E_*},E_*)$ is
invertible with continuous inverse as an operator restricted to even
functions. The implicit function theorem applied to \eqref{stationary}
gives a unique $C^1$ in $H^2(\R)$ branch of symmetric (even in $x$)
solutions in a neighborhood of $(\psi_{E_*},E_*).$ By uniqueness,
this branch must contain $(\psi_E, E),\ E\in(E_1,E_*).$ In
particular:
$$\lim_{E\rightarrow E_*}\|\psi_E-\psi_{E_*}\|_{H^2}=0,$$
and Lemma \ref{lm:lqtoh2} is completely proven.
\end{proof}

We now return to the proof of Theorem \ref{th:comp}. Based on
\eqref{h1b} and on the weak relative compactness of bounded sets
in the Hilbert space $H^1(\R)$ we can construct a sequence
$E_n\nearrow E_*$ and $\psi_{E_*}\in H^1(\R)$ such that
\eqref{econv} and \eqref{weakconv} are satisfied. It suffices now
to show that at least for a subsequence \eqref{lqconv} is
satisfied. Define
\begin{equation}\label{def:Psi}
\Psi_n=\frac{\psi_{E_n}}{\|\psi_{E_n}\|_{L^2}},
\end{equation}
and the concentration function:
$$\rho(\phi,t)=\sup_{y\in\R}\int_{y-t}^{y+t}|\phi(x)|^2dx,\quad \phi\in L^2(\R),$$
see \cite[Section 1.7]{caz:bk} for its main properties. Let
$$\mu=\lim_{t\rightarrow\infty}\liminf_{n\rightarrow\infty}\rho(\Psi_n,t).$$
Then, from \eqref{h1b} and part (i) of Theorem \ref{th:comp}, the
sequence $\Psi_n$ is bounded in $H^1(\R)$ and normalized in
$L^2(\R).$ According to the concentration compactness theory, see
for example \cite[Section 1.7]{caz:bk}, we have the following
three possible cases.
 \begin{enumerate}
 \item {\em - Vanishing:} if $\mu=0,$ then there is a subsequence $\Psi_{n_k}$ convergent to zero in
 $L^q(\R),\ 2< q\leq \infty,$ and, using \eqref{def:Psi} and part(i):
     $$ \psi_{E_{n_k}}\stackrel{L^q}{\rightarrow}0,\quad \mbox{for all}\ 2\leq q\leq \infty.$$
     Via Lemma \ref{lm:lqtoh2} we now get
     $$ \psi_{E_{n_k}}\stackrel{H^2}{\rightarrow}0 $$
     which contradicts the fact that $\lim_{E\rightarrow E_*}\|\psi_{E_n}\|_{L^2}^2=N_*\not= 0,$ see part (i) of the
     theorem.
 \item {\em - Splitting:} if $0<\mu<1,$ then there is a subsequence $\Psi_{n_k},$ the sequences
 $v_k,\ w_k,\ z_k\in H^2(\R)$ all of them bounded in $H^1,$ the sequence $\ y_k\in\R$ and the function
 $v_*\in H^1(\R)$ such that:
     \begin{eqnarray}
     \psi_{E_{n_k}}&=& \|\psi_{E_{n_k}}\|_{L^2}\Psi_{n_k}=v_k+w_k+z_k,\label{splitting}\\
     {\rm supp}\ v_k&\cap &{\rm supp}\ w_k =\emptyset,\quad \lim_{k\rightarrow\infty}{\rm dist}(y_k,{\rm supp}\ w_k)=\infty\label{separation}\\
     v_k(\cdot-y_k)&\stackrel{H^1}{\rightharpoonup}&v_*,\quad v_k(\cdot-y_k)\stackrel{L^q}{\rightarrow}v_*,\ 2\leq q\leq\infty,\label{vkconv}\\
     \|v_*\|^2_{L^2}&=&\mu N_*>0,\qquad \lim_{k\rightarrow\infty}\|w_k\|^2_{L^2}=(1-\mu)N_*>0\label{l2norm}\\
     z_k&\stackrel{H^1}{\rightharpoonup}&0,\quad z_k\stackrel{L^q}{\rightarrow}0,\ 2\leq q\leq\infty.\label{zkconv}
     \end{eqnarray}
 If, in addition, the sequence $y_k\in\R$ is unbounded, then, by possible passing to a subsequence, we have
 $\lim_{k\rightarrow\infty}y_k=\pm\infty.$ Both cases can be treated exactly the same so let us assume
 $\lim_{k\rightarrow\infty}y_k=\infty.$ Fix a compactly supported smooth function $\phi\in C_0^\infty(\R).$
 Then, from the weak formulation\eqref{stationary} we have:
     \begin{eqnarray}
     0&=&\langle \phi'(\cdot),(v_k+w_k+z_k)'(\cdot-y_k)\rangle + \langle \phi(\cdot),(V(\cdot-y_k)+E_{n_k})(v_k+w_k+z_k)(\cdot-y_k)\rangle\nonumber\\&&+\sigma\langle \phi(\cdot),|v_k+w_k+z_k|^{2p}(v_k+w_k+z_k)(\cdot-y_k)\rangle.\nonumber
     \end{eqnarray}
 By passing to the limit when $k\rightarrow\infty,$ the terms containing $v_k$ will converge to $v_*,$
 see \eqref{vkconv}, the terms containing $w_k$ respectively $z_k$ will converge to zero due to \eqref{separation}
 respectively \eqref{zkconv}, and the term containing $V(\cdot-y_k)$ converges to zero due to
 $\lim_{|x|\rightarrow\infty}V(x)=0.$ Hence
     $$\langle \phi ',v_*'\rangle + \langle \phi,E_*v_*\rangle+\sigma\langle \phi,|v_*|^{2p}v_*\rangle=0,$$
 and $v_*\in H^1(\R)$ is a weak solution of
     \begin{equation}\label{vstareq}
     -v_*''+E_*v_*+\sigma |v_* |^{2p}v_*=0,
     \end{equation}
 therefore a strong, exponentially decaying solution, with $v_*'$ also exponentially decaying. Let
     $$\xi_k(x)=\frac{v_*(x+y_k)-v_*(-x+y_k)}{\sqrt{2\mu N_*}}.$$
 $\xi_k(x)$ is obviously odd in $x,$ and the exponential decay of $v_*$ and its derivative implies:
     $$\lim_{k\rightarrow\infty}\langle v_*(x+y_k),v_*(-x+y_k)\rangle=0,\quad \lim_{k\rightarrow\infty}\langle v_*(x+y_k),L_+v_*(-x+y_k)\rangle=0,$$
 where $L_+$ is calculated at $E_{n_k}.$  Consequently:
     \begin{eqnarray}
     \lim_{k\rightarrow\infty}\|\xi_k\|^2_{L^2}&=&1,\nonumber\\
     \lim_{k\rightarrow\infty}\langle\xi_k,L_+\xi_k\rangle&=&\frac{1}{\mu N_*}\lim_{k\rightarrow\infty}\langle v_*(x+y_k),L_+v_*(x+y_k)\rangle,\nonumber
     \end{eqnarray}
 where we also used \eqref{l2norm} for the first identity, and the fact that $L_+$ is invariant under the
 transformation $x\mapsto -x$ because both $\psi_{E_{n_k}}(x)$ and $V(x)$ are even in $x,$ see also
 \eqref{operators}. Employing again the splitting \eqref{splitting}, the convergence \eqref{vkconv}, \eqref{zkconv},
 and the separation \eqref{separation} we have:
     $$\lim_{k\rightarrow\infty}\langle v_*(x+y_k),L_+v_*(x+y_k)\rangle=\langle v_*,-v_*''+E_*v_*+(2p+1)\sigma |v_*|^{2p}v_*\rangle=2p\ \sigma\int_\R |v_*|^{2p+2}(x)dx <0$$
 where we also used \eqref{vstareq}. All in all we have
     $$\lim_{k\rightarrow\infty}\langle\xi_k,L_+\xi_k\rangle=\frac{2p\ \sigma}{\mu N_*}\int_\R |v_*|^{2p+2}(x)dx <0$$
 which together with $\xi_k$ odd implies that for $k$ sufficiently large the second eigenvalue of
 $L_+$ at $E_{n_k}$ must become strictly negative, in contradiction with Remark \ref{rm:lplus}.

 If, on the other hand, the sequence $y_k\in\R$ is bounded, then, by possible passing to a subsequence, we have
 $\lim_{k\rightarrow\infty}y_k=y_*\in\R .$ Let  $\tilde v=v_*(\cdot+y_*).$ We will now rename
 $\tilde v$ to be $v_*$ and we get from \eqref{vkconv}
     \begin{equation}\label{vkconv1}
      v_k\stackrel{H^1}{\rightharpoonup}v_*,\quad v_k\stackrel{L^q}{\rightarrow}v_*,\ 2\leq q\leq\infty,
     \end{equation}
 and \eqref{l2norm} remains valid. As before the convergence implies that $v_*$ satisfies weakly, then strongly:
     \begin{equation}\label{vstareq1}
     -v_*''+V(x)v_*+E_*v_*+\sigma |v_* |^{2p}v_*=0,
     \end{equation}
 hence, both $v_*(x)$ and $v'_*(x)$ decay exponentially in $x.$ Consider
     $$\tilde\mu=\lim_{t\rightarrow\infty}\liminf_{k\rightarrow\infty}\rho(w_k,t),\quad \mbox{where}\ 0\leq\tilde\mu\leq (1-\mu)N_*.$$
 If $\tilde\mu >0$ then $w_k$ splits:
     $$w_k=\tilde v_k+\tilde w_k+ \tilde z_k$$
 where the sequences $\tilde v_k,\ \tilde w_k,\  \tilde z_k \in H^2(\R)$ have properties
 (\ref{separation})--(\ref{zkconv}). In particular there exist
 $\tilde v_*\in H^1(\R),\ \|\tilde v_*\|_{L^2}=\tilde\mu$ and the sequence $\tilde y_k\in\R$ such that
     $$\tilde v_k(\cdot-\tilde y_k)\stackrel{H^1}{\rightharpoonup}\tilde v_*,\quad \tilde v_k(\cdot-\tilde y_k)\stackrel{L^q}{\rightarrow}\tilde v_*,\ 2\leq q\leq\infty.$$
 But now, the sequence  $\tilde y_k$ is definitely unbounded since $\tilde y_k\in {\rm supp}\ w_k$ and $y_k$
 is bounded, see \eqref{separation}. Using
     $$\xi_k(x)=\frac{\tilde v_*(x+\tilde y_k)-\tilde v_*(-x+\tilde y_k)}{\sqrt{2\tilde\mu }},$$
 we get as before
     $$\lim_{k\rightarrow\infty}\langle\xi_k,L_+\xi_k\rangle=\frac{2p\ \sigma}{\tilde\mu }\int_\R |\tilde v_*|^{2p+2}(x)dx <0$$
 which contradicts the fact that the second eigenvalue of $L_+$ should remain non-negative, see Remark \ref{rm:lplus}.

 If, on the other hand, $\tilde\mu =0$ then, by possibly passing to a subsequence, we have
     $$w_k\stackrel{H^1}{\rightharpoonup}0,\quad w_k\stackrel{L^q}{\rightarrow}0,\ 2< q\leq\infty.$$
 Moreover, from \eqref{splitting}, \eqref{zkconv} and \eqref{vkconv1} we get
     $$\psi_{E_{n_k}}\stackrel{H^1}{\rightharpoonup}v_*,\quad \psi_{E_{n_k}}\stackrel{L^q}{\rightarrow}v_*,\ 2< q\leq\infty,$$
 which, via Lemma \ref{lm:lqtoh2}, implies $\lim_{k\rightarrow\infty}\|\psi_{E_{n_k}}-v_*\|_{H^2}=0.$
 The latter is in contradiction with
     $$\|v_*\|_{L^2}^2=\mu N_*<N_*=\lim_{k\rightarrow\infty}\|\psi_{E_{n_k}}\|_{L^2}.$$
 \item {\em - Compactness} If $\mu=1$ then there is a subsequence $\psi_{E_{n_k}},$ a sequence $y_k\in\R$ and a
 function $\psi_*\in H^1(\R),\ \|\psi_*\|_{L^2}=N_*$ such that:
     $$\psi_{E_{n_k}}(\cdot -y_k)\stackrel{H^1}{\rightharpoonup}\psi_*,\quad \psi_{E_{n_k}}(\cdot -y_k)\stackrel{L^q}{\rightarrow}\psi_*,\ 2\leq q\leq\infty.$$
 But the sequence $y_k$ must be bounded, otherwise, by possibly passing to a subsequence,
     $\lim_{k\rightarrow\infty}y_k=\pm\infty.$
 Both cases are treated in the same way, so let us assume $\lim_{k\rightarrow\infty}y_k=\infty.$
 Then, by the symmetry of $\psi_{E_{n_k}}$ we have:
     $$\psi_{E_{n_k}}(x)=\frac{\psi_{E_{n_k}}(x)+\psi_{E_{n_k}}(-x)}{2}\stackrel{L^2}{\rightarrow}\frac{\psi_*(x+y_k)+\psi_*(-x+y_k)}{2}=\xi_k(x),$$
 and we get the contradiction:
     $$N_*=\lim_{k\rightarrow\infty}\|\psi_{E_{n_k}}(x)\|^2_{L^2}\not=\frac{N_*}{2}=\lim_{k\rightarrow\infty}\|\xi_k\|^2_{L^2},$$
 where the last identity is a consequence of $\lim_{k\rightarrow\infty}\langle \psi_*(x+y_k),\psi_*(-x+y_k)\rangle=0.$

 So, the sequence $y_k\in\R$ must be bounded, and, by possible passing to a subsequence, we have
 $\lim_{k\rightarrow\infty}y_k=y_*,$ and
     $$\psi_{E_{n_k}}\stackrel{H^1}{\rightharpoonup}\psi_{E_*},\quad \psi_{E_{n_k}}\stackrel{L^q}{\rightarrow}\psi_{E_*},\ 2\leq q\leq\infty,$$
 where $\psi_{E_*}(x)=\psi_*(x+y_*).$ Therefore the hypotheses of Lemma \ref{lm:lqtoh2} are verified and part (ii)
 is proven.
 \end{enumerate}
Theorem \ref{th:comp} is now completely proven.
\end{proof}

\begin{remark}\label{rmk:slim} The proof of Theorem \ref{th:comp} can be greatly simplified if one knows apriori that, with the exception of the two lowest eigenvalues, the spectrum of $L_+$ is bounded away from zero.
\end{remark}
\noindent Indeed, in this case only the second eigenvalue of $L_+$ can approach zero, see Lemma \ref{le:fev}. Hence $L_+$ restricted to even functions is invertible with uniformly bounded inverse. In particular
$$\partial_E\psi_E=-L_+^{-1}\psi_E,\quad \|\partial_E\psi_E\|_{L^2}\leq K\|\psi_E\|_{L^2}$$
The argument in Theorem \ref{cor:max} can now be repeated to show directly that $\psi_E$ has a limit in $L^q,\ q\geq 2,$ and the proof is finished by applying Lemma \ref{lm:lqtoh2}.

\begin{corollary}\label{cor:sev} Under the assumptions of Theorem \ref{th:comp} we have
 \begin{itemize}
 \item[(i)] the second eigenvalue of $L_+(\psi_E,E),\ E<E_*,$ denote it by $\lambda(E),$ and only the second eigenvalue
approaches $0$ as $E\nearrow E_*;$
 \item[(ii)] if in addition $p>1/2$ and the derivative of the second eigenvalue
 of $L_+$ satisfies:
 $$\lambda'(E_*)=\lim_{E\rightarrow
 E_*}\frac{d\lambda}{dE}(E)\not=0$$
 then the set of real valued solutions of \eqref{stationary} in a $H^2\times\R $ neighborhood of $(\psi_{E_*},E_*)$
 consists of exactly two curves of class at least $C^{[2p]-1}$ intersecting only at $(\psi_{E_*},E_*).$
 \end{itemize}

\end{corollary}
\begin{proof} For part (i), Theorem \ref{th:comp} part (ii) implies that $L_+$ depends continuously on
$E\in [E_0,E_*],$ hence
its isolated eigenvalues will depend continuously on $E.$ In
particular, at $E_*,$ $L_+$ will have the lowest eigenvalue strictly
negative, followed by $0$ as the second eigenvalue, see Lemma
\ref{le:fev} and case (b) in Theorem \ref{cor:max}. If any other
continuous branch of eigenvalues of $L_+$ approaches zero as
$E\nearrow E_*,$ then $0$ becomes a multiple eigenvalue for $L_+$ at
$E_*$ in contradiction with Sturm-Liouville theory.

Part (ii) is a standard bifurcation result which uses
Lyapunov-Schmidt decomposition and Morse Lemma, see for example
\cite{Nir}. More precisely, we continue working with the functional
$F(\phi,E) : H^2(\R) \times \mathbb R\mapsto L^2(\R)$ given by
(\ref{def:F}) which has the Frechet derivative:
$$
D_{\phi} F(\phi,E) = -\partial_x^2 + V(x) + E + (2p+1) \sigma
|\phi|^{2p}=L_+(\phi,E).
$$
For $p>0$ the functional is $C^1$ while for $p>\frac{1}{2}$ it
is $C^2.$ By part (i) $D_{\phi} F(\psi_{E_*},E_*)$ has zero as a
simple eigenvalue. Let $\phi_*$ be the $L^2$-normalized
eigenfunction of $L_*=D_{\phi} F(\psi_{E_*},E_*)$ corresponding to
its zero eigenvalue. Then $L_*$ is a Fredholm operator of index zero
with ${\rm Ker}L_* = {\rm span}\{ \phi_* \}$ and ${\rm Ran}L_* =
\left[ {\rm Ker}L_* \right]^{\perp}$. Let
$$
P_\parallel \phi = \langle \phi_*,\phi\rangle_{L^2} \phi_*,\quad
P_\perp \phi = \phi - \langle \phi_*,\phi\rangle_{L^2} \phi_*
$$
be the two orthogonal projections associated to the decomposition
$L^2={\rm Ker}L_*\oplus {\rm Ran}L_*.$ By the standard
Lyapunov-Schmidt procedure, see for example \cite{Nir}, we get the
following result.
\begin{lemma}\label{lm:ls}(Lyapunov-Schmidt decomposition) There
exists a neighborhood $W\subset H^2\times\R$ of $(\psi_{E_*},E_*),$
a neighborhood $U\subset \R^2$ of $(0,E_*),$ and an unique $C^1$ map
$h:U\mapsto L^2\cap \{\phi_*\}^\perp$ such that for any solution
$(\phi,E)\in W$ of $F(\phi,E)=0$ there exists a unique $a\in\R$
satisfying:
$$(a,E)\in U,\qquad \phi =\psi_{E_*}+ a \phi_* + h(a,E)$$
and
\begin{equation}
F_\parallel(a,E)=\langle
\phi_*,F(\psi_{E_*}+a\phi_*+h(a,E),E)\rangle =0\label{parallel*}
\end{equation}
In addition, for all $(a,E)\in U$ we have:
\begin{eqnarray}
h(0,E_*)&=&0,\label{h0*}\\
\frac{\partial h}{\partial a}(a,E)&=&-(P_\perp L_+)^{-1}P_\perp
L_+\phi_*,\label{pah}\\
\frac{\partial h}{\partial E}(a,E)&=&-(P_\perp L_+)^{-1}P_\perp
[\psi_{E_*}+h],\label{peh}
\end{eqnarray}
where $L_+=L_+(\psi_{E_*}+a\phi_*+h(a,E),E).$

Moreover, if $p>\frac{1}{2}$ then $h$ is $C^2$ and for all $(a,E)\in
U$ we have:
\begin{eqnarray}
\frac{\partial^2 h}{\partial a^2}(a,E)&=&-(P_\perp L_+)^{-1}P_\perp
(\partial_aL_+)\left[\phi_*+\frac{\partial h}{\partial a}\right],\label{papah}\\
\frac{\partial^2 h}{\partial a\partial E}(a,E)&=&-(P_\perp
L_+)^{-1}P_\perp \left[\frac{\partial h}{\partial
a}+(\partial_aL_+)\frac{\partial h}{\partial E}\right],\label{papeh}\\
\frac{\partial^2 h}{\partial E^2}(a,E)&=&-(P_\perp L_+)^{-1}P_\perp
\left[\frac{\partial h}{\partial E}+(\partial_EL_+)\frac{\partial
h}{\partial E}\right],\label{pepeh}
\end{eqnarray}
where
\begin{eqnarray}
\partial_aL_+=(2p+1)2p\sigma |\psi_{E_*}+a\phi_*+h|^{2p-1}{\rm
sign}(\psi_{E_*}+a\phi_*+h)\left(\phi_*+\frac{\partial h}{\partial
a}\right),\nonumber\\
(\partial_EL_+)\frac{\partial h}{\partial
E}=\frac{\partial h}{\partial
E}+(2p+1)2p\sigma |\psi_{E_*}+a\phi_*+h|^{2p-1}{\rm
sign}(\psi_{E_*}+a\phi_*+h)\left(\frac{\partial h}{\partial E}\right)^2.\nonumber
\end{eqnarray}
\end{lemma}

So, the solutions of \eqref{stationary} in $W$ are given by the
solutions of \eqref{parallel*}. From \eqref{h0*} and $F(\psi_{E_*},E_*)=0$ we have $F_\parallel(0,E_*)=0.$ Differentiating \eqref{parallel*} we get:
\begin{eqnarray}
\frac{\partial F_\parallel}{\partial a}(a,E)&=&\langle \phi_*,\
L_+[\phi_*+\frac{\partial h}{\partial a}]\rangle,\label{pafpa}\\
\frac{\partial F_\parallel}{\partial E}(a,E)&=&\langle \phi_*,\
\psi_{E_*}+a\phi_*+h+L_+\frac{\partial h}{\partial
E}\rangle=a+\langle \phi_*,\ L_+\frac{\partial h}{\partial
E}\rangle.\label{pefpa}
\end{eqnarray}
In particular, from \eqref{pah}-\eqref{peh}, $L_*\phi_*=0,$ and $L_*$ self adjoint, we have
 \begin{eqnarray}
\frac{\partial F_\parallel}{\partial a}(0,E_*)&=&0,\label{pafpa0*}\\
\frac{\partial F_\parallel}{\partial E}(0,E_*)&=&0.\label{pefpa0*}
\end{eqnarray}

Since the gradient of $F_\parallel$ at $(0,E_*)$ is zero, the number
of solutions of \eqref{parallel*} in a small neighborhood of
$(0,E_*)$ is determined by the Hessian at $(0,E_*)$. For
$p>\frac{1}{2}$ we can calculate:
\begin{eqnarray}
\frac{\partial^2 F_\parallel}{\partial a^2}(a,E)&=&\langle \phi_*,\
(2p+1)2p\sigma |\psi_{E_*}+a\phi_*+h|^{2p-1}{\rm
sign}(\psi_{E_*}+a\phi_*+h)(\phi_*+\frac{\partial h}{\partial
a})^2+L_+\frac{\partial^2 h}{\partial
a^2}\rangle,\nonumber\\
\frac{\partial^2 F_\parallel}{\partial a\partial E}(a,E)&=&1+\langle
\phi_*,\ (2p+1)2p\sigma |\psi_{E_*}+a\phi_*+h|^{2p-1}{\rm
sign}(\psi_{E_*}+a\phi_*+h)(\phi_*+\frac{\partial h}{\partial
a})\frac{\partial h}{\partial E}\rangle\nonumber\\
&&+\langle\phi_*,\ L_+\frac{\partial^2 h}{\partial a\partial E}\rangle\nonumber\\
\frac{\partial^2 F_\parallel}{\partial E^2}(a,E)&=&\langle \phi_*,\
\frac{\partial h}{\partial E}+(2p+1)2p\sigma
|\psi_{E_*}+a\phi_*+h|^{2p-1}{\rm
sign}(\psi_{E_*}+a\phi_*+h)(\frac{\partial h}{\partial
E})^2+L_+\frac{\partial^2 h}{\partial E^2}\rangle,\nonumber
\end{eqnarray}
In particular
\begin{eqnarray}
\frac{\partial^2 F_\parallel}{\partial a^2}(0,E_*)&=&\langle
\phi_*,\ (2p+1)2p\sigma
\psi_{E_*}^{2p-1}\phi_*^2\rangle+\langle\phi_*,\ L_*\frac{\partial^2
h}{\partial
a^2}\rangle=0,\label{papafp}\\
\frac{\partial^2 F_\parallel}{\partial a\partial
E}(0,E_*)&=&1+\langle
\phi_*,\ (2p+1)2p\sigma \psi_{E_*}^{2p-1}\phi_*\frac{\partial h}{\partial E}\rangle+\langle\phi_*,\ L_*\frac{\partial^2 h}{\partial a\partial E}\rangle\nonumber\\
&=&1+(2p+1)2p\sigma\int_\R \psi_{E_*}^{2p-1}\phi_*^2\frac{dh}{dE}(0,E_*)dx\label{papefp}\\
\frac{\partial^2 F_\parallel}{\partial E^2}(0,E_*)&=&\langle
\phi_*,\ \frac{\partial h}{\partial E}\rangle+\langle\phi_*,\
(2p+1)2p\sigma \psi_{E_*}^{2p-1}(\frac{\partial h}{\partial
E})^2\rangle+\langle\phi_*,\ L_*\frac{\partial^2 h}{\partial
E^2}\rangle=0,\label{pepefp}
\end{eqnarray}
where we used \eqref{h0*}-\eqref{peh}, the fact that $\psi_{E_*}$ is even while $\phi_*$ is odd, the fact that $h$ and its
partial derivatives are in $\{\phi_*\}^\perp,$ and the fact that
$L_*$ is self adjoint with $L_*\phi_*=0.$ We next show:
\begin{equation}\label{eq:lape*}
\lambda'(E_*)=1+(2p+1)2p\sigma\int_\R
\psi_{E_*}^{2p-1}\phi_*^2\frac{dh}{dE}(0,E_*)dx \equiv
\frac{\partial^2 F_\parallel}{\partial a\partial E}(0,E_*).
\end{equation}
The second eigenvalue of $L_+(\psi_E,E)$ along the $C^1$ symmetric
branch $(\psi_E,E)$ satisfies the equation
$$
L_+(\psi_E,E) \phi_E =\lambda(E) \phi_E,\quad
\|\phi_E\|_{L^2}\equiv 1.
$$
Differentiating with respect to $E,$
we get:
$$\phi_E+(2p+1)2p\sigma\psi_E^{2p-1}\frac{d\psi_E}{dE}\phi_E+L_+\frac{d\phi_E}{dE}=\lambda'(E)\phi_E+\lambda(E)\frac{d\phi_E}{dE},$$
and, by taking the scalar product with $\phi_E,$ we obtain
\begin{equation}\label{eq:lape}1+(2p+1)2p\sigma\int_\R \psi_{E}^{2p-1}\phi_E^2\frac{d\psi_E}{dE}dx=\lambda'(E).\end{equation}
Using the continuous dependence of the spectral decomposition of
$L_+(\psi_E,E)$ with respect to $E\in (E_0,E_*]$, we have
$\lim_{E\nearrow E_*}\|\phi_E-\phi_*\|_{H^2}=0.$ Moreover, from
\eqref{eq:dpsie} and
\eqref{peh} we have
$$\lim_{E\nearrow E_*}\|\frac{d\psi_E}{dE}-\frac{dh}{dE}(0,E_*)\|_{H^2}=\lim_{E\nearrow E_*}\|L_+^{-1}\psi_E-L_*^{-1}\psi_{E_*}\|_{H^2}=0.$$ Passing now to the limit $E\nearrow E_*$ in the identity above we get \eqref{eq:lape*}.

From \eqref{papafp}-\eqref{eq:lape*} we have
$$\nabla^2F_\parallel(0,E_*)=\left[\begin{array}{cc} 0 & \lambda'(E_*)\\ \lambda'(E_*) & 0\end{array}\right]$$
Since by hypothesis $\lambda'(E_*)\not=0,$ the Hessian of
$F_\parallel$ is nonsingular and negative definite at $(0,E_*),$
hence by Morse Lemma, see \cite[Theorem 3.1.1 and Corollary
3.1.2]{Nir}, the set of solutions of $F_\parallel(a,E)=0$ in a
neighborhood of $(0,E_*)$ consists of exactly two curves of class
$C^{[2p]-1}$ intersecting only at $(0,E_*).$ This finishes the proof
of the corollary.
\end{proof}

Unfortunately, for $1/2\leq p <3/2,$ the corollary does not
guarantee that the two curves of solutions are $C^2$ which turns out
to be necessary for determining their orbital stability with respect
to the full dynamical system \eqref{GP}. However a more careful
analysis of equation \eqref{parallel*} recovers the full regularity
and stability of these solutions:

\begin{theorem}\label{th:bif}
Let $\sigma < 0$, $p \geq 1/2$, and consider the symmetric (even in
$x$) branch of solutions $(\psi_E,E)$ of {\em\eqref{stationary}}
which bifurcates from the lowest eigenvalue $-E_0$ of $L_0.$ Let
$(E_0,E_*)$ be the maximal interval on which this branch can be
uniquely continued. Assume $E_*<\infty ,$ and
\begin{equation}\label{ndegen}
\lambda'(E_*) = \lim_{E\nearrow E_*}\frac{d\lambda}{dE}(E)\not=0,
\end{equation}
where $\lambda(E)$ is the second eigenvalue of $L_+(\psi_E,E).$ Then, the set of real valued solutions
$(\phi,E)\in H^2 \times \R$ of the stationary NLS equation
\eqref{stationary} in a small neighborhood of $(\psi_{E_*}, E_*)
\in H^2 \times \R$ consists of exactly two $C^2$ curves
intersecting only at $(\psi_{E_*}, E_*):$
\begin{itemize}
\item[(i)] the first curve can be parameterized by $E\mapsto
\phi=\psi_E,\ E\in(E_0,E_*+\epsilon)$ for some small $\epsilon>0,$ it
is a $C^2$ continuation past the bifurcation point $E=E_*$ of the
symmetric branch, it has $\psi_E$ even for all $E$ and orbitally
unstable for $E>E_*;$

\item[(ii)] the second curve is of the form $(\phi(a),E(a)),\ a\in\R$ small, where the parameter can be chosen to be the
projection of $\phi-\psi_{E_*}$ onto $\ker L_*={\rm
span}\{\phi_*\}$ i.e. $\exists\epsilon>0$ such that for
$|a|<\epsilon:$
    \begin{equation}\label{asbranch}
    E=E(a)=E_*+\frac{Q}{2}a^2+o(a^2),\ \phi(a)=\psi_{E_*}+a\phi_*+\tilde h(a),
    \end{equation}
where $\tilde h(a)=O(a^2)\in\{\phi_*\}^\perp,$ and \begin{equation}\label{defQ} Q=-\frac{2p(2p+1)
\sigma^2}{\lambda'(E_*)} \left( \frac{2p-1}{3\sigma} \langle
\phi_*^2, \psi_{E_*}^{2p-2} \phi_*^2 \rangle_{L^2} -
2p(2p+1)\langle \psi_{E_*}^{2p-1} \phi_*^2, L_*^{-1}
\psi_{E_*}^{2p-1} \phi_*^2 \rangle_{L^2}\right)
\end{equation}
with $L_*=L_+(\psi_{E_*},E_*),$ along this curve
$\phi$ is neither even nor odd with respect to $x,$ and is
orbitally stable if
$$
Q>0,\quad {\rm and}\quad R > 0
$$
and orbitally unstable if
$$
Q<0,\quad {\rm or}\quad Q>0\ {\rm and}\ R<0,
$$
where
\begin{equation}
\label{oscond} R=\lim_{E\rightarrow E_*}\frac{d\|\phi\|_{L^2}^2}{dE} = 2\frac{\lambda'(E_*)}{Q} + N'(E_*),
\quad N(E) = \|\psi_E\|_{L^2}^2.
\end{equation}
\end{itemize}
\end{theorem}

\begin{proof}We continue to rely on the Lyapunov-Schmidt decomposition described Lemma \ref{lm:ls},
but we remark that $\psi_{E_*}=\lim_{E\nearrow E_*}\psi_E$ given by Theorem \ref{th:comp} is even, since $\psi_E,\ E<E_*,$ were even.
Hence the Frechet derivative of $F(\psi,E)$ with respect to $\psi$ at $(\psi_{E_*},E_*):$
$$D_\psi F(\psi_{E_*},E_*)[\phi]=L_*[\phi]=(-\partial_x^2+V+E_*)\phi+(2p+1)|\psi_{E_*}|^{2p}\phi,$$
transforms even functions into even functions. By Corollary \ref{cor:sev} part (i), $0$ is the second eigenvalue of $L_*$ hence its eigenfunction $\phi_*$ is odd via Sturm-Liouville theory. Consequently
$$L_*|_{even}:H^2_{even}\mapsto L^2_{even}$$
is an isomorphism. Implicit function theorem for $F(\psi,E)=0$ at
$(\psi_{E_*},E_*)$ implies that the set of even, real valued
solutions of \eqref{stationary} in a neighborhood of
$(\psi_{E_*},E_*)$ consists of a unique $C^1$ curve, $E\mapsto
\psi_E.$ Moreover $\psi_E>0,$ because $\psi_E$ is also the
eigenvector corresponding to the lowest eigenvalue of
$$L_-=-\partial_x^2+V+E_*+|\psi_{E_*}|^{2p},$$ and,
consequently, $F(E)=F(\psi_E,E)$ becomes $C^2$ even for $p=1/2.$
Hence $L_+=D_\psi F$ is $C^1$ in $E$ along this curve.
Differentiating once $F(\psi_E,E)\equiv 0$ we get:
$$\frac{d\psi_E}{dE}=-(L_+)^{-1}\psi_E$$
and the curve $E\mapsto\psi_E$ is $C^2$ because the right hand side
is $C^1.$ Moreover, the second eigenvalue of $L_+$ along the curve,
$\lambda(E),$ is $C^1$ in $E,$ and, because:
$$\lambda(E_*)=0,\qquad \frac{d\lambda}{dE}(E_*)<0,$$
see Corollary \ref{cor:sev} part (i) and \eqref{ndegen}, we deduce
that $\lambda(E)<0,\ E>E_*.$

For $E>E_*$ we now have that $L_+$ has two strictly negative
eigenvalues while $L_-$ has none along the symmetric (even in $x$)
branch of solutions of \eqref{stationary}. These imply that
$\exp(iEt)\psi_E$ is an orbitally unstable solution of \eqref{GP},
see \cite{Gr}, and finishes part (i) of the theorem.

For part (ii) we will rely on the curve of solutions discovered in
part (i) to do a sharper analysis of equation \eqref{parallel*}
compared to the one provided by Morse Lemma. The solutions
$(\psi_E,E)$ of \eqref{stationary} discovered in part (i) satisfy
$a=\langle\phi_*,\psi_E\rangle=0$ because $\phi_*$ is odd while
$\psi_E$ is even. Consequently one set of solutions of
\eqref{parallel*} is given by $a\equiv 0,$ in particular:
\begin{eqnarray}
\psi_E&=&\psi_{E_*}+h(0,E)>0\label{pozpe}\\
F_\parallel(0,E)&=&\langle\phi_*,F(\psi_{E_*}+h(0,E),E)\rangle\equiv
0\label{afactor}
\end{eqnarray}
Hence, ``$a$" can be factored out in the left hand side of equation
\eqref{parallel*}, and, solutions $a\not=0$ of this equation
satisfy:
\begin{equation}\label{eq:g}
g(a,E)=0
\end{equation}
where
\begin{equation}\label{defg}
g(a,E)=\left\{\begin{array}{ll}\frac{F_\parallel(a,E)-F_\parallel(0,E)}{a},&
\mbox{if }a\not=0\\ \frac{\partial F_\parallel}{\partial a}(0,E),&  \mbox{if
}a=0\end{array}\right.\end{equation}

We will show that:\begin{itemize}
   \item[(a)] $g(0,E_*)=0,$
   \item[(b)] $g$ is $C^1$ in a $\R\times\R$ neighborhood of $(a=0,E=E_*),$
   \item[(c)] $\frac{\partial g}{\partial E}(0,E_*)=\lambda'(E_*)\not=0,\quad \frac{\partial g}{\partial a}(0,E_*)=0$
  \end{itemize}
Implicit function theorem will then imply that the set of solutions
of \eqref{eq:g} in a neighborhood of $(0,E_*)$ consist of a unique
$C^1$ curve: $a\mapsto E(a),\ |a|<\epsilon$ for some $\epsilon >0,$
with $E(0)=E_*$ and
$$\frac{dE}{da}(0)=-\frac{\frac{\partial g}{\partial a}(0,E_*)}{\frac{\partial g}{\partial E}(0,E_*)}=0.$$
Then we will show that the curve $a\mapsto E(a)$ is $C^2$ with
\begin{equation}\label{sdea}
\frac{d^2E}{da^2}(0)=Q
\end{equation}
where $Q$ is given by \eqref{defQ}. Hence the nonsymmetric solutions
of \eqref{stationary}, are given by: $a\mapsto (\phi(a),E(a))$ with:
\begin{equation}\label{eq:asymb} E=E(a)=E_*+\frac{Q}{2}a^2+o(a^2),\quad \phi(a)=\psi_{E_*}+a\phi_*+h(a,E(a))>0\end{equation}
and $a\mapsto\phi(a)$ is $C^2$ (from $\R$ with values in $H^2$)
because from \eqref{eq:asymb} and \eqref{h0*}:
$$\frac{d\psi_E}{da}(a)=\phi_*+\frac{dh}{da}+E'(a)\frac{dh}{dE}=\phi_*-(P_\perp L_+)^{-1}P_\perp[L_+\phi_*+E'(a)(\phi_*+h)]$$
and the right hand side is $C^1$ in $a$ because $E,\ h$ are $C^1,$
and $L_+$ is also $C^1$ when calculated along $(\phi(a)>0,E(a)).$
The orbital stability of this curve of solutions follows from the
theory developed in \cite{Gr}, \cite{GSS} and:
\begin{eqnarray}
\lambda_1(a)&=&-\lambda'(E_*)Qa^2+o(a^2)\label{eq:la1}\\
\|\phi(a)\|_{L^2}^2&=&N(E_*)+\frac{1}{2}\left(2\lambda'(E_*)+Q N'(E_*)\right)
a^2+o(a^2)\label{eq:Na}
\end{eqnarray}
where $\lambda_1(a)$ is the second eigenvalue of $L_+(\phi(a), E(a)).$ Indeed, for $Q<0,$ \eqref{eq:la1} shows
that $L_+$ has two strictly negative eigenvalues while we know that
$L_-$ has none, hence the result in \cite{Gr} implies $\phi(a)$ is
orbitally unstable. For $Q>0,$  $L_+$ has exactly one strictly
negative eigenvalue, $L_-$ has none, and the result in \cite{GSS}
implies that $\phi(a)$ is orbitally stable if $\frac{d\|\phi\|_{L^2}}{dE}>0$ and
unstable if $\frac{d\|\phi\|_{L^2}}{dE}<0$ which via \eqref{eq:Na} and
\eqref{eq:asymb} is equivalent to $R>0,$
respectively $R<0$, where $R$ is defined by (\ref{oscond}).

Now \eqref{eq:Na} follows from \eqref{eq:asymb}, \eqref{eq:lape*}
and properties \eqref{h0*}-\eqref{pepeh} of $h(a,E),$ while
\eqref{eq:la1} follows from the same relations by differentiating
twice the equation for the second eigenvalue: $$L_+(\psi_E(a),
E(a))[\phi_1(a)]=\lambda_1(a)\phi_1(a),\qquad
\|\phi_1(a)\|_{L^2}\equiv 1.$$

It remains to prove (a)-(c) and \eqref{sdea}. (a) follows from
\eqref{defg} and \eqref{pafpa0*}. For $a\not=0$ (b) follows from $F$
being $C^1.$ For $a=0$ it suffices to prove:
$$\lim_{\stackrel{a\not=0}{a\rightarrow 0},E\rightarrow E_0}\frac{\partial g}{\partial a}(a,E)\quad\mbox{exists},\quad \mbox{and}\quad \lim_{\stackrel{a\not=0}{a\rightarrow 0},E\rightarrow E_0}\frac{\partial g}{\partial E}(a,E)=\frac{\partial^2F_\parallel}{\partial E\partial a}(0,E_0).$$
Note that $\frac{\partial^2F_\parallel}{\partial E\partial
a}(0,E_0)$ exists and it is continuous in $E_0$ because the partial
derivative with respect to $E$ of the right hand side of
\eqref{pafpa} is the derivative along the symmetric branch
$\psi_E=\psi_{E_*}+h(0,E)>0$ which we already know is $C^2.$ We will
prove that the first limit exists, the argument for the second is
similar. In what follows we use the shortened notation:
$$L_+(a,E)=L_+(\psi_{E_*}+a\phi_*+h(a,E),E)=D_\psi F(\psi_{E_*}+a\phi_*+h(a,E),E).$$

For $a\not=0$ we have
$$\frac{\partial g}{\partial a}(a,E)=-\frac{1}{a^2}\langle\phi_*,\ F(\psi_{E_*}+a\phi_*+h(a,E),E)-F(\psi_{E_*}+h(0,E),E)\rangle +\frac{1}{a}\langle\phi_*,\ L_+(a,E)[\phi_*+\frac{\partial h}{\partial a}(a,E)]\rangle.$$
We add and subtract $\frac{1}{a}\langle\phi_*,\
L_+(0,E)[\phi_*+\frac{\partial h}{\partial a}(0,E)]\rangle$ to get
\begin{eqnarray}\lefteqn{\lim_{\stackrel{a\not=0}{a\rightarrow 0},E\rightarrow E_0}\frac{\partial g}{\partial a}(a,E)}\nonumber\\&=&\lim_{\stackrel{a\not=0}{a\rightarrow 0},E\rightarrow E_0}-\frac{ F_\parallel(a,E)-F_\parallel (0,E)+a\langle\phi_*,\ L_+(0,E)[\phi_*+\frac{\partial h}{\partial a}(0,E)]\rangle}{a^2}\nonumber\\
&+&\lim_{\stackrel{a\not=0}{a\rightarrow 0},E\rightarrow
E_0}\frac{\langle\phi_*,\ \ L_+(a,E)[\phi_*+\frac{\partial
h}{\partial a}(a,E)]-L_+(0,E)[\phi_*+\frac{\partial h}{\partial
a}(0,E)]\rangle}{a}=I_1+I_2\nonumber\end{eqnarray} The differential
form of the intermediate value theorem gives:
$$I_1=-\frac{1}{2}\lim_{\stackrel{a\not=0}{a\rightarrow 0},E\rightarrow E_0}\frac{\langle\phi_*,\ \ L_+(a',E)[\phi_*+\frac{\partial h}{\partial a}(a',E)]-L_+(0,E)[\phi_*+\frac{\partial h}{\partial a}(0,E)]\rangle}{a'}$$
where $a'=a'(a,E)$ is between $0$ and $a.$ So $I_1=-\frac{1}{2}I_2$
provided $I_2$ exists.

For the limit $I_2,$ when $p>1/2$ we can use $F$ is $C^2$ hence
$L_+$ is $C^1$ and $h$ is $C^2$ to get:
$$I_2=\langle\phi_*,\ \ \partial_a L_+(0,E_0)[\phi_*+\frac{\partial h}{\partial a}(0,E_0)]+L_+(0,E_0)\frac{\partial^2h}{\partial a^2}(0,E_0)\rangle=0, $$
see \eqref{papah} for an explicit expression for both
$\partial_aL_+$ and $\frac{\partial^2h}{\partial a^2},$ and notice
that they are both even at $(0,E_0)$ while $\phi_*$ is odd. When
$p=1/2$ we rewrite $I_2$ as two limits:
\begin{eqnarray}
I_2&=&\lim_{\stackrel{a\not=0}{a\rightarrow 0},E\rightarrow E_0}\frac{\langle\phi_*,\ \ (L_+(a,E)-L_+(0,E))[\phi_*+\frac{\partial h}{\partial a}(a,E)]\rangle}{a}\nonumber\\
&+&\lim_{\stackrel{a\not=0}{a\rightarrow 0},E\rightarrow
E_0}\frac{\langle\phi_*,\ \ L_+(0,E)[\frac{\partial h}{\partial
a}(a,E)-\frac{\partial h}{\partial
a}(0,E)]\rangle}{a}=I_3+I_4\nonumber\end{eqnarray} But
$$(L_+(a,E)-L_+(0,E))[v]=2\sigma(|\psi_{E_*}+a\phi_*+h(a,E)|-|\psi_{E_*}+h(0,E)|)v$$
and by $-|a-b|\leq |a|-|b|\leq |a-b|$ and for each $x\in\R:$
$$|h(a,E)-h(0,E)|(x)\leq \left|\frac{\partial h}{\partial a}(a',E)(x)\right|\ |a|,\mbox{ for some }|a'|<|a|,$$
we get that the integrand in the expression for the limit $I_3$ is
bounded by:
$$\left|\frac{\phi_*(x)(L_+(a,E)-L_+(0,E))[\phi_*+\frac{\partial h}{\partial a}(a,E)](x)}{a}\right|\leq
 |\phi_*|^2(x)+|\phi_*|(x)\left|\frac{\partial h}{\partial a}(a',E)\right|(x)\left|\phi_*+\frac{\partial h}{\partial a}(a,E)\right|$$
where the right hand side can further be bounded by an integrable
function, since $h(a,E)\in H^2$ is $C^1$ in a neighborhood of
$(0,E_*).$

In addition, since $(a,E)\mapsto h(a,E)\in H^2(\R)\hookrightarrow
L^\infty(\R)\cap C(\R)$ is continuous and $\psi_{E_*}+h(0,E_0))>0,$
see \eqref{pozpe}, we get that for each $x\in\R $ there exists a
$\delta(x)>0$ such that $\psi_{E_*}+a\phi_*+h(a,E))>0,$ for
$|a|,|E-E_0|<\delta.$ So, we have the pointwise convergence:
\begin{eqnarray}
\lefteqn{\lim_{\stackrel{a\not=0}{a\rightarrow 0},E\rightarrow E_0}\frac{|\psi_{E_*}+a\phi_*+h(a,E)|(x)-|\psi_{E_*}+h(0,E)|(x)}{a}}\nonumber\\
&=&\lim_{\stackrel{a\not=0}{a\rightarrow 0},E\rightarrow
E_0}\frac{a\phi_*(x)+(h(a,E)-h(0,E))(x)}{a}
=\phi_*(x)+\frac{\partial h}{\partial a}(0,E_0)(x)\ \forall
x\in\R,\nonumber\end{eqnarray} which combined with the Lebesque
Dominated Convergence Theorem implies
$$I_3=\langle\phi_*,\ 2\sigma (\phi_*+\frac{\partial h}{\partial a}(0,E_0))^2\rangle =0,$$
since $\phi_*$ is odd and $(\phi_*+\frac{\partial h}{\partial
a}(0,E_0))^2$ is even.

Similarly, from \eqref{pah} we get the pointwise convergence:
$$\lim_{\stackrel{a\not=0}{a\rightarrow 0},E\rightarrow E_0}\frac{\frac{\partial h}{\partial a}(a,E)|(x)-\frac{\partial h}{\partial a}(0,E)(x)}{a}=
-(P_\perp L_+)^{-1}P_\perp(\partial_a L)[\phi_*+\frac{\partial
h}{\partial a}(0,E_0)](x)\forall x\in\R,$$ and, again by Lebesque
Dominated Convergence Theorem:
$$I_4=-2\sigma\langle\phi_*,\ L_+(0,E_0)(P_\perp L_+)^{-1}P_\perp[\phi_*+\frac{\partial h}{\partial a}(0,E_0)]^2\rangle=0,$$
since $\phi_*$ is odd while the other factor is even.

In conclusion, for $p\geq 1/2$ we have
$$\lim_{\stackrel{a\not=0}{a\rightarrow 0},E\rightarrow E_0}\frac{\partial g}{\partial a}(a,E)=\frac{1}{2}\langle\phi_*,\ \ \partial_a L_+(0,E_0)[\phi_*+\frac{\partial h}{\partial a}(0,E_0)]+L_+(0,E_0)\frac{\partial^2h}{\partial a^2}(0,E_0)\rangle=0.$$
A similar argument shows
$$
\lim_{\stackrel{a\not=0}{a\rightarrow 0},E\rightarrow
E_0}\frac{\partial g}{\partial
E}(a,E)=\frac{\partial^2F_\parallel}{\partial E\partial a}(0,E_0)$$
where
$$\frac{\partial^2F_\parallel}{\partial E\partial a}(0,E_0)=1+(2p+1)2p\sigma\langle\phi_*,\ (\psi_{E_*}+h(0,E_0))^{2p-1}(\phi_*+\frac{\partial h}{\partial a}(0,E_0))\frac{\partial h}{\partial E}(0,E_0)+L_+\frac{\partial^2 h}{\partial E\partial a}(0,E_0)\rangle.$$

So, $g(a,E)$ is $C^1$ with
$$g(0,E_*)=0,\quad\mbox{and}\quad\frac{\partial g}{\partial E}(0,E_*)=1+(2p+1)2p\sigma\int_\R\psi_{E_*}^{2p-1}\phi_*^2\frac{\partial h}{\partial E}(0,E_*)dx=\lambda'(E_*)\not=0$$
see \eqref{eq:lape*} and \eqref{ndegen}. Hence the set of solutions
of $g(a,E)=0$ in a neighborhood of $(0,E_*)$ consists of a unique
$C^1$ curve $a\mapsto E(a)$ with $E(0)=E_*$ and
$$\frac{dE}{da}(0)=-\frac{\frac{\partial g}{\partial a}}{\frac{\partial g}{\partial E}}(0,E_*)=0.$$
The curve is in fact $C^2,$ because for $a\not=0,\ p>1/2$ we have
that $g(a,E)$ is $C^2$ while for $a\not=0,\ p=1/2$ a similar
argument as the one above involving pointwise convergence to
$\psi_E(a)=\psi_{E_*}+a\phi_*+h(a,E(a))>0$ can be employed. For
$a=0$ we have from definition of the derivative:
$$E''(0)=\lim_{\stackrel{a\not=0}{a\rightarrow 0}}-\frac{\frac{\partial g}{\partial a}(a,E(a))}{a\frac{\partial g}{\partial E}(a,E(a))}=
\frac{1}{-\lambda'(E_*)} \lim_{\stackrel{a\not=0}{a\rightarrow
0}}\frac{1}{a}\frac{\partial g}{\partial a}(a,E(a)),$$ where
$$\lim_{\stackrel{a\not=0}{a\rightarrow 0}}\frac{1}{a}\frac{\partial g}{\partial a}(a,E(a))=\frac{1}{3}(2p+1)2p(2p-1)\sigma
\langle \phi_*^2, \psi_{E_*}^{2p-2} \phi_*^2
\rangle-(2p+1)^2(2p)^2\sigma^2\langle \psi_{E_*}^{2p-1}
\phi_*^2, L_*^{-1} \psi_{E_*}^{2p-1} \phi_*^2 \rangle . $$ The last
identity follows from the same argument as for $\lim_{a\rightarrow
0,E\rightarrow E_0}\frac{\partial g}{\partial a}(a,E)$ with the only
difference that for $1/2<p<1$ Lebesque Dominated Convergence Theorem
requires:
$$|\psi_{E_*}+a\phi_*+h(a,E(a))|^{2p-2}|\phi_*|\in L^2\bigcap L^\infty.$$
But since both $\phi_*$ and
$\psi_E(a)=\psi_{E_*}+a\phi_*+h(a,E(a))>0$ are solutions of the
uniform elliptic equations:
$$L_*\phi_*=0,\qquad L_-\psi_E(a)=0,$$
we have via upper and lower bounds for uniform elliptic equations:
$$|\phi_*(x)|\leq C(\delta)\exp(-\sqrt{E_*-\delta}|x|),\mbox{ for }\delta>0,\ \mbox{and}\ \psi_E(a)\geq C(\epsilon)\exp(-\sqrt{E_*+\epsilon}|x|),\mbox{ for }\epsilon>E(a)-E_*.$$ Hence, for $1/2<p<1$ we get:
$$|\psi_E(a)|^{2p-2}\leq C\exp(-\tilde\epsilon |x|)$$
for any $\tilde\epsilon >0$ such that:
$$E_*+\tilde\epsilon<\frac{E_*}{2-2p}.$$

This finishes (a)-(c) and \eqref{sdea}, consequently Theorem
\ref{th:bif}.
\end{proof}
We note that Theorem \ref{th:bif} shows that the branch of
symmetric states goes through a pitchfork bifurcation at
$(\psi_{E_*},E_*),$ which is classified, based  on stability
analysis, as supercritical when $Q > 0$ and $R > 0$, and
subcritical when either $Q < 0$ or $Q > 0$ and $R < 0$. We do not
have a general result determining which one occurs {\em except for the
double-well potentials with large separation,} e.g. \eqref{double-potential} with $s$ large:
\begin{corollary}\label{cor:dwbif} Consider the equation {\em \eqref{stationary}} with $\sigma<0$ and potential of the form:
$$V\equiv V_s(x)=V_0(x+s)+V_0(-x+s),\qquad x\in\R,\ s>0.$$
Assume $V_0$ satisfies {\em (H1), (H2)} and {\em (H4).} Then there exists $s_*>0$ such that for all $s \geq s_*$ the branch
of real valued, even in $x$ solutions $(\psi_E,E)$ bifurcating
from the lowest eigenvalue of $-\Delta+V_s$ undergoes a pitchfork bifurcation at $E=E_*,$ where $
\lim_{s\rightarrow\infty}E_*=E_0.$ Moreover, the asymmetric branch
emerging at the bifurcation point is orbitally stable if
$p<p_*=\frac{3+\sqrt{13}}{2}$ and orbitally unstable if $p>p_*,$
while the symmetric branch $(\psi_E,E)$ continues past the
bifurcation point but becomes orbitally unstable.
\end{corollary}

\begin{remark} We remark that the case $p=1$ has already been obtained in \cite{Kirr}. In what follows we present a more direct argument to obtain the same result for all $p\geq 1/2.$ The argument can be easily adapted to higher space dimensions, see \cite{kn:bif}, and to the case $\sigma >0$ in which we can rigorously show that a pitchfork bifurcation occurs along the first excited state (the branch bifurcating from the second lowest eigenvalue of $-\Delta+V_s$). This result has been predicted in \cite{Sacchetti2}.
\end{remark}

\begin{proof} Under hypotheses (H1), (H2) and (H4) for $V_0$, it is known that the spectrum $\Sigma$ of $-\partial_x^2+V_s$ has the lowest eigenvalues $-E_{0,s}<-E_{1,s}$ satisfying
\begin{equation}\label{dw:ev}
\lim_{s\rightarrow\infty}|E_{k,s}-E_0|=0,\ k=0,1,\quad \mbox{and}\quad\exists\ s_*,d>0:\ \mbox{dist}(E_{0,s},\Sigma\setminus\{E_{1,s}\})\geq d\ \forall s\geq s_*.
\end{equation}
Moreover, the normalized eigenfunctions $\psi_{0,s}, \psi_{1,s}$ corresponding to these eigenvalues satisfy:
\begin{eqnarray}
\lim_{s\rightarrow\infty}\left\|\psi_{0,s}(x)-\frac{\psi_0(x+s)+\psi_0(-x+s)}{\sqrt{2}}\right\|_{H^2}&=&0,\label{dw:ef0}\\
\lim_{s\rightarrow\infty}\left\|\psi_{1,s}(x)-\frac{\psi_0(x+s)-\psi_0(-x+s)}{\sqrt{2}}\right\|_{H^2}&=&0,\label{dw:ef1}
\end{eqnarray}
where $\psi_0$ is the eigenfunction of $-\partial_x^2+V_0$
corresponding to its lowest eigenvalue $-E_0,$ see
\cite[Appendix]{Kirr} and references therein. Proposition
\ref{th:ex} shows that for each $s>0,$ a unique curve $(\psi_E,E)$
of real valued, nontrivial   solutions of
$$-\partial_x^2\phi+V_s\phi+E\phi+\sigma |\phi |^{2p}\phi=0,$$
bifurcates from $(0,E_0)$ and can be parametrized by $a=\langle\psi_{0,s},\psi_E\rangle,$ i.e. there exists $\varepsilon>0,$ such that for $|a|<\varepsilon$ we have:
\begin{eqnarray}
E&=&E_0-\sigma\|\psi_{0,s}\|_{L^{2p+2}}^{2p+2}|a|^{2p}+O(|a|^{4p}),\mbox{ i.e. }|E-E_0-\sigma\|\psi_{0,s}\|_{L^{2p+2}}^{2p+2}|a|^{2p}|\leq C_1|a|^{4p}\label{dw:ea}\\
\psi_E&=&a\psi_{0,s}+O(|a|^{2p+1}),\mbox{ i.e. }\|\psi_E-a\psi_{0,s}\|_{H^2}\leq C_2 |a|^{2p+1}.\label{dw:psia}
\end{eqnarray}
Moreover, $\psi_E(x)$ is even in $x.$ We will rely on the fact that the distance between the two lowest eigenvalues of $\partial_x^2+V_s$ converges to zero as $s\rightarrow\infty,$ see \eqref{dw:ev}, to show that there exists $s_*> 0$ such that for each $s\geq s_*$ the second eigenvalue $\lambda (a,s)$ of $L_+(\psi_E,E)$ must cross zero at $a=a_*=a_*(s),$ where $0<a_*<\varepsilon,$ and \begin{equation}\label{dw:a*infty}
\lim_{s\rightarrow\infty}a_*(s)=0.\end{equation}
Moreover, for each $s\geq s_*$ we have
\begin{equation}\label{dw:la*}
\frac{d\lambda}{dE}(a_*,s)\leq -p<0.\end{equation}
Hence the hypotheses of Theorem \ref{th:bif} are satisfied and a pitchfork bifurcation occurs at $E_*=E(a_*)$ for each $s\geq s_*.$ We will also calculate $Q=Q(s),\ R=R(s)$ at these bifurcation points and show they are continuous on the interval $s\in[s_*,\infty)$ with
\begin{eqnarray}\lim_{s\rightarrow\infty}a_*^{2-2p}(s)Q(s)&=&-\sigma\frac{2^{2-p}}{3}(2p+1)(p+1)\|\psi_0\|_{L^{2p+2}}^{2p+2}\label{dw:Qinfty}\\
\lim_{s\rightarrow\infty}a_*^{2p-2}(s)R(s)&=&\frac{2^p(-p^2+3p+1)}{-\sigma (2p+1)(p+1)p\|\psi_0\|_{L^{2p+2}}^{2p+2}}.\label{dw:Rinfty} \end{eqnarray}
Therefore, by choosing a larger $s_*$ if necessary, we have $Q(s)>0,$ for all $s\geq s_*$ and, if $p<p_*=\frac{3+\sqrt{13}}{2}$ then $R(s)>0,$ for all $s\geq s_*$ while if $p>p_*$ then $R(s)<0,$ for all $s\geq s_*.$ The proof of the corollary is now finished.

It remains to prove \eqref{dw:a*infty}-\eqref{dw:Rinfty}. They follow from rather tedious calculations involving the spectral properties \eqref{dw:ev}-\eqref{dw:ef1}, and bifurcation estimates \eqref{dw:ea}-\eqref{dw:psia}. We include them for completeness. First we note that there exists $s_*>0$ such that the estimates \eqref{dw:ea}--\eqref{dw:psia} are uniform in $s\geq s_*,$ i.e. the constants $C_1,\ C_2$ can be chosen independent of $s\geq s_*.$ The reason is that the estimates rely on contraction principle applied to the operator $K(\phi,E,s):H^2\mapsto H^2$ given by:
$$K(\phi,E,s)=\mathbb{I}-(-\partial_x^2+V_s+E_{0,s})^{-1}P_{\perp , s}(-\partial_x^2+V_s+E+(2p+1)\sigma |\phi |^{2p}),$$
where $P_{\perp , s}$ denotes the orthogonal (in $L^2$) projection onto $\psi_{0,s}^\perp.$ Since the spectrum of $(-\partial_x^2+V_s+E_{0,s})P_{\perp , s}$ {\em restricted to even functions} remains bounded away from zero for $s$ sufficiently large and $V_s:H^2\mapsto L^2$ is uniformly bounded, we can choose the Lipschitz constant for $K,$ hence the constants $C_1,\ C_2$ above, independent of $s.$

Now, if $\lambda(a,s)$ denotes the second eigenvalue of $L_+(\psi_E,E),$ then we have:
\begin{equation}\label{dw:las}\lambda(a,s)=E_{0,s}-E_{1,s}+\frac{d\lambda}{dE}(0,s)(E-E_{0,s})+O(|E-E_{0,s}|^2),\end{equation}
where, as before, the constant hidden in the $O(|E-E_{0,s}|^2)$ term can be chosen independent of $s$ for large $s.$ Using now \eqref{eq:lape} we get:
$$\frac{d\lambda}{dE}(0,s)=\lim_{a\rightarrow 0}\left[1+(2p+1)\sigma\int_R\phi_E^2\frac{d}{dE}|\psi_E|^{2p}dx\right]=
1-\frac{2p+1}{\|\psi_{0,s}\|_{L^{2p+2}}^{2p+2}}\int_\R\psi_{1,s}^2(x)\psi_{0,s}^{2p}(x)dx$$ where we relied on the expansions \eqref{dw:ea}, \eqref{dw:psia}, and on the continuous dependence with respect to $a$ of the spectral decomposition of $L_+$ which implies that the eigenfunction $\phi_E$ corresponding to the second eigenvalue of $L_+$
converges to $\psi_{1,s}.$ Moreover, using the fact that the expansions of $\psi_E,\ \phi_E$ are uniform in $s$ for large $s,$ we get  from \eqref{dw:ef0}--\eqref{dw:ef1}:
\begin{equation}\label{dw:psiphiinfty}
\lim_{\stackrel{s\rightarrow\infty}{a\rightarrow 0}}a^{-q}\int_\R\psi_E^q(x)\phi_E^{2k}(x)dx=2^{1-k-q/2}\|\psi_0\|_{L^{q+2k}}^{q+2k},
\quad\mbox{for all }q\geq 0,\mbox{ and }k=1,2,\ldots ,\end{equation}
which in particular implies
\begin{equation}\label{dw:l0infty}\lim_{\stackrel{s\rightarrow\infty}{a\rightarrow 0}}\frac{d\lambda}{dE}(a,s)=-2p<0.\end{equation}
From \eqref{dw:ev}, \eqref{dw:las} and \eqref{dw:l0infty} we now have
$$\lim_{s\rightarrow\infty}\lambda(a,s)=0+2p\sigma 2^{-p}\|\psi_0\|_{L^{2p+2}}^{2p+2}|a|^{2p}+O(|a|^{4p}).$$
Hence, because $\sigma<0,$ there exists $s_*>0$ and $\epsilon>0$ such that:
$$\lambda(\epsilon,s)<0,\quad \mbox{and}\quad \frac{d\lambda}{dE}(a,s)\leq -p<0,\quad\mbox{for all }s\geq s_*\mbox{ and }|a|\leq\epsilon,\ a\not=0.$$
Using now $\lambda(0,s)=E_{0,s}-E_{1,s}>0,\ s>0$ we infer that for each $s\geq s_*,$ $\lambda(a,s)$ changes sign exactly once in the interval $a\in [0,\epsilon]$ at
$$a=a_*(s)\approx \sqrt[2p]{\frac{E_{0,s}-E_{1,s}}{-\lambda'(0,s)}}.$$
Hence \eqref{dw:a*infty} and \eqref{dw:la*} hold.

We now compute $Q$ and $R$ in the limit $s\rightarrow\infty,$ relying on \eqref{dw:a*infty}. In formula \eqref{defQ} which defines $Q$ we have already showed:
\begin{equation}\label{dw:lp*infty}\lim_{s\rightarrow\infty}\lambda'(E_*)=\lim_{\stackrel{s\rightarrow\infty}{a\rightarrow 0}}\frac{d\lambda}{dE}(a,s)=-2p,\end{equation}
and
$$\lim_{s\rightarrow\infty}\langle\phi_*^2,\psi_{E_*}^{2p-2}\phi_*^2\rangle=\lim_{\stackrel{s\rightarrow\infty}{a\rightarrow 0}}a^{2-2p}\int_\R\psi_E^{2p-2}(x)\phi_E^{4}(x)dx=2^{-p}\|\psi_0\|_{L^{2p+2}}^{2p+2},$$
see \eqref{dw:l0infty} and \eqref{dw:psiphiinfty}. For the remaining scalar product we use
$$L_*\psi_{E_*}=L_+(\psi_{E_*},E_*)\psi_{E_*}=\sigma 2p \psi_{E_*}^{2p+1}$$
which for even functions is equivalent to:
$$L_*^{-1}\psi_{E_*}^{2p+1}=\frac{1}{2p\sigma}\psi_{E_*}.$$ We note that from \eqref{dw:psiphiinfty} we have
$$\psi_{E_*}^{2p-1}\phi_*^2=\psi_{E_*}^{2p-1}\left[\frac{\psi_{E_*}^2}{\|\psi_{E_*}\|_{L^2}^2}+
\left(\phi_*^2-\frac{\psi_{E_*}^2}{\|\psi_{E_*}\|_{L^2}^2}\right)\right]=\frac{\psi_{E_*}^{2p+1}}{\|\psi_{E_*}\|_{L^2}^2}+o(\frac{1}{s}),$$
hence
\begin{eqnarray}
\langle\psi_{E_*}^{2p-1}\phi_*^2,L_*^{-1}\psi_{E_*}^{2p-1}\phi_*^2\rangle &=&
\frac{1}{\|\psi_{E_*}\|_{L^2}^4}\langle\psi_{E_*}^{2p+1},L_*^{-1}\psi_{E_*}^{2p+1}\rangle+\frac{2}{\|\psi_{E_*}\|_{L^2}^2}\langle L_*^{-1}\psi_{E_*}^{2p+1},\psi_{E_*}^{2p-1}
\left(\phi_*^2-\frac{\psi_{E_*}^2}{\|\psi_{E_*}\|_{L^2}^2}\right)\rangle\nonumber\\
&&+\langle\psi_{E_*}^{2p-1}
\left(\phi_*^2-\frac{\psi_{E_*}^2}{\|\psi_{E_*}\|_{L^2}^2}\right),L_*^{-1}\psi_{E_*}^{2p-1}
\left(\phi_*^2-\frac{\psi_{E_*}^2}{\|\psi_{E_*}\|_{L^2}^2}\right)\rangle\nonumber\\
&=&\frac{1}{2p\sigma\|\phi_{E_*}\|_{L^2}^4}\int_R\phi_{E_*}^{2p+2}(x)dx+
\frac{1}{p\sigma\|\phi_{E_*}\|_{L^2}^2}\int_R\psi_E^{2p}\left(\phi_*^2-\frac{\psi_{E_*}^2}{\|\psi_{E_*}\|_{L^2}^2}\right)dx\nonumber\\
&&+\langle\psi_{E_*}^{2p-1}
\left(\phi_*^2-\frac{\psi_{E_*}^2}{\|\psi_{E_*}\|_{L^2}^2}\right),L_*^{-1}\psi_{E_*}^{2p-1}
\left(\phi_*^2-\frac{\psi_{E_*}^2}{\|\psi_{E_*}\|_{L^2}^2}\right)\rangle.\nonumber
\end{eqnarray}
Passing to the limit when $s\rightarrow\infty$ and using \eqref{dw:psiphiinfty} we have:
\begin{eqnarray}
\lim_{s\rightarrow\infty}a_*^{2-2p}\langle\psi_{E_*}^{2p-1}\phi_*^2,L_*^{-1}\psi_{E_*}^{2p-1}\phi_*^2\rangle &=&
\frac{2^{-p}}{2p\sigma}\|\psi_0\|_{L^{2p+2}}^{2p+2}\nonumber\\
&&+\lim_{s\rightarrow\infty}a_*^{2-2p}\langle\psi_{E_*}^{2p-1}
\left(\phi_*^2-\frac{\psi_{E_*}^2}{\|\psi_{E_*}\|_{L^2}^2}\right),L_*^{-1}\psi_{E_*}^{2p-1}
\left(\phi_*^2-\frac{\psi_{E_*}^2}{\|\psi_{E_*}\|_{L^2}^2}\right)\rangle.\nonumber\end{eqnarray}
Because the first two and only the first two eigenvalues of $L_*$ approach zero as $s\rightarrow\infty$ we need to expand the quadratic form in the last limit in terms of the associated spectral projections. Since the quadratic form involves only even functions and the eigenfunction corresponding to the second eigenvalue of $L_*$ is odd we only need to worry about the projection onto the eigenfunction $\xi_*$ corresponding to the first eigenvalue $\lambda_0(E_*)$ of $L_*.$ We have:
$$\lim_{s\rightarrow\infty}a_*^{2-2p}\langle\psi_{E_*}^{2p-1}
\left(\phi_*^2-\frac{\psi_{E_*}^2}{\|\psi_{E_*}\|_{L^2}^2}\right),L_*^{-1}\psi_{E_*}^{2p-1}
\left(\phi_*^2-\frac{\psi_{E_*}^2}{\|\psi_{E_*}\|_{L^2}^2}\right)\rangle
=\lim_{s\rightarrow\infty}a_*^{2-2p}\frac{|\langle\psi_{E_*}^{2p-1}
\left(\phi_*^2-\frac{\psi_{E_*}^2}{\|\psi_{E_*}\|_{L^2}^2}\right),\xi_*\rangle|^2}{\lambda_0(E_*)}$$
The latter can be calculated via L'Hospital where, as in \eqref{dw:l0infty},
$$\lim_{s\rightarrow\infty}\lambda_0(E_*)=\lim_{\stackrel{s\rightarrow\infty}{a\rightarrow 0}}\frac{d\lambda_0}{dE}(a,s)=-2p<0$$
and the derivative of the denominator converges to zero. All in all we get \eqref{dw:Qinfty}.

Finally, to compute $\lim_{s\rightarrow\infty}R$ we use the definition \eqref{oscond}. We have
$$N'(E_*)=2\langle\frac{d\psi_E}{da}\left(\frac{dE}{da}\right)^{-1},\psi_{E_*}\rangle
=\frac{2a_*+O(|a_*|^{2p+1})}{-\sigma 2p\|\psi_{0,s}\|_{L^{2p+2}}^{2p+2}a_*^{2p-1}+O(|a_*|^{4p-1})},$$
where we used \eqref{dw:ea}-\eqref{dw:psia}. Consequently
$$\lim_{s\rightarrow\infty}a_*^{2p-2}N'(E_*)=\frac{2^{p}}{-\sigma p\|\psi_0\|_{L^{2p+2}}^{2p+2}}$$
which combined with \eqref{oscond}, \eqref{dw:lp*infty} and \eqref{dw:Qinfty} gives \eqref{dw:Rinfty}.
\end{proof}

\section{Behavior of the symmetric and asymmetric states for large $E$}\label{se:large}

In this section we show that if the branch of symmetric states
$(\psi_E,E)$ bifurcating from $(0,E_0)$ can be uniquely continued
on the interval $E\in(E_0,\infty),$ i.e. case (a) in Theorem
\ref{cor:max} holds, then, modulo re-scaling, this branch must
bifurcate from a nontrivial, even solution of the
constant--coefficient NLS equation:
\begin{equation}
\label{uinf}
 -u_{\infty}''(x) +\sigma |u_{\infty}|^{2p}u_\infty(x) + u_{\infty}(x) =
 0,\quad u_\infty\in H^1.
\end{equation}
Since the above equation has exactly one such solution:
\begin{equation}
\label{explicit-solution} u_{\infty}(x) =
\left(\frac{1+p}{-\sigma}\right)^{\frac{1}{2p}} \; {\rm
sech}^{\frac{1}{p}}(p x),
\end{equation}
we infer essential properties of the branch $(\psi_E,E)$ via
bifurcation theory. In particular we show that when $V(x)$ has a
non-degenerate local maximum at $x=0$ then $L_+$ computed at
$(\psi_E,E)$ has two negative eigenvalues for $E$ large,
contradicting Remark \ref{rm:lplus}. This finishes the proof of our
main theorem.

However the arguments developed in this section tell much more about
all solutions $(\psi_E, E)$ of the stationary NLS equation
\eqref{stationary} for large $E.$ Certain scaling of the
$\|\psi_E\|_{L^2},\ \|\psi_E\|_{L^{2p+2}}$ and
$\|\nabla\psi_E\|_{L^2}$ norms for large $E$ emerges from Theorem
\ref{th:brelarge}. Combined with the concentration compactness
arguments, these norms imply that the stationary solutions either
bifurcate from solutions \eqref{explicit-solution} translated to be
centered at a critical point of $V(x),$ or centered at infinity, see
Remarks \ref{rm:main1} and \ref{rm:main2}. The stationary solutions
bifurcating from a finite translation of \eqref{explicit-solution}
are localized near a critical point of $V(x),$ and if the latter is
non-degenerate the orbital stability of these solutions can be
determined, see Theorem \ref{th:bfelarge} and Remark \ref{rm:main}.

\begin{theorem}\label{th:brelarge} Let $\sigma <0, $ and consider a
$C^1$ branch of stationary solutions $(\psi_E,E)$ for $E\in
(E_1,\infty)$. If $V(x)$ satisfies
\begin{equation}\label{potelarge}V(x)\in L^{\infty}(\R),\quad
\lim_{|x|\rightarrow\infty}V(x)=0,\quad x V'(x)\in
L^\infty(\R),
\end{equation}
then
\begin{itemize}
\item[(i)] there exists $0<b<\infty$ such that
 \begin{eqnarray}
 \lim_{E\rightarrow\infty}\frac{\|\psi_E\|_{L^{2p+2}}^{2p+2}}{E^{\frac{1}{2}+\frac{1}{p}}}&=&b\label{2pnormelarge}\\
 \lim_{E\rightarrow\infty}\frac{\|\psi_E\|_{L^{2}}^{2}}{E^{\frac{1}{p}-\frac{1}{2}}}&=&\frac{-\sigma}{2}\frac{p+2}{p+1}b\label{2normelarge}\\
 \lim_{E\rightarrow\infty}\frac{\|\nabla
 \psi_E\|_{L^2}^{2}}{E^{\frac{1}{2}+\frac{1}{p}}}&=&\frac{-\sigma}{2}\frac{p}{p+1}b,\label{gradnormelarge}
 \end{eqnarray}
and, after the change of variables:
\begin{equation}\label{uelarge}
 u_E(x)=R^{1/p}\psi_E(Rx),\qquad R=\frac{1}{\sqrt{E+V(0)}},\end{equation}
$u_E$ satisfies:
\begin{eqnarray}
 \lim_{E\rightarrow\infty}\|u_E\|_{H^1} = -\sigma b > 0\label{ueh1}\\
 -u_E''(x)+R^2(V(Rx)-V(0))u_E(x)+u_E(x)+\sigma |u_E|^{2p}u_E(x) = 0.\label{eq:ue}
 \end{eqnarray}
\item[(ii)] if in addition $\psi_E(x)$ is even in $x$ and $L_+$
computed at $(\psi_E ,E)$ has exactly one negative eigenvalue for
all $E\in (E_1,\infty )$ then $u_E$ defined above converges:
\begin{equation}\label{ueh2}
\lim_{E\rightarrow\infty}\|u_E-u_\infty\|_{H^2}=0
\end{equation}
where $u_\infty\not= 0$ satisfies \eqref{uinf} and is given by
\eqref{explicit-solution}.
 \end{itemize}
\end{theorem}

Before we prove the theorem let us note that it implies that, for
the case (a) of Theorem \ref{cor:max}, the branch of symmetric
states, under the re-scaling \eqref{uelarge}:
$(\psi_E,E)\mapsto(u_E,R),$ bifurcates from the solution
$(u_E=u_\infty,R=0)$ of equation \eqref{eq:ue}, where $u_\infty$
is given by \eqref{explicit-solution}. This bifurcation can be
analyzed in detail:

\begin{theorem}\label{th:bfelarge}
Consider $x_0\in\R ,$ the equation:
\begin{equation}\label{eq:bfrsmall}
G(u,R)=-u''+R^2(V(Rx+x_0)-V(x_0))u+u+\sigma |u|^{2p}u=0,
\end{equation}
and the solution $(u=u_\infty,R=0),$ with $u_\infty$ given by
\eqref{explicit-solution}.
\begin{itemize}
 \item[(i)] If $V$ is differentiable at $x_0$ but $V'(x_0)\not=0$ then the set of solutions of
 $G(u,R) = 0$ in a neighborhood of $(u_\infty,0) \in H^1\times\R$ is given by $R=0$ and
 translations of $u_\infty:$
     $$\{(u_\infty (\cdot -s),0)\ :\ s\in\R\}.$$
 \item[(ii)] If $V$ is twice differentiable at $x_0$ and $x_0$ is a non-degenerate critical point of
 $V,$ i.e. $V'(x_0)=0,\ V''(x_0)\not=0,$ then the set of solutions of $G(u,R) = 0$ in a
 neighborhood of $(u_\infty,0) \in H^1\times\R$ consists of two orthogonal $C^1$ curves:
     $$\{(u_\infty (\cdot -s),0)\ :\ s\in\R\}\quad\mbox{and}\quad \{(u_R,R)\ :\ R\in\R ,\ |R| \mbox{ small}\},$$
 where
 $$
 \|u_R\|_{L^2}^2=\|u_\infty\|_{L^2}^2+\left(\frac{1}{2p}-\frac{3}{4}\right)V''(x_0)\|xu_\infty(x)\|_{L^2}^2R^4+o(R^4).
 $$
 Moreover, if $x_0$ is a local maximum (respectively local minimum) for $V$ then $L_+$ computed at $(u_R,R)$
 has exactly two (respectively exactly one) negative eigenvalues.
\end{itemize}
\end{theorem}

We now outline a few remarks.

\begin{remark}
\label{rm:main} The last two theorems combined with {\em Remark \ref{rm:lplus}}
finish the proof of our main theorem. Moreover, for the generic
potential \eqref{double-potential} with $s>s_*$, which has exactly
two non-degenerate local minima at $x=\pm x_0$ and one
non-degenerate local maximum at $x=0,$ the above theorem gives two
more branches of asymmetric states:
\begin{equation}\label{genscalling}E=R^{-2}-V(\pm x_0),\qquad \psi_E(x)=R^{-\frac{1}{p}}u_R\left(\frac{x\mp x_0}{R}\right)\end{equation}
localized near the two minima. Both branches are orbitally stable
for $p\leq 2$ and orbitally unstable for $p>2.$ This is because
the operator $L_+(\psi_E,E)$ has exactly one
negative eigenvalue, see part (ii) of {\em Theorem \ref{th:bfelarge},}
and, according to the general theory in \cite{GSS}, the sign of
$\partial_E\|\psi_E\|_{L^2}^2$ determines the orbital stability.
Using \eqref{genscalling} we have
\begin{eqnarray}
\partial_E\|\psi_E\|_{L^2}^2&=&\left(\frac{1}{p}-\frac{1}{2}\right)
(E+V(\pm x_0))^{\frac{1}{p}-\frac{3}{2}}\|u_R\|_{L^2}^2-\frac{1}{2}(E+V(\pm
x_0))^{\frac{1}{p}-2}\partial_R\|u_R\|_{L^2}^2\nonumber\\
&=&E^{\frac{1}{p}-\frac{3}{2}}\left[\left(\frac{1}{p}-\frac{1}{2}\right)
\|u_\infty\|_{L^2}^2+O(E^{-\frac{1}{2}})\right]\nonumber
\end{eqnarray}
Hence $\partial_E\|\psi_E\|_{L^2}^2>0,$ for $p<2,$ implying
stability, $\partial_E\|\psi_E\|_{L^2}^2<0,$ for $p>2,$ implying
instability, while for $p=2$ we have
$$
\|\psi_E\|_{L^2}^2=\|u_R\|_{L^2}^2=\|u_\infty\|_{L^2}^2-\frac{1}{2}V''(\pm
x_0)\|xu_\infty(x)\|_{L^2}^2R^4+o(R^4),
$$
i.e. $\|\psi_E\|_{L^2}^2$ is increasing with $E\sim R^{-2},$
implying stability.
\end{remark}

\begin{remark}\label{rm:main1} The proof of {\em Theorem \ref{th:brelarge} part (ii)}
shows that in the absence of hypothesis $\psi_E$ even, the re-scaled $u_E$ may
concentrate at $\pm\infty,$ i.e.
$$
\exists y_E\in\R \mbox{ such that }\lim_{E\rightarrow\infty}y_E=\pm\infty
\mbox{ and }\lim_{E\rightarrow\infty}\|u_E(\cdot-y_E)-u_\infty\|_{H^2}=0.
$$
This possibility prevents us to claim that for the double well potential \eqref{double-potential} with $s>s_*$ the two branches of asymmetric states concentrated near its two minima, which exist for large $E$ via {\em Theorem \ref{th:bfelarge},}
are in fact the continuation of the two branches of
asymmetric states emerging from the pitchfork bifurcation along
the symmetric branch at $E=E_*<\infty,$ via {\em Theorem \ref{th:bif}.}
Numerical simulations strongly support this claim but a proof of {\em Theorem \ref{th:bfelarge}} which includes the case $x_0=\pm\infty$ is needed for a rigorous resolution of the problem. Note that, for usual potentials V with $\lim_{|x|\rightarrow\infty}V(x)=0,$ $x_0=\pm\infty$ is actually a degenerate critical point, i.e.
$$\lim_{x\rightarrow\pm\infty}\frac{d^kV}{dx^k}(x)=0,\qquad k=1,2,3,\ldots .$$
\end{remark}

\begin{remark}\label{rm:main2} Moreover, in the absence of the spectral hypothesis that $L_+(\psi_E,E)$ has exactly one negative eigenvalue, the proof of {\em Theorem \ref{th:brelarge} part (ii)} shows that
$u_E$ may split in two or more functions, depending on the number
of negative eigenvalues of $L_+,$ which concentrate at points
further and further apart, i.e there exist $u_E^1,u_E^2,\ldots
,u_E^N\in H^2,$ and $y_E^1,y_E^2,\ldots ,y_E^N\in\R$ such that:
$$
\lim_{E\rightarrow\infty}|y_E^j-y_E^k|=\infty \mbox{ for }j\not=k,
\quad
\lim_{E\rightarrow\infty}\|u_E^k(\cdot-y_E^k)-u_\infty\|_{H^2}=0,
\mbox{ for }k=1,2\ldots N,
$$
and
$$
u_E=\sum_{k=1}^{n}u_E^k+v_E,\qquad \mbox{where }\lim_{E\rightarrow\infty}\|v_E\|_{L^q}=0,\ \mbox{for all } 2\leq q\leq\infty.
$$
Such nonlocal bifurcations are excluded for
ground states but are relevant for excited states of the
stationary NLS equation (\ref{stationary}), see \cite{kn:bif} for partial results in this direction.
\end{remark}

We now proceed with the proofs of the above theorems.

\begin{proof1}{\em of Theorem \ref{th:brelarge}.}\ \ \
Recall from \eqref{def:energy}:
$${\cal E}(E)=\int_\mathbb{R}| \nabla\psi_E(x)|^2dx+
\int_{\mathbb{R}}V(x)|\psi_E
(x)|^2dx+\frac{\sigma}{p+1}\int_\mathbb{R}|\psi_E (x)|^{2p+2}dx,$$
and from \eqref{eq:denergy}
$$\frac{d{\cal E}}{dE}=-E\frac{dN}{dE},$$
where $N(E)=\|\psi_E\|_{L^2}^2.$ Also, from \eqref{eq:spstat}:
$$\| \nabla\psi_E \|_{L^2}^2+ \int_{\mathbb
R}V(x)|\psi_E(x)|^2dx+\sigma\|\psi_E\|_{L^{2p+2}}^{2p+2}+
E\|\psi_E\|_{L^2}^2=0$$ which can be rewritten:
$${\cal E}(E)+\frac{\sigma p}{p+1}\|\psi_E\|_{L^{2p+2}}^{2p+2}=-EN(E).$$
Differentiating the latter with respect to $E$ we get:
\begin{equation}\label{eq:dnormp}
\frac{d}{dE}\|\psi_E\|_{L^{2p+2}}^{2p+2}=\frac{p+1}{-\sigma p}N.
\end{equation}

Now, from $\psi_E$ a weak solution of \eqref{stationary} using $x
\psi_E'(x)$ as a test function and integrating by parts we get the
Pohozaev type identity:
\begin{equation}\label{eq:poho}
-\|\nabla\psi_E \|_{L^2}^2+ \int_{\mathbb R}(V(x)+x V'(x))
|\psi_E(x)|^2dx+\frac{\sigma}{p+1}\|\psi_E\|_{L^{2p+2}}^{2p+2}=
-E\|\psi_E\|_{L^2}^2
\end{equation}
which added to \eqref{eq:spstat} implies:
$$
2EN+\int_{\mathbb R}(2V(x)+x V'(x))
|\psi_E(x)|^2dx=-\sigma\frac{p+2}{p+1}\|\psi_E\|_{L^{2p+2}}^{2p+2}.
$$
The integral term in the identity above can be bounded via H\"
older inequality:
$$
-\left[2\|V\|_{L^{\infty}}+\|x\cdot\nabla
V\|_{L^\infty}\right]N\leq \int_{\mathbb R}(2V(x) + x
V'(x))|\psi_E(x)|^2dx\leq \left[2\|V\|_{L^{\infty}}+\|x V'
\|_{L^\infty}\right]N,
$$
hence, for $C=2\|V\|_L^{\infty}+\|x V'\|_{L^\infty},$ we have:
\begin{equation}\label{ineq:nnormp}
2(E-C)N(E)\leq
-\sigma\frac{p+2}{p+1}\|\psi_E\|_{L^{2p+2}}^{2p+2}\leq 2(E+C)N(E).
\end{equation}
Plugging \eqref{ineq:nnormp} in \eqref{eq:dnormp} and using the
notation $Q(E)=\|\psi_E\|_{L^{2p+2}}^{2p+2}$ we get:
$$\frac{p+2}{2p}\frac{Q(E)}{E+C}\leq \frac{dQ}{dE}(E)\leq \frac{p+2}{2p}\frac{Q(E)}{E-C}.$$
Fix $E_2>\max\{C,E_1\}$ and integrate on $[E_2,E]:$
$$\frac{Q(E_2)}{(E_2+C)^{1/2+1/p}}(E+C)^{1/2+1/p}\leq Q(E)\leq \frac{Q(E_2)}{(E_2-C)^{1/2+1/p}}(E-C)^{1/2+1/p},$$
hence:
$$\frac{Q(E_2)}{(E_2+C)^{1/2+1/p}}\leq\frac{\|\psi_E\|_{L^{2p+2}}^{2p+2}}{E^{1/2+1/p}}\leq \frac{Q(E_2)}{(E_2-C)^{1/2+1/p}},$$
which implies \eqref{2pnormelarge} since
$$\lim_{E_2\rightarrow\infty}\frac{\frac{Q(E_2)}{(E_2-C)^{1/2+1/p}}}{\frac{Q(E_2)}{(E_2+C)^{1/2+1/p}}}=1.$$

Now, \eqref{2normelarge} follows from dividing \eqref{ineq:nnormp}
by $E^{1/2+1/p}$ and passing to the limit $E\rightarrow\infty,$
while \eqref{gradnormelarge} follows from dividing \eqref{eq:spstat}
by $E^{1/2+1/p}$ and passing to the limit $E\rightarrow\infty.$

Note that \eqref{uelarge} and the fact that $\psi_E$
solves \eqref{stationary} already implies \eqref{eq:ue}. Moreover
\eqref{uelarge} combined with
\eqref{2pnormelarge}-\eqref{gradnormelarge} shows that:
\begin{eqnarray}
\|u_E\|_{L^{2p+2}}^{2p+2}&=&R^{1+2/p}\|\psi_E\|_{L^{2p+2}}^{2p+2}\stackrel{E\rightarrow\infty}{\rightarrow}b\nonumber\\
\|u_E\|_{L^2}^2&=&R^{2/p-1}\|\psi_E\|_{L^2}^2\stackrel{E\rightarrow\infty}{\rightarrow}-\frac{\sigma}{2}\frac{p+2}{p+1}b\nonumber\\
\|\nabla
u_E\|_{L^2}^2&=&R^{2/p+1}\|\nabla\psi_E\|_{L^2}^2\stackrel{E\rightarrow\infty}{\rightarrow}-\frac{\sigma}{2}\frac{p}{p+1}b.\nonumber
\end{eqnarray}
Adding the last two we get \eqref{ueh1}. Part
(i) is now completely proven.

For part (ii) we use concentration compactness:
 \begin{enumerate}
 \item {\em -Vanishing,} i.e. $\lim_{E\rightarrow\infty}\|u_E\|_{L^q}=0,\ 2<q\leq\infty ,$ cannot happen.
 Indeed, assuming the contrary, from \eqref{eq:ue} we get:
    \begin{equation}\label{eq:uefp}u_E=(-\Delta+1)^{-1}[-R^2(V(Rx)-V(0))u_E-\sigma |u_E|^{2p}u_E].\end{equation}
 Hence, using that $(-\Delta+1)^{-1}:L^2\mapsto H^2$ is unitary, we have:
    $$\|u_E\|_{H^2}\leq R^2\|V(Rx)-V(0)\|_{L^\infty}\|u_E\|_{L^2}+|\sigma|\|u_E\|_{L^{4p+2}}^{2p+1}.$$
 Since the right hand side converges to zero we get
 $\|u_E\|_{H^1}\leq \|u_E\|_{H^2}\stackrel{E\rightarrow\infty}{\rightarrow}0$ which contradicts \eqref{ueh1}.
 \item {\em -Splitting} cannot happen. The argument is a slight adaptation of the one we used to exclude
 splitting in the proof of Theorem \ref{th:comp} part (ii) and relies on the hypothesis that $L_+(\psi_E ,E)$ has
 exactly one negative eigenvalue for all $E\in(E_1,\infty ).$
 \item {\em -Compactness} is the only case possible and implies that for any sequence $E_n\rightarrow\infty$
 there exists a subsequence $E_{n_k}$ and $y_k\in\R,\ \tilde u_\infty\in H^1$ such that:
   $$\lim_{k\rightarrow\infty}\|u_{E_k}(\cdot-y_k)-\tilde u_\infty\|_{H^1}=0.$$
As in the compactness part of the proof of Theorem \ref{th:comp} the
symmetry of $u_{E_k}$ implies that $y_k\in\R$ must be a bounded
sequence. By possibly choosing a subsequence we have
$\lim_{k\rightarrow\infty}y_k=y_\infty\in\R,$ hence
$$\lim_{k\rightarrow\infty}\|u_{E_k}-\tilde u_\infty(\cdot+y_\infty)\|_{H^1}=0.$$
By plugging $u_{E_k}$ in \eqref{eq:uefp} and passing to the limit
$k\rightarrow\infty$ we infer that $u_\infty(\cdot+y_\infty)$ is an
even $H^1$ solution of \eqref{uinf}, hence
$u_\infty(\cdot+y_\infty)=u_\infty$ given by
\eqref{explicit-solution}.

We have just showed that for each sequence $E_n\rightarrow\infty$
there exists a subsequence $E_{n_k}$ such that
$$\lim_{k\rightarrow\infty}\|u_{E_k}-u_\infty\|_{H^1}=0.$$
Since the limit is unique we have:
$$\lim_{E\rightarrow\infty}\|u_E-u_\infty\|_{H^1}=0$$
and \eqref{eq:uefp} now shows that the convergence is in $H^2.$
\end{enumerate}
Theorem \ref{th:brelarge} is now completely proven.
\end{proof1}

\begin{proof1}{\em of Theorem \ref{th:bfelarge}:} Since the Frechet derivative with respect to $u$ of $G(u,R):$
$$D_uG(u,R)[v]=L_+(u,R)[v]=-v''+R^2(V(Rx+x_0)-V(x_0))v+v+(2p+1)\sigma |u|^{2p}v$$
at $u=u_\infty,\ R=0$ has kernel spanned by $u'_\infty$ we could use
the standard Lyapunov-Schmidt decomposition with respect to
$L_+(u_\infty,0)$ to reduce \eqref{eq:bfrsmall} to finding the
zeroes of a map from $\R^2$ to $\R.$ However the latter will have
zero gradient and Hessian at $(0,0)$ because $G(0,R)=O(R^3)$ in case
(i) and $G(0,R)=O(R^4)$ in case (ii). A generalization of the Morse
Lemma and calculations of derivatives up to order four will be
required to fully analyze the reduced problem. In particular $V(x)$
will need to be $C^4$ or in $W^{4,\infty}(\R)$ to be able to carry
on the analysis. We avoid this unnecessary complication by using a
decomposition similar to the one in \cite{FW}:
\begin{lemma}\label{lm:dec} There exists $\epsilon,\delta>0$ such that for any
$u\in L^2$ with $\|u-u_\infty\|_{L^2}<\epsilon$ there exists a unique
$s\in\R,\ |s|<\delta,$ with the property that:
$$u=u_\infty(\cdot-s)+v,\qquad v\perp u'_\infty(\cdot-s).$$
Moreover the map $u\mapsto s$ is $C^2$ from $L^2$ to $\R$ and there
exists $C>0$ such that
$$|s|\leq C\|u-u_\infty\|_{L^2}.$$
\end{lemma}
The Lemma follows directly from applying the implicit function
theorem to the problem of finding the zeroes of the $C^2$ map
$F:L^2\times \R\mapsto \R$ given by
$$F(u,s)=\langle u'_\infty(\cdot-s),u-u_\infty(\cdot-s)\rangle$$
in a neighborhood of $(u_\infty,0),$ because:
$$\frac{\partial F}{\partial s}(u_\infty,0)=-\langle u''_\infty,u_\infty-u_\infty\rangle+\langle u'_\infty,u'_\infty\rangle=\|u'_\infty\|_{L^2}^2\not=0.$$

Returning now to the equation \eqref{eq:bfrsmall} we note that
$(u_\infty(\cdot-s),0),\ s\in\R$ is the set of all solutions when
$R=0.$ To find other solutions in a $H^1\times\R$ neighborhood  of
$(u_\infty,0)$ we decompose them according to the above lemma:
\begin{equation}\label{elargedec}
u(x)=u_\infty(x-s)+v(x),\qquad v\perp u'_\infty(\cdot-s)
\end{equation}
and rewrite \eqref{eq:bfrsmall} in the equivalent form:
\begin{eqnarray}
G_\perp(v,R,s)&=&-v''+P_\perp(s)R^2[V(Rx+x_0)-V(x_0)][u_\infty(x-s)+v]+v\nonumber\\&&+(2p+1)\sigma P_\perp(s)|u_\infty(x-s)|^{2p}v+P_\perp(s)N(s,v)=0\label{perpelarge}\\
G_\parallel(v,R,s)&=&\langle
u'_\infty(\cdot-s),R^2[V(Rx+x_0)-V(x_0)][u_\infty(x-s)+v]+N(s,v)\rangle=0,\label{parelarge}
\end{eqnarray}
where $P_\perp(s)$ is the projection onto the orthogonal complement
of $u'_\infty(\cdot-s)$ in $L^2,$ and
\begin{equation}\label{def:Nsv}
N(s,v)=\sigma
|u_\infty(\cdot-s)+v|^{2p}(|u_\infty(\cdot-s)+v)-\sigma
|u_\infty(\cdot-s)|^{2p}u_\infty(\cdot-s)-(2p+1)\sigma
|u_\infty(\cdot-s)|^{2p}v.
\end{equation}
Note that, for all $x\in\R$ and some $0\leq t\leq 1,$ we have:
\begin{eqnarray}
|N(s,v_1)(x)-N(s,v_2)(x)|&\leq &(2p+1)2p|\sigma |\left(|u_\infty(x-s)|+\max\{|v_1(x)|,|v_2(x)|\}\right)^{2p-1}\nonumber\\
&&\times |tv_1(x) +(1-t)v_2(x)|\ |v_1(x)-v_2(x)|,\nonumber
\end{eqnarray}
hence
\begin{eqnarray}
\|N(s,v_1)-N(s,v_2)\|_{L^2}&\leq &(2p+1)2p|\sigma |\left(\|u_\infty\|_{L^\infty}+\max\{\|v_1\|_{L^\infty},\|v_2\|_{L^\infty}\}\right)^{2p-1}\nonumber\\
&&\times
\max\{\|v_1\|_{L^\infty},\|v_2\|_{L^\infty}\}\|v_1-v_2\|_{L^2}\nonumber
\end{eqnarray}
in particular, if we assume $\|v_{1,2}\|_{H^2}\leq 1$ then we can
find a constant $C_N>0$ such that:
\begin{equation}\label{Nsvest1}
\|N(s,v_1)-N(s,v_2)\|_{L^2}\leq
C_N\max\{\|v_1\|_{H^2},\|v_2\|_{H^2}\}\|v_1-v_2\|_{H^2}
\end{equation}
and
\begin{equation}\label{Nsvest2}
\|N(s,v_1)\|_{L^2}\leq C_N\|v_1\|_{H^2}^2.
\end{equation}

Using now the notation
$L_\infty(s)[v]=L_+(u_\infty(\cdot-s),0)[v]=-v''+v+(2p+1)\sigma
|u_\infty(\cdot-s)|^{2p}v$ we can rewrite \eqref{perpelarge} in the
fixed point form:
\begin{eqnarray}
v&=&-L^{-1}_\infty(s)P_\perp(s)R^2[V(Rx+x_0)-V(x_0)]u_\infty(\cdot-s)\nonumber\\
&&-L^{-1}_\infty(s)P_\perp(s)\left[R^2(V(Rx+x_0)-V(x_0))v+N(s,v)\right],\label{perpfp}\\
&=&v_0(s,R)+K_{s,R}(v)\nonumber
\end{eqnarray}
where:
 \begin{itemize}
 \item $L^{-1}_\infty(s):L^2\cap\{u'_\infty(\cdot-s)\}^\perp\mapsto H^2\cap\{u'_\infty(\cdot-s)\}^\perp$ is linear,
 bounded, with bound independent on $s\in\R ;$
 \item $P_\perp(s)R^2[V(Rx+x_0)-V(x_0)]:H^2\mapsto L^2\cap\{u'_\infty(\cdot-s)\}^\perp$ is linear and bounded
 by $2R^2\|V\|_{L^\infty}$ uniformly for $s\in\R ;$
 \item $P_\perp(s)N(s,\cdot):H^2\mapsto L^2\cap\{u'_\infty(\cdot-s)\}^\perp$ is locally Lipschitz with
 Lipschitz constant independent on $s\in\R,$ $Lip(r)\leq C_Nr$ in the ball of radius $r\leq 1$ centered
 at origin in $H^2,$ see \eqref{Nsvest1}.
 \end{itemize}
Contraction principle can now be applied to \eqref{perpfp} in the
ball: $B(0,r)=\{v\in H^2\ :\ \|v\|_{H^2}\leq r\}$ provided
$\|v_0(s,R)\|_{H^2}\leq r/2$ and $K_{s,R}$ is a contraction on
$B(0,r)$ with Lipschitz constant $Lip\leq 1/2.$ Based on the above
estimates it suffices to require $r\leq 1$ and:
\begin{eqnarray}
2R^2\|L_\infty^{-1}\|_{L^2\mapsto H^2}\|V\|_{L^\infty}\|u_\infty\|_{H^2}&\leq &\frac{r}{2}\nonumber\\
\|L_\infty^{-1}\|_{L^2\mapsto H^2}(2R^2\|V\|_{L^\infty}+C_Nr)&\leq
&\frac{1}{2}\nonumber
\end{eqnarray}
which can be accomplished by choosing:
$$r=\min\{1,(2\|L_\infty^{-1}\|_{L^2\mapsto H^2}C_N+1/\|u_\infty\|_{H^2})^{-1}\},\qquad R\leq \sqrt{r(4\|L_\infty^{-1}\|_{L^2\mapsto H^2}\|V\|_{L^\infty}\|u_\infty\|_{H^2})^{-1}}=R_0.$$
Hence \eqref{perpfp}, and consequently \eqref{perpelarge}, has a
unique solution $v=v(s,R)$ in $B(0,r)$ for each $s\in\R$ and $0\leq
R\leq R_0.$ This solution depends $C^1$ on $s$ and $R$ since
\eqref{perpfp} is $C^1$ in these parameters. $v(s,R)$ can be
obtained by successive approximations:
$$v_0(s,R),\ v_1(s,R)=v_0(s,R)+K_{s,R}(v_0(s,R)),\ v_2(s,R)=v_0(s,R)+K_{s,R}(v_1(s,R)),\ldots \stackrel{H^2}{\rightarrow}v(s,R)$$
Moreover, from the contraction principle we have:
$$\|v(s,R)-v_1(s,R)\|_{H^2}\leq \frac{Lip}{1-Lip}\|v_1-v_0\|_{H^2}\leq \|L_\infty^{-1}\|_{L^2\mapsto H^2} (2R^2\|V\|_{L^\infty}+C_N\|v_0\|_{H^2})\|v_0\|_{H^2}.$$
In what follows we will show that for $|s|\leq 1$ we have:
\begin{equation}\label{v0est}
\|v_0(s,R)\|_{H^2}=\left\{\begin{array}{ll} O(R^3) & \mbox{if }
V'(x_0)\not=0\\ O(R^4) & \mbox{if } V'(x_0)=0\end{array}\right.
\end{equation}
where in the second case we assume $V$ is twice differentiable at
$x_0.$ Consequently:
\begin{equation}\label{vest}
v(s,R)=v_0(s,R)+\left\{\begin{array}{ll} O(R^5) & \mbox{if }
V'(x_0)\not=0\\ O(R^6) & \mbox{if } V'(x_0)=0\end{array}\right.
\end{equation}
For \eqref{v0est} we use the definition \eqref{perpfp} of $v_0:$
$$\|v_0\|_{H^2}\leq R^2\|L_\infty^{-1}\|_{L^2\mapsto H^2}\|[V(Rx+x_0)-V(x_0)]u_\infty(\cdot-s)\|_{L^2},$$
and
\begin{eqnarray}\|[V(Rx+x_0)-V(x_0)]u_\infty(\cdot-s)\|_{L^2}^2&=&\int_\R [V(Rx+x_0)-V(x_0)]^2u^2_\infty(x-s)dx\nonumber\\
&=&\int_\R
u^2_\infty(x)[V(Rx+Rs+x_0)-V(x_0)]^2dx.\nonumber\end{eqnarray} Since
$V(x)$ is differentiable at $x=x_0$ and bounded for almost all $x,$
there exists $C_1>0$ such that almost everywhere: $|V(x)-V(x_0)|\leq
C_1 |x-x_0|.$ Moreover if $V(x)$ is twice differentiable at $x=x_0$
with $V'(x_0)=0$ then there exists $C_2>0$ such that almost
everywhere: $|V(x)-V(x_0)|\leq C_2 |x-x_0|^2.$ Plugging in above and
using $\int_\R |x|^nu^2_\infty(x)dx<\infty,\ n=0,1,2\ldots$ due to
the well known exponential decay of $u_\infty,$ we get
\eqref{v0est}.

Since \eqref{eq:bfrsmall} is equivalent with the system
\eqref{perpelarge}-\eqref{parelarge}, and \eqref{perpelarge} has the
unique solution $v(s,R)$ in a neighborhood of zero for all $s\in\R,\
0\leq R\leq R_0,$ all solutions of \eqref{eq:bfrsmall} in a small
neighborhood of $(u_\infty,R=0)$ will be given by the solutions of
\eqref{parelarge} with $v=v(s,R)$ and $(s,R)$ in a small
neighborhood of $(0,0):$
\begin{eqnarray}
G_\parallel(s,R)&=&G_\parallel(v(s,R),R,s)=\langle u'_\infty(\cdot-s),R^2[V(Rx+x_0)-V(x_0)]u_\infty(\cdot-s)\rangle\label{eq:par2}\\
 &&+ \langle u'_\infty(\cdot-s),R^2[V(Rx+x_0)-V(x_0)]v(s,R)+N(s,v(s,R))\rangle=0.\nonumber
\end{eqnarray}
Note that by the estimate \eqref{v0est}-\eqref{vest} we have, for
$V'(x_0)\not=0$ and $|s|\leq 1:$
$$\langle u'_\infty(\cdot-s),R^2[V(Rx+x_0)-V(x_0)]v(s,R)+N(s,v(s,R))\rangle=O(R^6)$$
while
\begin{eqnarray}
\langle u'_\infty(\cdot-s),R^2[V(Rx+x_0)-V(x_0)]u_\infty(\cdot-s)\rangle&=&R^2\int_\R u'_\infty(x)u_\infty(x)[V(Rx+Rs+x_0)-V(x_0)]dx\nonumber\\
&=&-\frac{R^3}{2}\int_\R u^2_\infty(x)V'(Rx+Rs+x_0)dx\nonumber\\
&=&-\frac{V'(x_0)\|u_\infty\|^2_{L^2}}{2}R^3+o(R^3)\nonumber\end{eqnarray}
where in the last step we used $\lim_{R\rightarrow
0}V'(Rx+Rs+x_0)=V'(x_0)$ and Lebesque dominated convergence theorem.
Hence, for $V'(x_0)\not=0$ and $|s|\leq 1:$
$$0=G_\parallel(s,R)=-\frac{V'(x_0)\|u_\infty\|^2_{L^2}}{2}R^3+o(R^3)$$
which has only the $R=0,\ s\in\R$ in a small neighborhood of
$(s,R)=(0,0).$ However, for $V'(x_0)=0$ and $|s|\leq 1,$ we have
\begin{eqnarray}
\langle u'_\infty(\cdot-s),R^2[V(Rx+x_0)-V(x_0)]u_\infty(\cdot-s)\rangle&=&-\frac{R^3}{2}\int_\R u^2_\infty(x)V'(Rx+Rs+x_0)dx\nonumber\\
&=&-\frac{V''(x_0)}{2}R^4\int_\R u^2_\infty(x)(x+s)dx+o(R^4)\nonumber\\
&=&-R^4\frac{V''(x_0)}{2}\|u_\infty\|_{L^2}^2s+o(R^4),\nonumber\end{eqnarray}
hence
$$\tilde G_\parallel(s,R)=R^{-4}G_\parallel(s,R)=-\frac{V''(x_0)}{2}\|u_\infty\|_{L^2}^2s+o(1)=0$$
for which the implicit function theorem can be applied at
$(s,R)=(0,0)$ where $\partial_s\tilde
G_\parallel(0,0)=-\frac{V''(x_0)}{2}\|u_\infty\|_{L^2}^2\not=0.$
Note that the $o(1)$ term is differentiable with respect to $s$
because $v(s,R)$ is and it remains $o(1)$ in $R.$

The estimate
$$
\|u_R\|_{L^2}^2=\|u_\infty\|_{L^2}^2+\left(\frac{1}{2p}-\frac{3}{4}\right)V''(x_0)R^4\|xu_\infty(x)\|_{L^2}^2+o(R^4).
$$
can be obtained from $u_R=u_\infty(\cdot-s(R))+v(s(R),R),$ see
\eqref{elargedec}, which implies:
$$\|u_R\|_{L^2}^2=\|u_\infty\|_{L^2}^2+2\langle u_\infty(\cdot-s),v\rangle+\|v\|_{L^2}^2=\|u_\infty\|_{L^2}^2+2\langle u_\infty(\cdot-s),v_0\rangle+O(R^6)$$
From \eqref{perpfp} we have:
\begin{eqnarray}
\langle u_\infty(\cdot-s),v_0\rangle&=&-\langle L^{-1}_\infty u_\infty(\cdot-s), R^2[V(Rx+x_0)-V(x_0)]u_\infty(\cdot-s)\rangle\nonumber\\
&=&-\frac{1}{2}R^4V''(x_0)\int_\R (x+s)^2u_\infty(x)L^{-1}_\infty
[u_\infty] dx+o(R^4),\nonumber\end{eqnarray} and the integral term
can be computed using $-L^{-1}_\infty
[u_\infty]=\frac{1}{2p}u_\infty+\frac{1}{2}xu'_\infty.$

The main part of the theorem is now finished. It remains to prove
the spectral properties of the linear operator:
$$L_+(R)[\phi]=D_uG(u(R),R)[\phi]=-\phi''+R^2(V(Rx+x_0)-V(x_0))\phi+\phi+(2p+1)\sigma |u(R)|^{2p}\phi,$$
where $u(R)=u_\infty(\cdot-s(R))+v(s,R).$ It is well known that:
$$L_+(0)[\phi]=L_\infty(0)[\phi]=-\phi''+\phi+(2p+1)\sigma |u_\infty|^{2p}\phi,$$
has exactly one, strictly negative eigenvalue which is simple and
zero is the next eigenvalue which is also simple with corresponding
eigenvector $u'_\infty.$ Since $u(R)$ depends continuously on $R,$
$L_+(R)$ is continuous with respect to $R$ in the resolvent sense,
hence the isolated eigenvalues and eigenvectors depend continuously
on $R.$ To establish the sign of the second eigenvalue we use the
following expansion for its eigenvector:
$$\phi(R)=(1+a(R))u'_\infty(\cdot-s(R))+\Phi(R),\qquad \Phi(R)\perp u'_\infty(\cdot-s(R)),$$
and the eigenvalue equation:
$$L_+(R)[\phi(R)]=\lambda(R)\phi(R)$$
which is equivalent to:
\begin{eqnarray}
\lefteqn{\lambda(R)\phi(R)=L_\infty(s)\phi(R)}\nonumber\\
&&+\underbrace{\left[R^2(V(Rx+x_0)-V(x_0))+(2p+1)2p\sigma
|u_\infty(\cdot
-s)|^{2p-1}v(s,R)\right]}_{V(s,R)=O(R^4)}\phi(R)+DN(R)\phi(R)
\nonumber\end{eqnarray} where
\begin{eqnarray}
DN(R)&=&(2p+1)\sigma \left(|u_\infty(\cdot-s)+v(s,R)|^{2p}-|u_\infty(\cdot-s)|^{2p}-2p|u_\infty(\cdot -s)|^{2p-1}v(s,R)\right)\nonumber\\
&=&\left\{\begin{array}{ll} v(s,R) & \mbox{if } p=1/2 \\ O(R^{8p}) &
\mbox{if } 1/2<p<1 \\ O(R^8) & \mbox{if } p\geq
1.\end{array}\right.\nonumber\end{eqnarray} We project the
eigenvalue equation onto $u'_\infty(\cdot-s(R))$ and its orthogonal
complement in $L^2$ to obtain the equivalent system of two
equations:
\begin{eqnarray}
\lefteqn{\left[L_\infty(s)-\lambda(R)+P_\perp(s)V(s,R)+P_\perp(s)DN(R)\right]\Phi(R)}\nonumber\\
&=&-P_\perp(s)\left[V(s,R)+DN(R)\right](1+a(R))u_\infty(\cdot-s)\nonumber\\
\lambda(R)(1+a(R))\|u'_\infty\|_{L^2}^2&=&\langle
u'_\infty(\cdot-s(R)),V(s,R)\phi(R)\rangle+o(R^4)\nonumber
\end{eqnarray}
As before, from the first equation we deduce $\Phi(R)=O(R^4),$ while
replacing now in the second one
$\phi(R)=(1+o(1))u'_\infty(\cdot-s(R))+O(R^4),$ we get:
$$\lambda(R)=\|u'_\infty\|_{L^2}^{-2}\langle u'_\infty(\cdot-s),V(s,R)u'_\infty(\cdot-s)\rangle+o(R^4).$$
To calculate
\begin{eqnarray}
\langle u'_\infty(\cdot-s),V(s,R)u'_\infty(\cdot-s)\rangle&=&\langle u'_\infty(\cdot-s),R^2[V(Rx+x_0)-V(x_0)]u'_\infty(\cdot-s)\rangle\label{eq:Vsr}\\
&&+\langle u'_\infty(\cdot-s),(2p+1)2p\sigma |u_\infty(\cdot
-s)|^{2p-1}v(s,R)u'_\infty(\cdot-s)\rangle\nonumber\end{eqnarray} we
use $v(s,R)=v_0(s,R)+O(R^6)$ and the equation satisfied by $v_0,$
see \eqref{perpfp}:
$$L_\infty(s)v_0=-P_\perp(s)R^2[V(Rx+x_0)-V(x_0)]u_\infty(\cdot-s).$$
Taking its space derivative:
\begin{eqnarray}
\lefteqn{L_\infty(s)v'_0+(2p+1)2p\sigma |u_\infty(\cdot -s)|^{2p-1}u'_\infty(\cdot-s)v_0}\nonumber\\
&=&-R^3V'(Rx+x_0)u_\infty(\cdot-s)-R^2[V(Rx+x_0)-V(x_0)]u'_\infty(\cdot-s)\nonumber\\
&&+\langle
u'_\infty(\cdot-s),R^2[V(Rx+x_0)-V(x_0)]u_\infty(\cdot-s)\rangle
u''_\infty(\cdot-s)\nonumber
\end{eqnarray}
then its scalar product with $u'_\infty(\cdot-s)$ we get
\begin{eqnarray}
\lefteqn{\langle u'_\infty(\cdot-s),(2p+1)2p\sigma |u_\infty(\cdot -s)|^{2p-1}v_0(s,R)u'_\infty(\cdot-s)\rangle}\nonumber\\
&=&-R^3\langle u'_\infty(\cdot-s),V'(Rx+x_0)u_\infty(\cdot-s)\rangle
-R^2\langle
u'_\infty(\cdot-s),V(Rx+x_0)-V(x_0)]u'_\infty(\cdot-s)\rangle\nonumber\end{eqnarray}
which plugged into \eqref{eq:Vsr} leads to:
\begin{eqnarray}
\langle u'_\infty(\cdot-s),V(s,R)u'_\infty(\cdot-s)\rangle&=&-R^3\langle u'_\infty(\cdot-s),V'(Rx+x_0)u_\infty(\cdot-s)\rangle+O(R^6)\nonumber\\
&=&\frac{R^4}{2}V''(x_0)\|u_\infty\|_{L^2}^2+o(R^4)\nonumber
\end{eqnarray}
All in all we have:
$$\lambda(R)=\frac{1}{2}V''(x_0)R^4+o(R^4),$$
which shows that the second eigenvalue of $L_+(R)$ becomes negative
(respectively positive) if $x_0$ is a local maxima (respectively
local minima) for the potential $V(x).$ The theorem is now
completely proven.
\end{proof1}

\section{Numerical results}\label{se:num}

To illustrate the results and further investigate the behavior of
the ground state branches we have performed a series of numerical
computations on the equation
 $$-\frac{1}{2}\psi''(x) + \frac{1}{2}V(x) \psi(x) -|\psi(x)|^{2p} \psi(x)  + E \psi(x) =
 0,$$
which is equivalent with \eqref{stationary} with $\sigma =-2$ if one
regards the eigenvalue parameter $E$ in this section as being half
the parameter used in the previous sections. The potential $V\equiv
V_s$ is given by \eqref{double-potential}. When $s < s_* \approx
0.6585$, the potential is a single well but it becomes a double well
for $s > s_*$. We also recall that $p = p_* \approx 3.3028$ is the
theoretical threshold power for the nonlinearity that separates
different pitchfork bifurcations in the regime $s\rightarrow\infty,$
see Corollary \ref{cor:dwbif}. The principal conclusions of our
investigations for $p=1$, $p=3$, and $p = 5$ can be summarized as
follows:

\begin{enumerate}
\item When the potential is a single well (that is $s < s_*$),
the symmetric ground state exists for all $E > E_0$ and the operator $L_+$
at this states has a single negative
eigenvalue for all $E > E_0$. There are no bifurcations along this branch which is consistent with the result in \cite{js:ubs}.

\item When the potential is a double well (that is $s > s_*$),
the symmetric ground state exists for all $E > E_0$ and the second negative
eigenvalue of $L_+$ along the branch of symmetric states emerges for
$E > E_*$, where $E_*$ depends on $s$ and $E_* > E_0$. The
asymmetric states bifurcate at $E = E_*$ and exist for all $E > E_*$. The second
eigenvalue of $L_+$ along the branch of asymmetric states is
positive for all $E > E_*$. The numerical results are in agreement with Theorem \ref{th:main}. Furthermore they show that there are no other bifurcations along these branches past $E_*,$ that, as $E\rightarrow\infty,$ one branch of asymmetric states localizes in the left well while the other localizes in the right well, and, modulo re-scaling, they both converge to the NLS soliton localized in the left, respectively right, minima of the potential, see Remark \ref{rm:main}.

\item When $p < p_*$, the pitchfork bifurcation is supercritical and
the branch of asymmetric states has bigger $L^2$ norm than the one for the
symmetric state at $E = E_*$. When $p > p_*$, the pitchfork bifurcation is subcritical and
the branch of asymmetric states has smaller $L^2$ norm for $E \gtrapprox E_*$
than the one for the symmetric state at $E = E_*$. The numerical results are consistent with Corollary \ref{cor:dwbif} but also suggest that the separation between wells does not have to be large, i.e., as soon as the potential has two wells, the supercritical/subcritical character of the bifurcation is controlled by the nonlinearity.
\end{enumerate}

\begin{figure}
\begin{center}
\includegraphics[height=6.5cm]{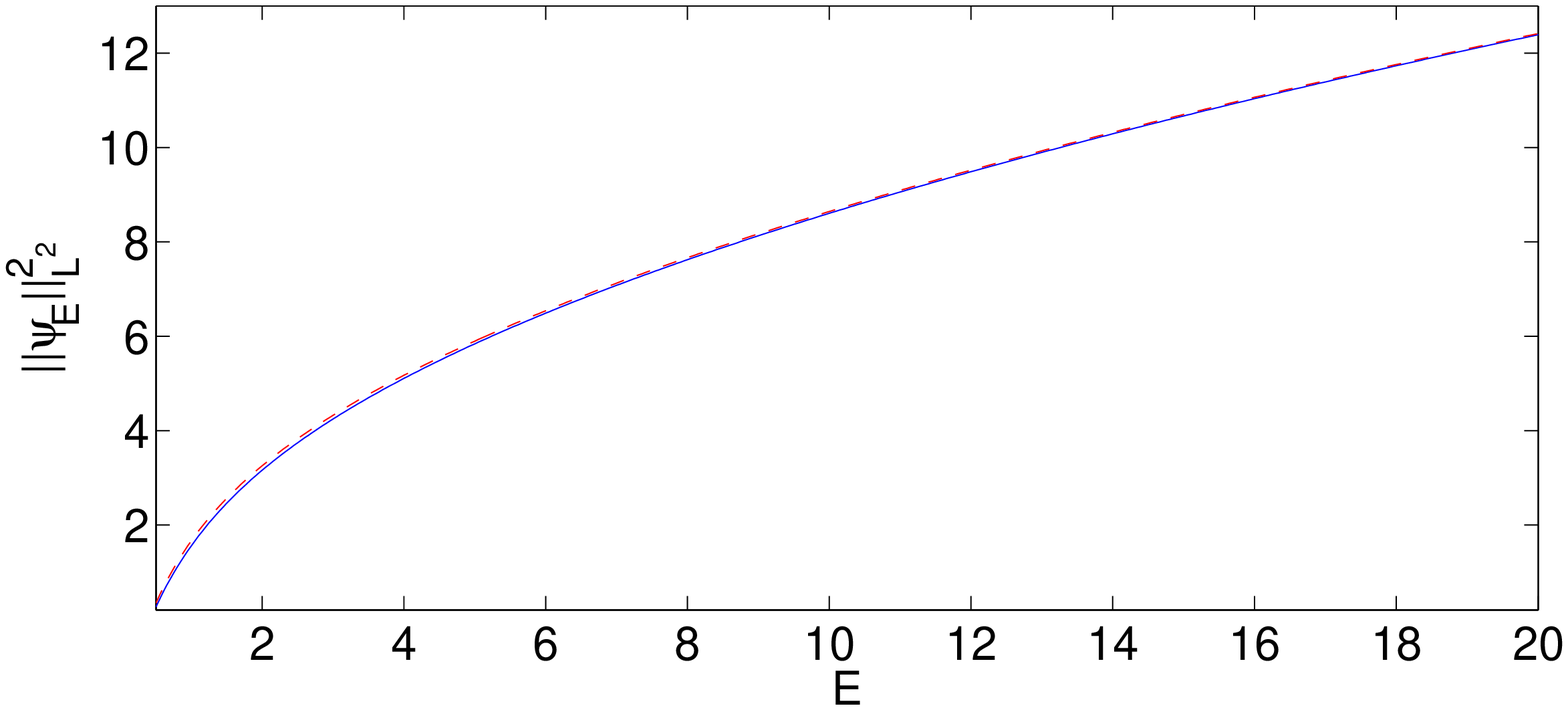}
\includegraphics[height=5.5cm]{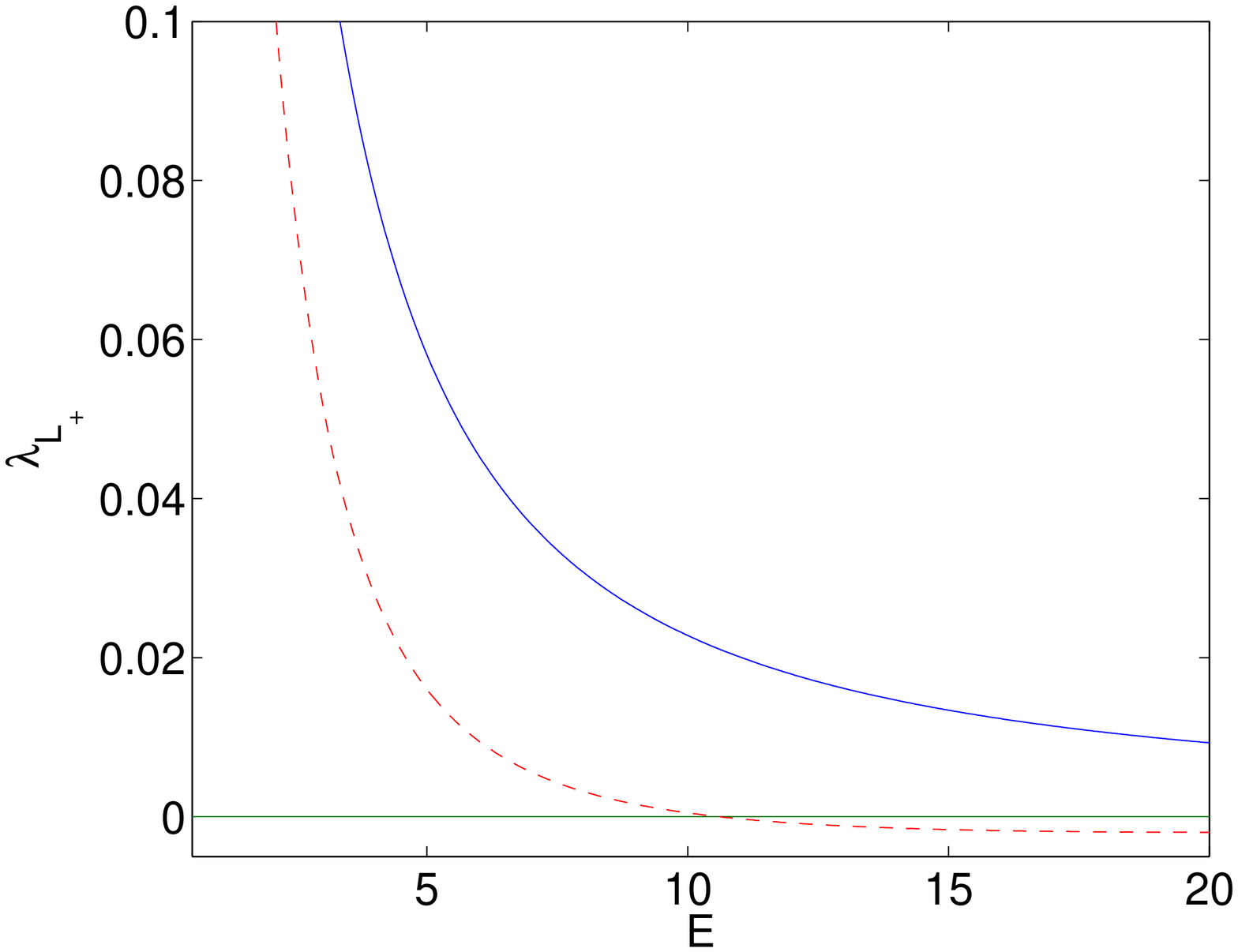}
\includegraphics[height=5.5cm]{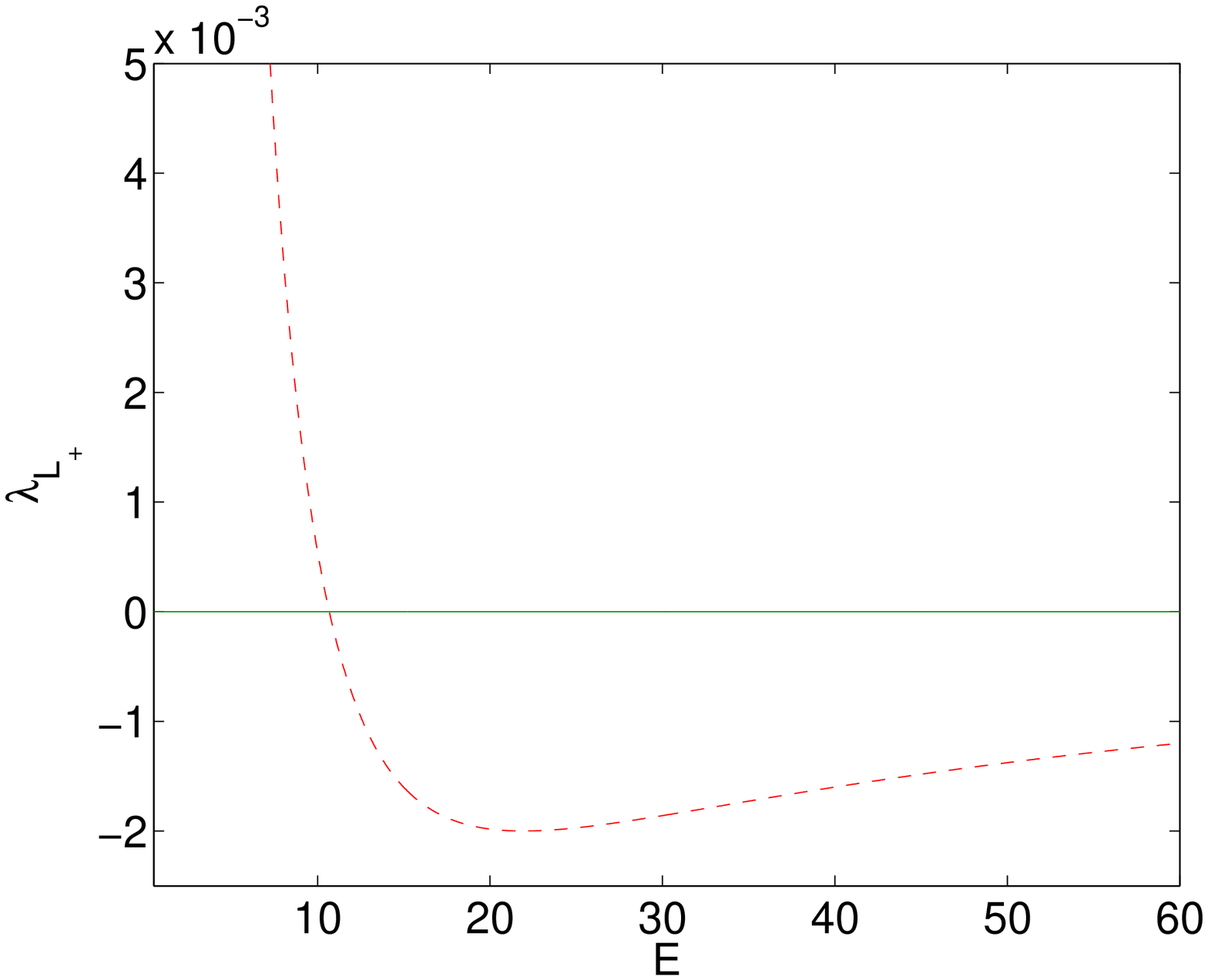}
\end{center}
\caption{The top panel shows the dependence of the squared $L^2$ norm of
the symmetric state on the parameter $E$ for $p = 1$. The bottom left panel
shows the trajectory of the second eigenvalue of $L_+$ for the cases
of $s=0.6<s_*$ (blue solid line) and $s=0.7>s_*$ (red dashed line).
The bottom right panel shows an expanded plot of the second case for
considerably larger values of $E$.} \label{efig1}
\end{figure}

The conclusions are showcased in the following five figures. Figure
\ref{efig1} illustrates the cases of $s=0.6$ and $s=0.7$, for $p=1$,
that straddle the critical point $s_* \approx 0.6585$. The top panel
presents the dependence of the squared $L^2$ norm of the symmetric
state on $E$, while the bottom panel presents the second eigenvalue
of the operator $L_+$ as a function of $E$. It is clear that for $s
< s_*$, the second eigenvalue of $L_+$ tends asymptotically to 0,
without ever crossing over to negative values (solid line), while
for $s>s_*$, such a crossing exists (dashed line), occurring for
$E_* \approx 10.68$. On the other hand, to examine whether a
secondary crossing may exist for larger values of $E$, we have
continued the $s=0.7$ branch to considerably higher values of $E$ in
the bottom right panel of the figure, observing the eventual
convergence of the eigenvalue to $\lambda=0$, without any trace of a
secondary crossing to positive values.

\begin{figure}
\begin{center}
\includegraphics[height=5.5cm]{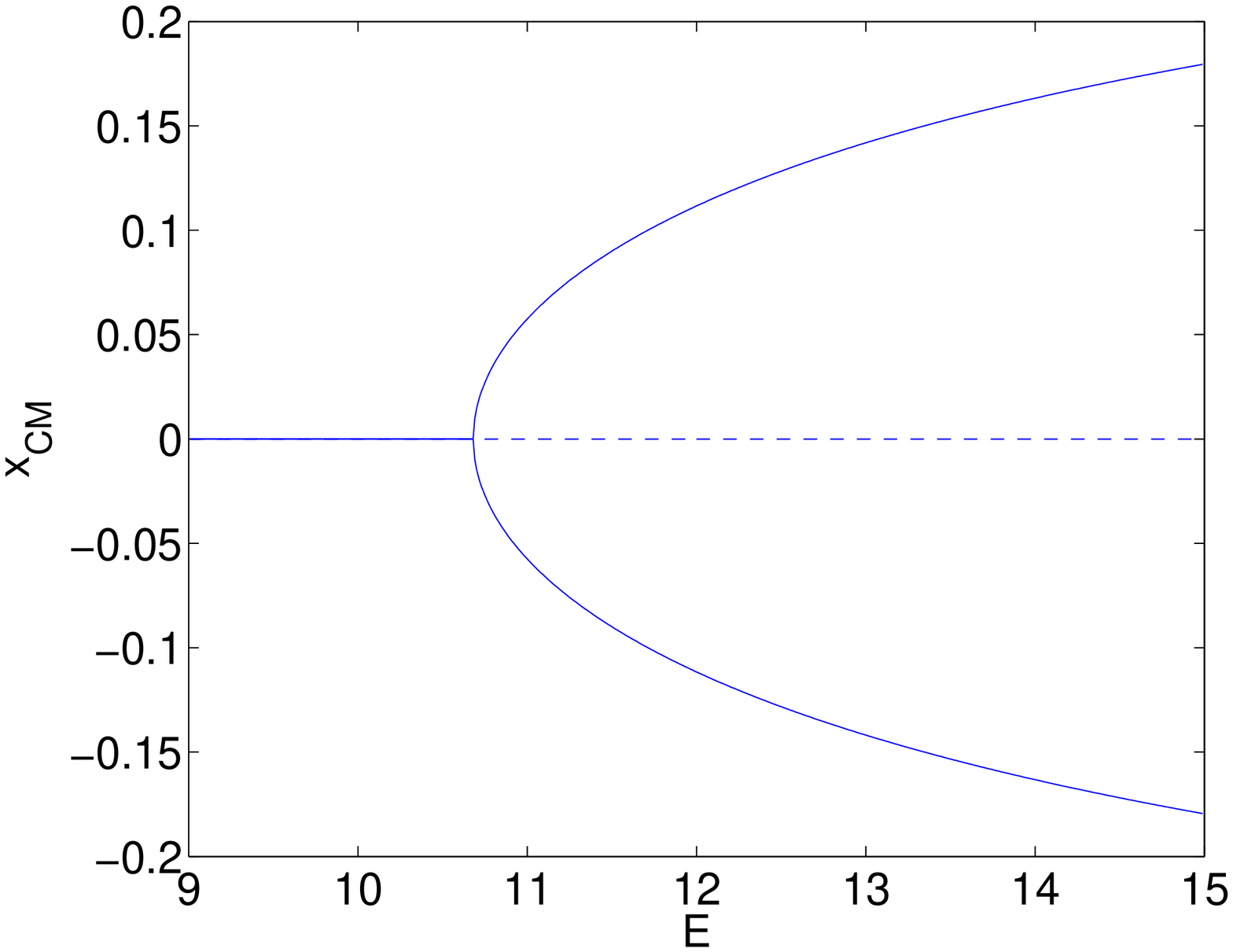}
\includegraphics[height=5.5cm]{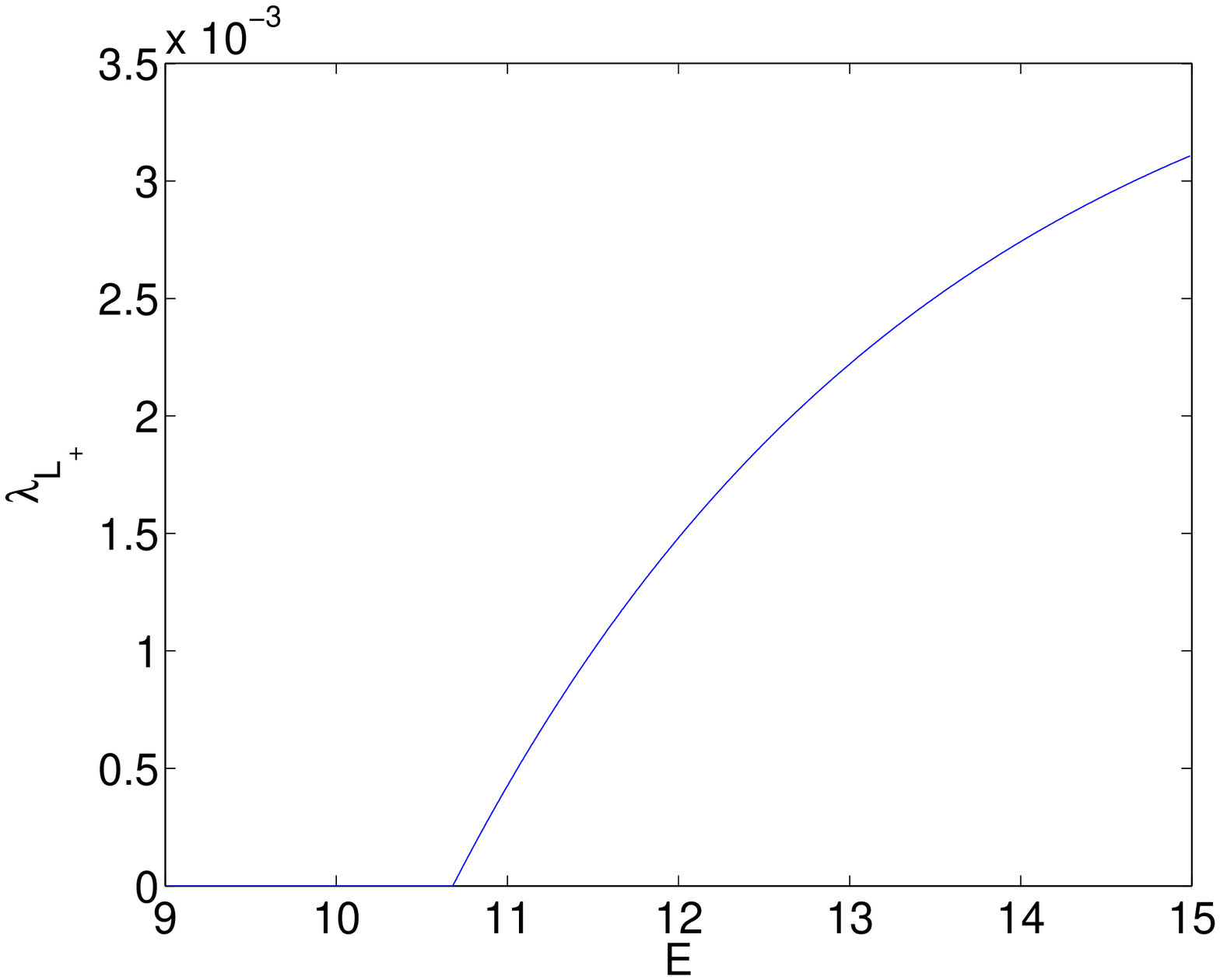}
\includegraphics[height=5.5cm]{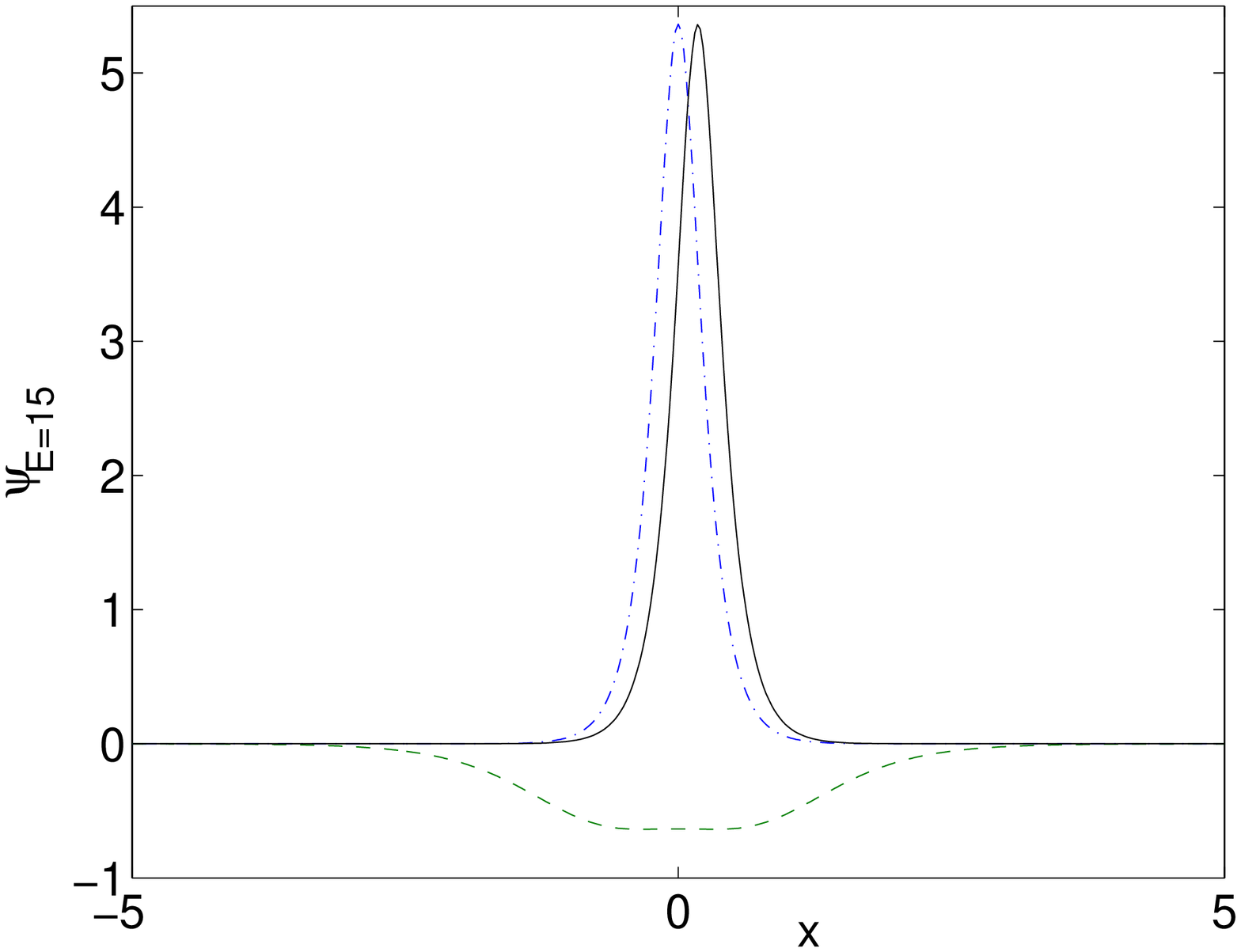}
\end{center}
\caption{The top left panel shows the pitchfork bifurcation of
asymmetric states for $s=0.7>s_*$ and $p=1$. Two asymmetric ($x_{CM}
\neq 0$) states emerge for $E > E_* \approx 10.68$. The second
eigenvalue of $L_+$ for such asymmetric states is positive as shown
in the top right panel. The bottom panel shows symmetric and
asymmetric states versus $x$ for $E = 15$.} \label{efig1a}
\end{figure}

Figure \ref{efig1a} further clarifies the bifurcation structure of
the asymmetric states for $s=0.7 > s_*,\ p=1.$ As a relevant diagnostic,
we monitor the location of the center of mass of the solution
$$
x_{CM}= \frac{\int_{\mathbb{R}} x |\psi_E|^2 dx}{\int_{\mathbb{R}}
|\psi_E|^2 dx}.
$$
We can clearly see from the top left panel that beyond the critical
threshold of $E_* \approx 10.68$, two asymmetric states with $x_{CM}
\neq 0,$ corresponding to $a<0$ and $a>0$ in Theorem \ref{th:main}, bifurcate out of the symmetric state with $x_{CM}=0$, with
the latter becoming unstable as per the crossing of the second
eigenvalue of $L_+$ to the negative values. For the asymmetric
branches with $x_{CM} \neq 0$ emerging past the bifurcation point,
the second eigenvalue of $L_+$ is shown in the top right panel of
the figure, with its positivity indicating the stability of
asymmetric states. These panels corroborate the supercritical
pitchfork bifurcation scenario $Q>0,\ R>0$ in Theorem \ref{th:bif}. The bottom panel shows both
symmetric and asymmetric states for $E = 15$.

\begin{figure}
\begin{center}
\includegraphics[height=6cm]{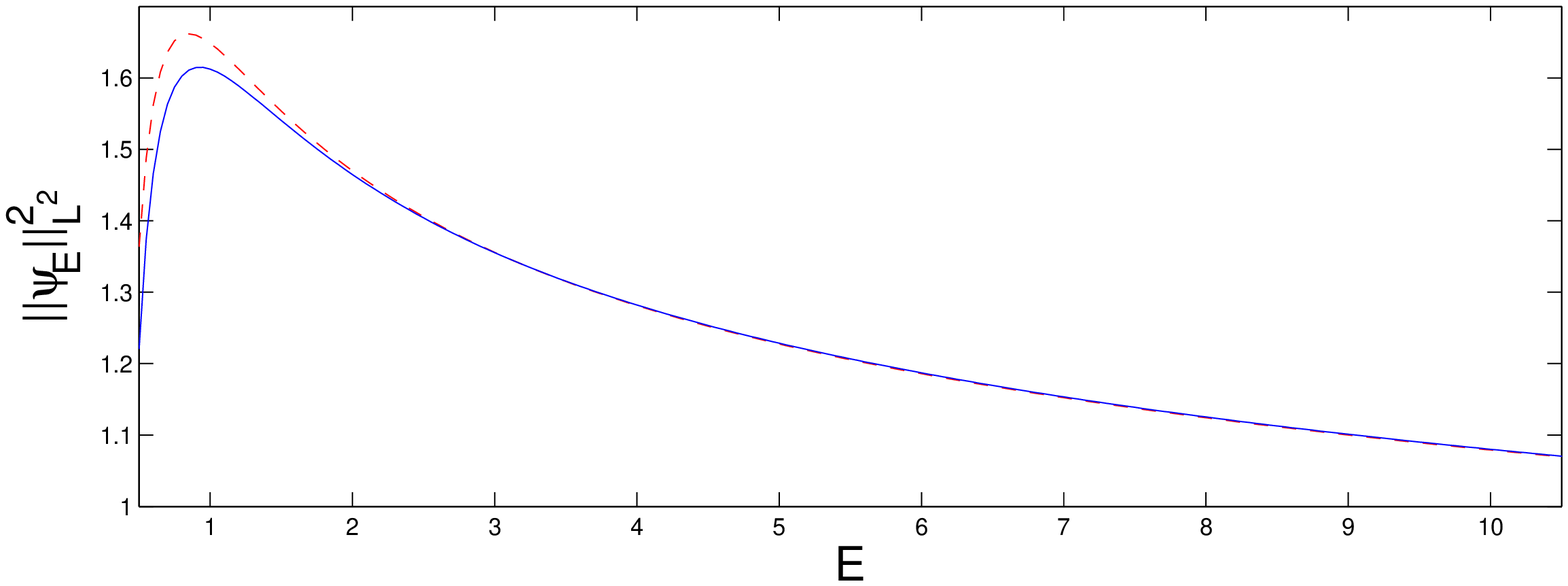}
\includegraphics[height=5.5cm]{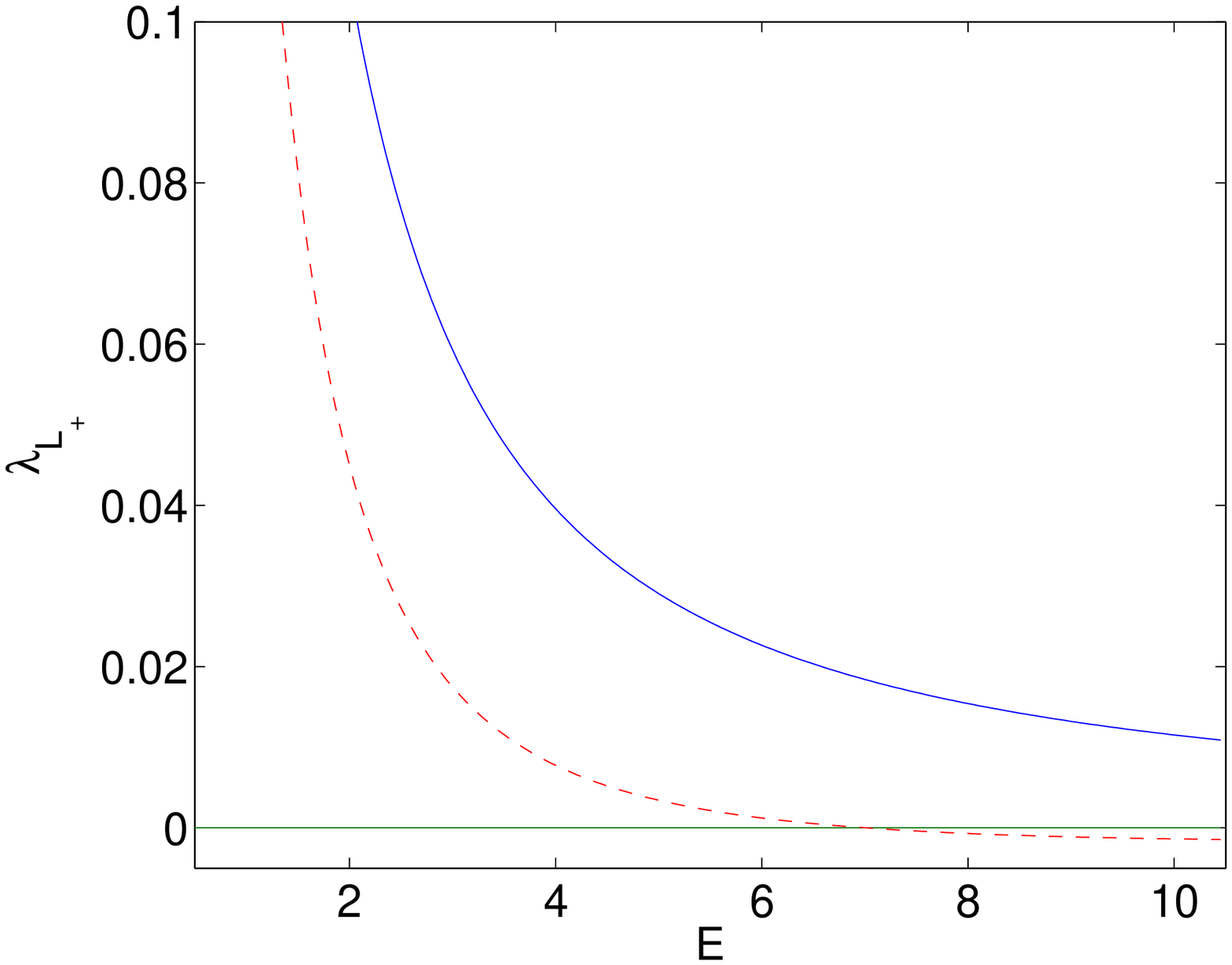}
\includegraphics[height=5.5cm]{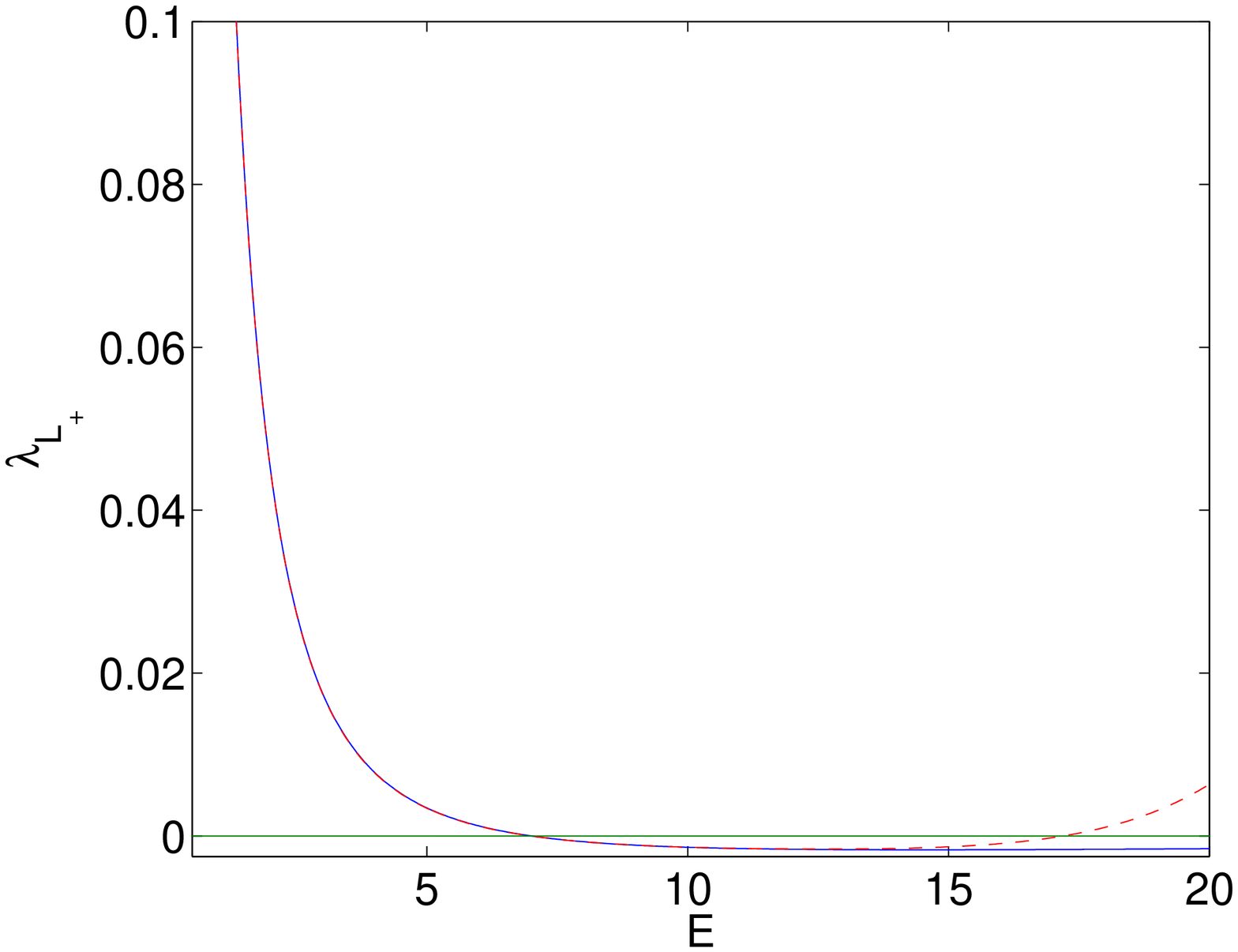}
\end{center}
\caption{The top and bottom left panels are similar to Fig.
\ref{efig1}, but now for $p=3$. The blue solid line corresponds to
the case of $s=0.6<s_*$, while the red dashed to $s=0.7>s_*$. The
bottom right panel shows the importance of sufficiently fine
discretization in resolving this second eigenvalue for large $E$.
Here the red dashed line corresponds to a spatial grid spacing of
$\Delta x=0.025$, while the solid blue line is obtained for $\Delta
x=0.0125$.} \label{efig2}
\end{figure}

The relevant computations are repeated for higher values of $p$.
The corresponding numerical results for $p = 3$ are shown in Figure \ref{efig2}.
Again illustrating the cases $s=0.6 < s_*$ and $s=0.7 > s_*$, we
observe that a crossing of the relevant eigenvalue occurs in the
latter but not in the former. Notice that in the latter case of
$s=0.7$, as shown in the bottom left panel of Fig. \ref{efig2}, the
second eigenvalue crossing occurs for $E_* \approx 7$, i.e., for a
smaller value of $E$ than in the $p=1$ case. Generally, we have
found that the higher the $p$, the earlier the relevant crossing
occurs and also the more computationally demanding the relevant
numerical problem becomes, as the solution narrows and it becomes
challenging to appropriately resolve it even with a fairly fine
spatial grid for large values of $E$. This is clearly illustrated in
the bottom right panel of Fig. \ref{efig2}, where it can be seen
that in the absence of sufficient resolution (dashed line for larger
spacing of the spatial discretization),  a {\it spurious} secondary
crossing is observed for the second eigenvalue of $L_+$. This
secondary crossing is eliminated by finer discretizations (solid
line).

\begin{figure}
\begin{center}
\includegraphics[height=10cm]{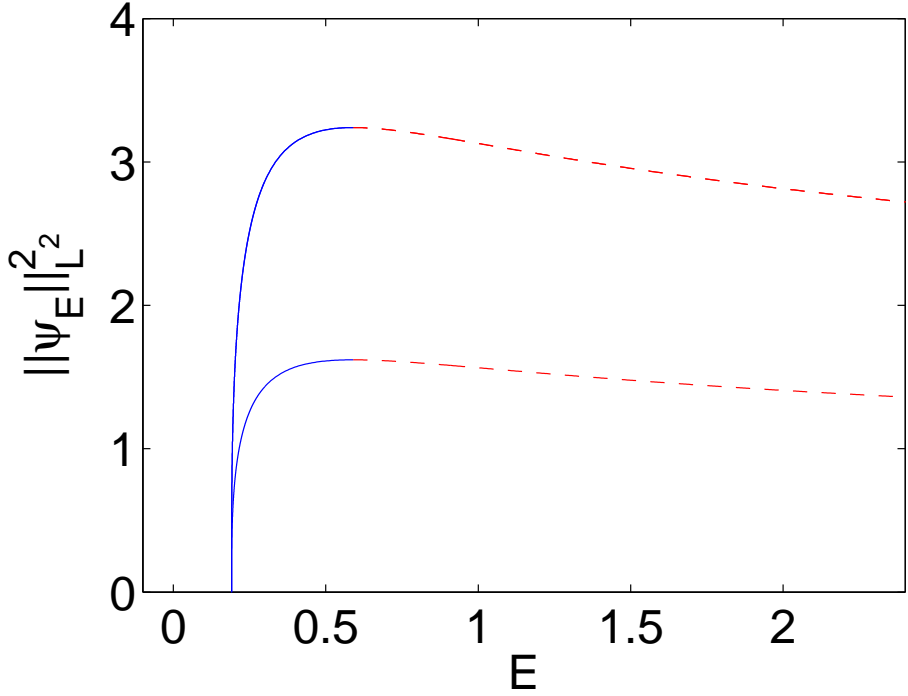}
\end{center}
\caption{The graph shows the dependence of the squared $L^2$ norm of
the stationary states on the parameter $E$ for $p = 3$ and $s=10.$ The left most (blue solid) line are the anti-symmetric (exited) states emerging from zero at the second lowest eigenvalue $E_1=0.191042.$ Almost on top of it is the (blue solid) line of symmetric ground states emerging from zero at $E_0=0.191046.$ The latter bifurcates at $E_*\gtrapprox E_0$ into the dashed red line (symmetric states) and the right most blue solid line (asymmetric states). At $E\approx 0.5908$ the slope of the $L^2$ norm becomes negative for all branches.The solid lines denote linearly stable branches, while
dashed ones denote unstable branches.} \label{efig2a}
\end{figure}

It should be pointed out that the symmetric states
become unstable in the supercritical case $p = 3$ for $E \approx
0.95,\ s=0.6$ and $E \approx 0.85,\ s=0.7$, due to the change of the
slope of $N(E)=\|\psi_E\|_{L^2}^2$ (see the top
panel of Fig. \ref{efig2}). Therefore, the asymmetric states bifurcating from the symmetric ones at $E_* \approx 7$ will also be orbitally unstable, because $Q>0, \lambda'(E_*)<0,$ and $N'(E_*)<0$ will imply $R<0$ in Theorem \ref{th:bif}. However, as $s$ becomes larger,
the value of $E_*$ becomes smaller and $E_* \to E_0$ as $s \to \infty,$ see Corollary \ref{cor:dwbif}. Figure \ref{efig2a} shows the dependence of the squared $L^2$ norm for the symmetric, asymmetric, and anti-symmetric stationary states for p = 3 and s = 10. The pitchfork bifurcation occurs while the slope of $\|\psi_E\|_{L^2}^2$ is still positive and leading to orbitally stable asymmetric states and a change of stability along the symmetric states from stable for $E<E_*$ to unstable for $E>E_*$ as stipulated in Corollary \ref{cor:dwbif}. Note that the numerical simulations suggest that the asymmetric branches can be continued for all $E>E_*,$ hence the slope of their $L^2$ norm square will change for large $E$ and they will become unstable, see Fig. \ref{efig2a} and Remark \ref{rm:main}.

\begin{figure}
\begin{center}
\includegraphics[height=5cm]{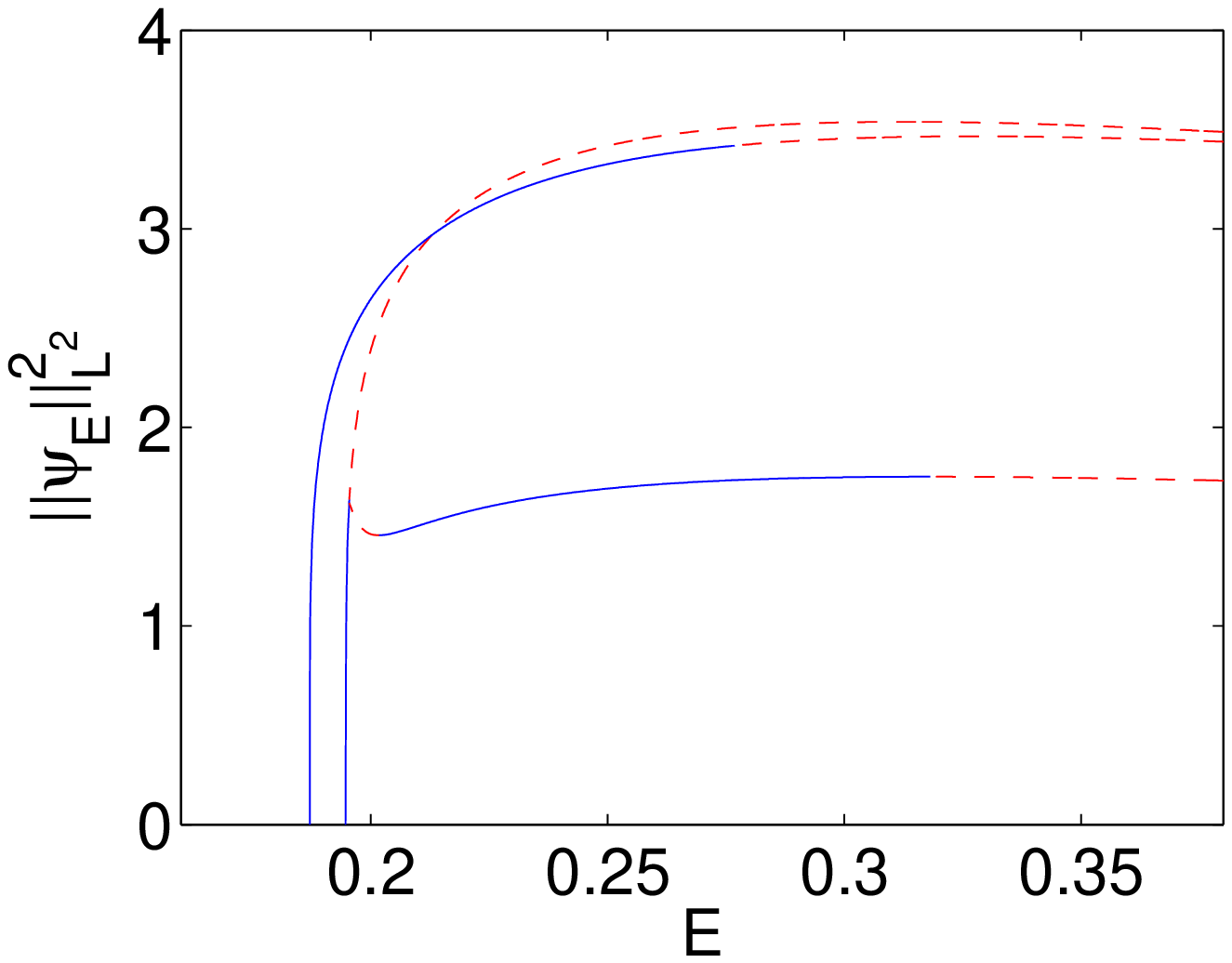}
\includegraphics[height=5cm]{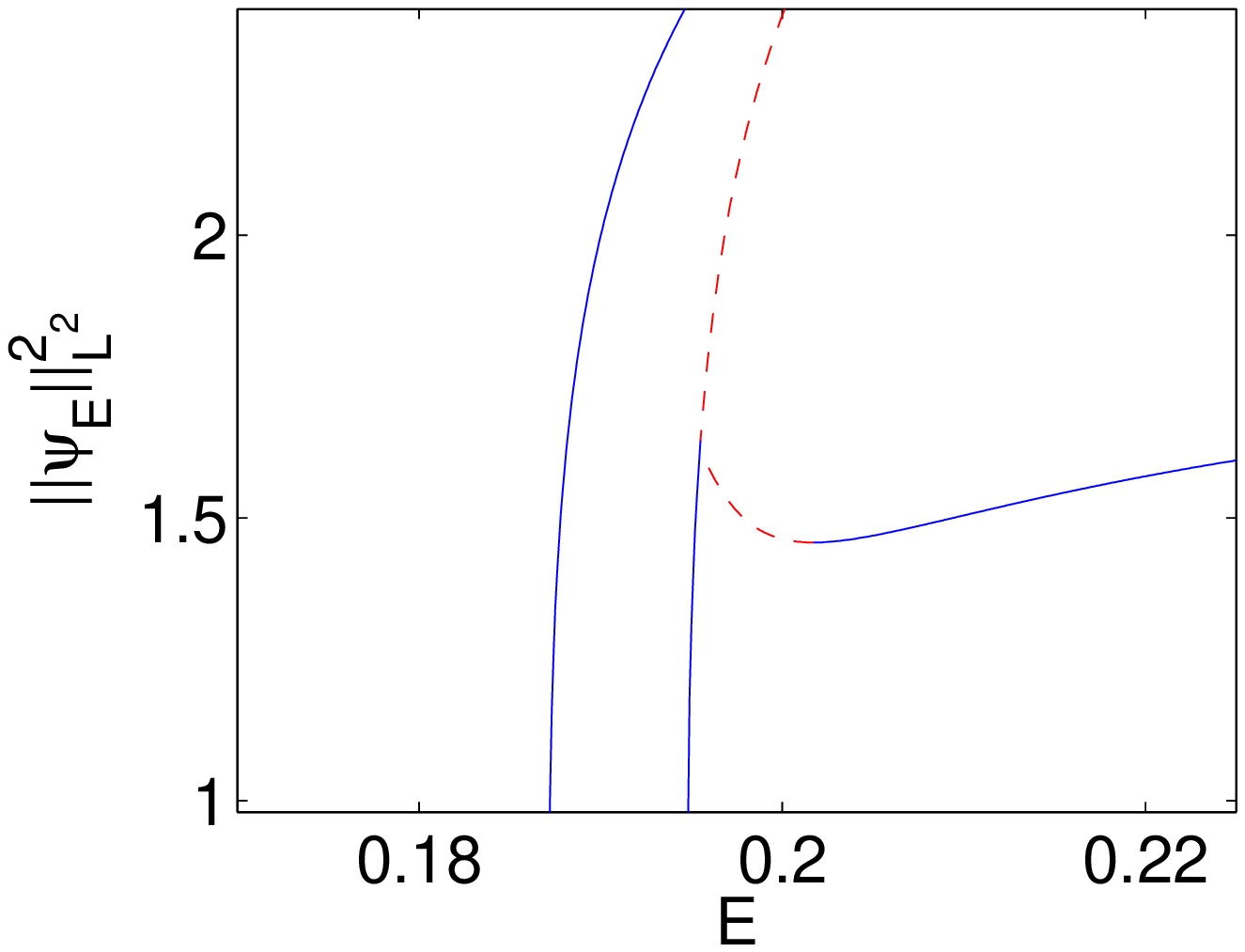}
\includegraphics[height=5cm]{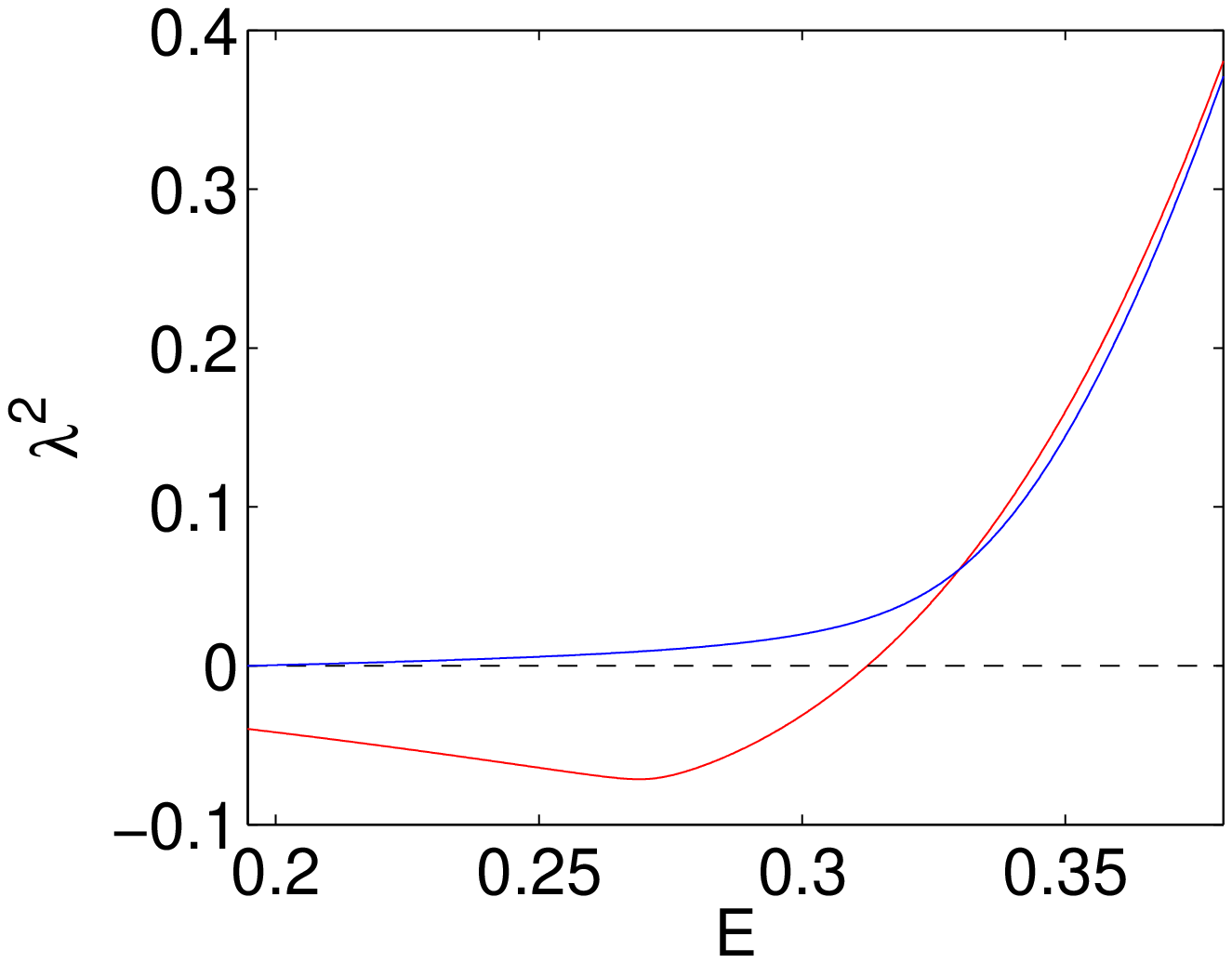}
\includegraphics[height=5cm]{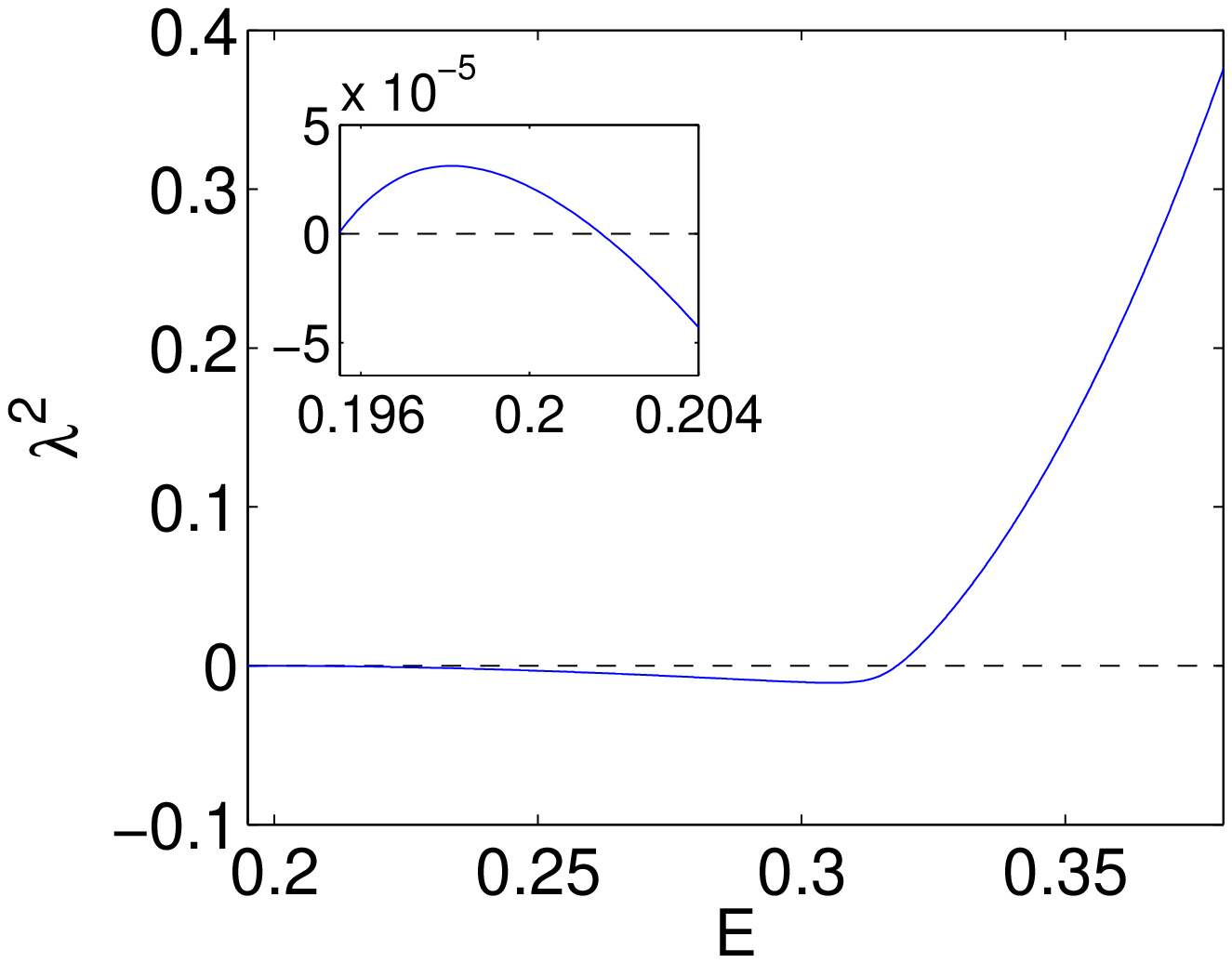}
\end{center}
\caption{The top panels show dependence of the squared $L^2$ norm of
the stationary states on the parameter $E$ for $p = 5,\ s=4$, where the right panel is a blowup
of the left panel. The leftmost branch is the anti-symmetric (excited) states  and
the other branch corresponds to the symmetric ground states which bifurcate into asymmetric states.
The bottom panels show the squared eigenvalue of the linearization spectrum
associated with the symmetric (left) and asymmetric (right) branches. The insert on the bottom
right panel gives a blowup of the figure to illustrate instability of asymmetric states near
the subcritical pitchfork bifurcation. The solid lines on the
top panels denote linearly stable branches, while
dashed ones denote unstable branches.} \label{efigNew}
\end{figure}

The numerical results for $p = 5 > p_*$ and $s = 4 > s_*$
are shown in Figure \ref{efigNew}. In this case, the subcritical pitchfork bifurcation
occurs at $E_* \approx 0.196$ at the positive slope of $\| \psi_E \|^2_{L^2}$ with respect to
$E$. The top panels show the behavior of squared $L^2$ norms for symmetric, anti-symmetric,
and asymmetric stationary states. The blowup on the top right panel shows that the asymmetric states have decreasing
$L^2$ norm for $E_* < E < E^*,$ corresponding to the case $Q>0,\ R<0$ in Theorem \ref{th:bif}, and in agreement with Corollary \ref{cor:dwbif}. However, their $L^2$ norm becomes increasing for $E > E^*$, where $E^* \approx 0.202,$ hence the asymmetric states are orbitally stable in this regime.
The bottom panels show the squared eigenvalues of the stability problem
associated with the symmetric (left) and asymmetric (right) states. The symmetric state
is unstable for any $E > E_*$ (because of the second negative eigenvalue of the operator $L_+$).
It becomes even more unstable for $E > \tilde{E}_*$, where $\tilde{E}_* \approx 0.32$,
when another unstable eigenvalue
appears (because of the negative slope of the $L^2$ norm). The
asymmetric state is unstable for $E_* < E < E^*$ (because of the negative slope of
the $L^2$ norm) and stable for $E > E^*$. We note that the asymmetric state becomes unstable
past $E\approx 0.32$ because of the negative slope of the $L^2$ norm,
similarly to the branch of symmetric states and consistent with Remark \ref{rm:main}.

While the results presented herein formulate a relatively
comprehensive
picture of the one-dimensional phenomenology in the context
of double wells, some questions still remain open for future
investigations. One of them, raised in Remark \ref{rm:main1}, is associated
with states of the form of $u_{\infty}$
spatially concentrated at $\pm \infty$. Our numerics for the potential \eqref{double-potential} has
not revealed such states presently, but it would be relevant
to lend this subject separate consideration. Additionally, we have not seen the case $Q<0$ in Theorem \ref{th:bif}, i.e. the eigenvalues $E$ are decreasing along the asymmetric branch.
Another important question concerns the generalization
of the results presented herein to higher dimensional
settings. There, the bifurcation picture is expected to be
more complicated when the eigenvalues crossing zero are not simple.

\end{document}